\documentclass[prb,superscriptaddress,amsmath,amssymb,twocolumn]{revtex4-2}
\pdfoutput=1
\usepackage{physics}
\usepackage{lipsum}
\usepackage{soul}
\usepackage{booktabs}
\usepackage{natbib}
\usepackage{float}
 \usepackage{subfloat}
\usepackage[english]{babel}
\makeatother
\usepackage{layouts} 
\usepackage{graphicx}
\usepackage{caption}
\usepackage{subcaption}
\usepackage[usenames,dvipsnames]{color}
\usepackage{stackengine}
\usepackage[T1]{fontenc}
\usepackage[colorlinks=true,allcolors=blue]{hyperref}

\begin{document}

\title{Weakly interacting one-dimensional topological insulators: a bosonization approach
}
\author{Polina Matveeva}
\affiliation{Department of Physics, Bar-Ilan University, Ramat Gan, 52900, Israel}
\author{Dmitri Gutman}
\affiliation{Department of Physics, Bar-Ilan University, Ramat Gan, 52900, Israel}
\author{Sam T.~Carr}
\affiliation{School of Physics and Astronomy,  University of Kent, Canterbury CT2 7NH, United Kingdom}

\begin{abstract}
We investigate the topological properties of one-dimensional weakly interacting topological insulators using bosonization.  To do that we study the topological edge states that emerge at the edges of a model realized by a strong impurity or at the boundary between topologically distinct phases.  In the bosonic model,  the edge states are manifested as degenerate bosonic kinks at the boundaries.  We first illustrate this idea on the example of the interacting Su-Schrieffer-Heeger (SSH) chain.   We compute the localization length of the edge states as the width of an edge soliton that occurs in the SSH model in the presence of a strong impurity.  Next,  we examine models of two capacitively coupled SSH chains that can be either identical or in distinct topological phases.  We find that weak Hubbard interaction reduces the ground state degeneracy in the topological phase of identical chains.  We then prove that similarly to the non-interacting model,  the degeneracy of the edge states in the interacting case is protected by chiral symmetry.   We then study topological insulators built from two SSH chains with inter-chain hopping,  that represent models of different chiral symmetric universality classes.   We demonstrate in bosonic language that the topological index of a weakly coupled model is determined by the type of inter-chain coupling,  invariant under one of two possible chiral symmetry operators.      Finally,  we show that a general one-dimensional model in a phase with topological index $\nu$ is equivalent at low energies to a theory of at least $\nu$ SSH chains.  We illustrate this idea on the example of an SSH model with longer-range hopping. 
\end{abstract}

\maketitle
\section{Introduction} 
Topological phases are the states of matter that do not obey standard Landau symmetry breaking classification \cite{landau_statistics}.  They are characterized by topological properties that are global and robust with respect to perturbations.  A prominent example is the Integer Quantum Hall Effect,  that occurs in a two dimensional metal in a perpendicular magnetic field. The Hall conductance is quantized in such a system and is determined by a non-trivial first Chern number  
related to the number of filled Landau levels \cite{Thouless1982}.   The value of the Hall conductance is  thus robust with respect to perturbations, that do not close the bulk gap.  

After the discovery of the Integer Quantum Hall effect there have been a number of the topological models that do not rely on the presence of a magnetic field.  In two dimensions these include the famous  Haldane model of the Anomalous Hall effect \cite{Haldane1988}  and the Kane Mele model of a $\mathbb{Z}_2$  topological insulator \cite{KaneMele2005}.  There have also been numerous theoretical \cite{Fu2007,Qi2008,Roy2009} and experimental \cite{Hsieh2009,Hsieh2008,Xia2009,Zhang2009,ZhangHaijun2009,Chen2009} studies  of three-dimensional topological insulators. 

 Turning to one-dimension,  one of the best-known topological models is the Su-Schrieffer-Heeger (SSH) model \cite{SSH1979} which was originally proposed to describe properties of polyacetylene molecules,  but was later realized to be a prototype of a symmetry-protected topological phase \cite{Ryu2006}.  Another famous one-dimensional topological model is the Kitaev chain \cite{Kitaev2001} that describes the simplest $p-$ wave superconductor.  It attracted much attention due to topologically protected Majorana edge modes that serve as a prospective qubit realization.  

The non-interacting phases are well studied and understood,  and their topological classification based on non-unitary symmetries is completely established \cite{Zirnbauer1996, Altland1997,Kitaev2009,Ryu2010}.  It turns out that the electron-electron interaction enriches the physics in many ways,  in particular:  
\begin{enumerate}
\item  In the weak interaction limit interactions may change the non-interacting classification,  although the states remain adiabatically connected to the non-interacting ones.   
\item  Interactions may lead to new topological phases that can not be continuously connected to non-interacting ones.  
\end{enumerate}
In the second class,  the most known example is the fractional quantum Hall effect  \cite{Tsui1982}.  
In this system  interactions give rise to  correlated states,  with  non-Abelian statistics and quasi-particles that carry fractional charge.   The states of FQHE can be classified in accordance with their topological order,  based on 
 their entanglement entropy \cite{Kitaev2006,Levin2006},  K-matrix \cite{Wen1992,Blok1990,Read1990} and group cohomology approach  \cite{Wen2013,Wen2014}.  This remains a rich area for contemporary research,  as the general classification of possible phases remains to be discovered.  In one dimension,  however,  these principles have been established for the  symmetry protected topological phases \cite{Gu2009,Turner2011}.  They are based on the symmetry  group of the fermionic model,  with the topological phases classified according to its projective representations.  Note,  however,  that the symmetry of the Hamiltonian of an interacting model is not necessarily  
the same as the symmetry of its low energy fixed point.   It can be lower (spontaneous symmetry breaking),  or  higher (dynamical symmetry enhancement) than that of the fermionic Hamiltonian.  In the later case the  emergent  symmetry may protect the topological phase  \cite{SantosGutmanCarr2019}.
 
We now specialize to the case of weakly interacting models, which have a non-trivial topology even before interactions are added.   In all dimensions the classification of fermionic topological phases is known in  
 this limit \cite{MorimotoMudry2015, You2014}.  In general terms, for the topological insulators in an odd number of dimensions that are in the $\mathbb{Z}$ classes in the non-interacting case is reduced to $\mathbb{Z}_n$,  where $n = \{2,4,8,16,32\}$ depending on the dimensionality and the symmetry class.  In an even number of dimensions including the Quantum Hall effect in $d=2$ weak interactions do not change the classification.

We now focus on the case of one dimension,  which is the topic of this paper.  The most know example of reduction of topological classification  are  weakly interacting stack of Majorana chains \cite{FidkowskiKitaev2010,FidkowskiKitaev2011}.   In particular,  it was demonstrated  that the non-interacting classification of Majorana chain model reduces from $\mathbb{Z}$ to $\mathbb{Z}_8$ which stimulated the search for the general classification described above.  There are also a handful of other interacting models such as the Haldane model \cite{Haldane19831, Haldane19832} and closely related AKLT model \cite{Affleck1987} which are known to have a $\mathbb{Z}_2$ topological classification.  At first sight these seem very different from the Majorana chains,  it was later shown that there is a mapping between them \cite{Verresen2017}.   

Even for weak interactions where this classification is formally established it is useful to have concrete interacting models where one can quantitatively derive properties rather than just relying on the adiabatic continuity and general topological principles.  A very standard way to treat interactions in one dimensions is to employ the bosonization technique \cite{gogolin2004bosonization,Giamarchi_book}.  Our aim is to distinguish between different topological phases based on the low energy properties of the model captured by this description.  

In our work we focus on examples built from coupled SSH chains. 
The  non-interacting limit of such models was previously studied in details 
in \cite{Matveeva2023},  which further allows us to test  the validity of the bosonization approach.  In particular, we are interested in the description of phases with winding numbers larger than one.   The description of the gapped phases is not trivial within bosonization even in the non-interacting limit,  and requires an accurate lattice regularization and a proper treatment of the ultraviolet  singularities.
However once a description of the non-interacting phases is accomplished within this description,  the incorporation of interaction effects becomes trivial.

The effect of interactions on a single SSH chain was first investigated in \cite{Sirker2014} via numerical techniques where they showed that the topological phase is stable up to a critical interaction.  This has been confirmed by various other studies \cite{Yahyavi_2018,Lin2020,Marques2017}.   A similar type of phase transition was found  in  \cite{Nersesyan2020} for the model of two strongly coupled interacting SSH chains in the vicinity of a gapless critical point.  In this limit the coupled chain model becomes equivalent to a single chain. 

The edge states in a single SSH chain were recently explored by bosonization in  \cite{Giamarchi2023}.   In this work it was shown that bosonization can indeed accurately describe the edge states in the topological phase of the SSH model,  by treating the open boundary conditions carefully.  However this lead to a rather complicated bosonic model that is not trivial to extend to coupled SSH chains. In our work we will show that we reproduce the same results by using simplified boundary conditions.  

The Hubbard SSH model may be thought of as the simplest generalization of an interacting SSH model.  It consists of two identical SSH chains coupled via Hubbard interaction.  This was first studied using perturbative Greens function methods in \cite{Gurarie2012},  where it was demonstrated that the interaction reduces the degeneracy of the edge states from 16 to 4.  Since this pioneering work this model has been studied extensively \cite{Tsuchiizu2004,Benthien2006,Ejima2016,Ejima2018,Verresen2017} agreeing with the  perturbative results of \cite{Gurarie2012} and exploring the myriad of possible strong coupling phases.  To the best of our knowledge this has never been studied using bosonization.

The models discussed above all fall into the BDI class of the topological classification.  By adding hopping between two SSH chains one can create models in the other topological classes in one dimension \cite{Matveeva2023}.  In this work it was also shown that all the $\mathbb{Z}$ classes are equivalent in the sense that one can adiabatically transform one into another without changing the number of topologically protected edge states.   One can ask the question whether it is also true when the interactions are present.

In this work we develop a bosonization technique to study weakly interacting SSH-like models.  In contrast to \cite{Giamarchi2023} we treat the open boundary conditions simply as a strong impurity which amounts to relatively straightforward boundary conditions in the bosonised language.  We show for a single SSH chain (winding number $\nu=1$) this reproduces the results found in  \cite{Giamarchi2023}.  We then extend these results to two SSH chains (winding number $\nu=2$) coupled first via interactions (the SSH Hubbard model) and then also including hopping between the chains.  The bosonization reproduces the known reduction of the degeneracy in the edge states for all of these cases.  Finally we look at an extended SSH chain which has a phase with winding number $\nu=2$ . Despite this being a single chain we show that it can be mapped onto the previously studied ladder models and thus interactions give the same degeneracy reduction.  

While the majority of the results of this paper have been derived in the past by different methods,  the demonstration that it can been seen with bosonization paves the way to study the topological properties of more exotic strongly interacting phases that are not easily linked to the non-interacting models.  It also allows a framework to understand topological metallic phases previously described in bosonization \cite{Kainaris2015, KainarisCarrGutman2017,SantosGutmanCarr2016,SantosGutmanCarr2019,Mirlin2018,Keselman2015}.   

Let us overview the structure and results of our paper in more detail.   
 \begin{itemize}
 \item In Section \ref{Section_InteractingSSH} we study the  interacting  SSH chain.    By using bosonization we show that for strong repulsive interactions,  the system undergoes an Ising phase transition to the charge density wave (CDW) phase with spontaneously broken $\mathbb{Z}_2$ symmetry.   This is consistent with the numerical results of \cite{Sirker2014}. 
 \item  In Section \ref{sec:edge states} we study the edge states in the interacting SSH model.   We find that the edge states are stable with respect to weak interactions.  We compute their localization length as a function of interactions and show that it is not monotonic.   These results are consistent with the results of \cite{Giamarchi2023} where the bosonization with open boundaries is used.   Furthermore, we incorporate the umklapp term into the analysis and demonstrate that in the case of repulsion,  it leads to an increase in the localization length.  This observation aligns with the fact that repulsive interactions tend to reduce the energy gap in the system.   We also show within a bosonic description that the degeneracy of the edge states is protected by chiral symmetry,  similar to the non-interacting case.  
 \item   In Section \ref{Sec:uncoupledSSH} we consider the properties of two SSH chains coupled by interaction in the cases when they are in the same topological phases and when one of the chains is in a trivial phase and the other is in a topological one.   By studying bosonic phase diagram we demonstrate how the many-body degeneracy of the edge states is reduced in the case of capacitively coupled chains, which is consistent with results obtained earlier.  By deriving the bosonic description of symmetry operators for a two-chain model we demonstrate that chiral symmetry protects the degeneracy of the edge states.  We also show that for a given sign of interactions topological degeneracy of the ground state may be protected by different symmetries rather than chiral.  
 \item In Section \ref{sec:coupled_chains} we add inter-chain hopping to the model.  In the absence of interactions,  such a model describes topological insulators in all chiral universality classes that we constructed before in \cite{Matveeva2023}.  We show in the bosonic language that winding number of a coupled model is determined by chiral symmetry only,  and breaking of other symmetries does not affect its value.  
 \item In the last Section \ref{Sec:extendedSSH} we return to a single SSH chain but with longer range hopping that has a phase with winding number of 2.  Unlike the coupled chain models with four bands this model has only two and therefore it is not obvious how to see additional edge states in bosonization.  The resolution to this is that this model has four Fermi points along the relevant phase transition line,  and hence maps at low energy to the coupled chain models we studied before.   We also show that this can be easily generalized to models with arbitrary winding numbers. 
 \end{itemize}

\section{Bosonic description of SSH model}\label{Section_InteractingSSH}
In this section,  we first review the topological properties of the non-interacting SSH model.   Next we add interactions,  bosonize the full model,  and analyze its bulk ground state in the absence of edges. 

\subsection{Non-interacting SSH model}
We consider a non-interacting SSH model,  described by the following Hamiltonian: 
\begin{align}
\label{SSH_tight_binding}
\hat{H}_{\text{SSH}}= w \sum_n c^{\dagger}_{n,\text{A}}c_{n, \text{B}} + v \sum_n c^{\dagger}_{n,\text{B}} c_{n+1, \text{A}}, 
\end{align}
where $w, v$ are real.  The SSH model describes a one-dimensional chain of atoms A and B,  connected by a dimerized hopping amplitude.  This model has chiral (sublattice) symmetry  because there is no hopping between atoms belonging to the same sublattice.   This allows us to define an integer topological invariant -- the winding number $\nu$ \cite{Ryu2010},  that distinguishes between two topologically inequivalent phases of the model.  In order to define it we write the Hamiltonian (\ref{SSH_tight_binding}) in momentum space as: 
\begin{align}
\label{SSH_tight_binding_k}
\hat{H}_{\text{SSH}}= \sum_k c^{\dagger}_k h_{\text{SSH}} c_k,   \nonumber \\
h_{\text{SSH}} = \begin{pmatrix}
0 & \Delta(k) \\ 
\Delta^*(k) &0
\end{pmatrix},  \Delta(k) = w+ve^{ik},
\end{align}
where the Hamiltonian is written in the basis  $c_k=\{c_{k,A},c_{k,B}\}^\text{T}$.
The winding number is given by the following integral over the Brillouin zone \cite{Matveeva2023}: 
\begin{equation}
\label{phase_detq}
\nu=\frac{-i}{2\pi}\int\limits_{\text{BZ}} \partial_k i\phi(k) dk,
\end{equation}
where $\phi(k)$ is defined as the complex phase  of $ \Delta (k)$ \footnote{For multiple bands $\phi(k)$ is defined as the complex phase $ \det \Delta (k)$.  }.

For the SSH model  (\ref{SSH_tight_binding}) there are two possible phases:  $\nu=1$ for $|w|>|v|$ known as the topological phase and $\nu=0$ for $|w|<|v|$ known as a trivial phase.  

In order to bosonize the model we require a metallic phase to begin with.  To do that,  we rewrite the Hamiltonian (\ref{SSH_tight_binding}) in terms of the parameters $t=(w+v)/2$ and $\delta t = (v-w)/2$ to identify the gapless part of the model,  proportional to $t$ and gap opening terms proportional to $\delta t$:    
\begin{align}
\label{SSH_t_deltat}
\hat{H}_{\text{SSH}}= t \sum_j c^{\dagger}_{j}c_{j+1} + \delta t \sum_j (-1)^j c^{\dagger}_{j} c_{j+1} + \text{h.c.}
\end{align}
If $\delta t=0$ the spectrum of the model $\epsilon(k)= 2t\cos(k)$ is gapless at half filling and can be linearized around the Fermi points $\pm k_F$,  where $ k_F =  \pi/2a_0$ and  $a_0$ is the lattice constant.  The effective fermionic operator can be expanded in terms of right- and left-moving modes smooth on the scales of the inverse Fermi momentum:  
\begin{align}
\label{RL_fermions}
c_j \rightarrow \Psi(x) = e^{ik_F x} R (x) + e^{-ik_F x} L(x).
\end{align}
As a detail , the interpretation of $R(x)$ and $L(x)$ as right- and left-movers requires $t<0$.  This is unimportant as the physics is equivalent for $t>0$.  

Next,  we use standard bosonisation conventions for spinless fermions, see Appendix \ref{bosonization_spinless},  and obtain the following Hamiltonian written in terms of the canonical bosonic fields $\phi(x)$ and $\Pi(x)$:      
\begin{align}
\label{SSH_bosonised}
\hat{H}_{\text{SSH}}= \frac{v_F}{2} \int dx \left[(\partial_x \phi (x) )^2+\Pi^2(x)\right]+V^{\text{SSH}}_{\text{gap}}, \\
V^{\text{SSH}}_{\text{gap}}= -\frac{\delta t }{\pi a_0} \int dx \cos[\sqrt{4\pi} \phi(x)], \nonumber 
\end{align}
where the Fermi velocity $v_F=2 |t| a_0$.  The non-linear term $V^{\text{SSH}}_{\text{gap}}$  is relevant,  therefore the ground state of the model (\ref{SSH_bosonised}) can be determined quasiclassically by minimizing the potential $V^{\text{SSH}}_{\text{gap}}$.  Depending on the sign of $\delta t$ one gets the following result: 
\begin{align}
\label{SSH_bosonic_GS}
\begin{cases}
\delta t>0:  \sqrt{4\pi} \phi(x)=0  \mod 2\pi  \\
\delta t<0: \sqrt{4\pi} \phi(x) = \pi  \mod 2\pi 
\end{cases}
\end{align}
 Now let us turn to the interacting model and study how the non-interacting ground state (\ref{SSH_bosonic_GS}) changes in the presence of interactions. 

\subsection{Interacting model}\label{SSH_interacting}
The generic interaction term for spinless fermions can be written in the continuum limit as: 
\begin{align}
\label{SSH_interaction_term}
H_{\text{int}}= \int dx V(x-x')  \rho(x)  \rho(x'), 
\end{align}
where the density $\rho(x) =\Psi^{\dagger}(x) \Psi(x)$.  After bosonizing this model by using Appendix \ref{bosonization_spinless},  we obtain the following bosonic Hamiltonian:
\begin{align}
\label{SSH_int_H}
H=H_{\text{LL}} +  V_{\text{int}}+ V^{\text{SSH}}_{\text{gap}}, \nonumber \\
H_{\text{LL}} = \frac{u}{2} \left[ \frac{1}{K} (\partial_x \phi)^2+ K (\partial_x \theta)^2 \right], \nonumber \\
V_{\text{int}}= \frac{g}{2(\pi a_0)^2} \cos[2\sqrt{4\pi} \phi],  \nonumber \\ 
 V^{\text{SSH}}_{\text{gap}}= -\frac{\delta t }{\pi a_0} \int dx \cos[\sqrt{4\pi} \phi(x)], 
\end{align}
where $K$ -- is the Luttinger parameter $K \simeq 1-2V_0/(\pi v_F)$,  and $u$ is the renormalized Fermi velocity $u=v_F(1+V_0/\pi v_F)$.   The coupling constant $g = a_0V(k=2k_F)$ determines the umklapp amplitude,  and $V_0 = a_0 V(k=0)$ describes the forward scattering. 

For weak interactions the umklapp term is irrelevant,  so the only difference from the non-interacting case is the presence of Luttinger Liquid parameter.  While this would completely change the metallic phase this is not important in the presence of the single particle gap $ V^{\text{SSH}}_{\text{gap}}$.  As the repulsive interaction strength increases,  the umklapp term becomes relevant for $K<1/2$.  The quasiclassical ground state of the umklapp term $V_{\text{int}}$ is given by: 
\begin{align}
\label{SSH_int_GS}
\begin{cases}
g>0:  \sqrt{4\pi} \phi(x)= \pi/2  \mod \pi  \\
g<0: \sqrt{4\pi} \phi(x) = 0 \mod \pi 
\end{cases} 
\end{align}
 Thus in the case $g<0$ the non-interacting ground state for the single-particle gap  (\ref{SSH_bosonic_GS})  is compatible with the  ground state determined by the umklapp term $V_{\text{int}}$  (\ref{SSH_int_GS}).  In this case the interactions will not lead to any phase transitions. 
 
 In the other case $g>0$ the non-interacting ground state (\ref{SSH_bosonic_GS}) is incompatible with the values of $\phi(x)$ fixed by interactions.  The strongly interacting phase $g \gg |\delta t|$ is characterized by a broken $\mathbb{Z}_2$ symmetry ($\phi \rightarrow -\phi$) and physically corresponds to a charge density wave.  

\section{Edge states in SSH model}\label{sec:edge states}
In this section, we discuss the bosonic description of the topological edge states in both the non-interacting SSH chain and the interacting model.  We compare different physical realizations of the boundary.  First, we show that there are edge states at the boundary between the two phases of the SSH model.  Secondly,  we consider the case when the boundary is realized by a strong impurity,  and argue that an additional quantum mechanical phase needs to be added to the bosonic field at the boundary to reproduce the topological edge states.    Finally,  we relate the localization length of the edge states to the width of a soliton in the Sine-Gordon model and compare it with the results obtained by bosonization with open boundaries without including umklapp.  We show that the two methods provide qualitatively similar results.   We next include the backscattering to the model and compute the localization length in the SSH chain with nearest-neighbor interactions and next-nearest neighbor interactions.

\subsection{Edge states at the interface between trivial and topological phases of non-interacting SSH chain  }

Before we discuss the edge states in the SSH model is important to note that there is no physical distinction between the two phases with $|w|>|v|$ and $|v|<|w|$ of the SSH model if one imposes periodic boundary conditions \cite{Fuchs2021}.  This is because of the ambiguity in defining the unit cell.  We can,  however,  discuss a boundary between one phase and the other.  This  was first discussed in the original paper of Su,  Schrieffer,  and Heeger \cite{SSH1979},  where the gap arises from spontaneously symmetry breaking (Peierls transition),  and therefore these boundaries are the elementary excitations going from one ground state to the other. 

We will consider the boundary as externally imposed:  
\begin{align}
\begin{cases}
\delta t(x)<0,  x>0 \\ 
\delta t(x)>0,  x<0 
\end{cases},
\end{align}
that fixes the location of the domain walls and allows us to interpret the physics here as the physics of edge states.  

The fact that $\delta t$ changes sign implies $\delta t(x=0)=0$ and thus at this point the gap is closed.   This alone is not sufficient however to ensure the edge states pinned at the zero energy.  This requires chiral symmetry which is present in the SSH model.   

\begin{figure}
\includegraphics[scale=0.3]{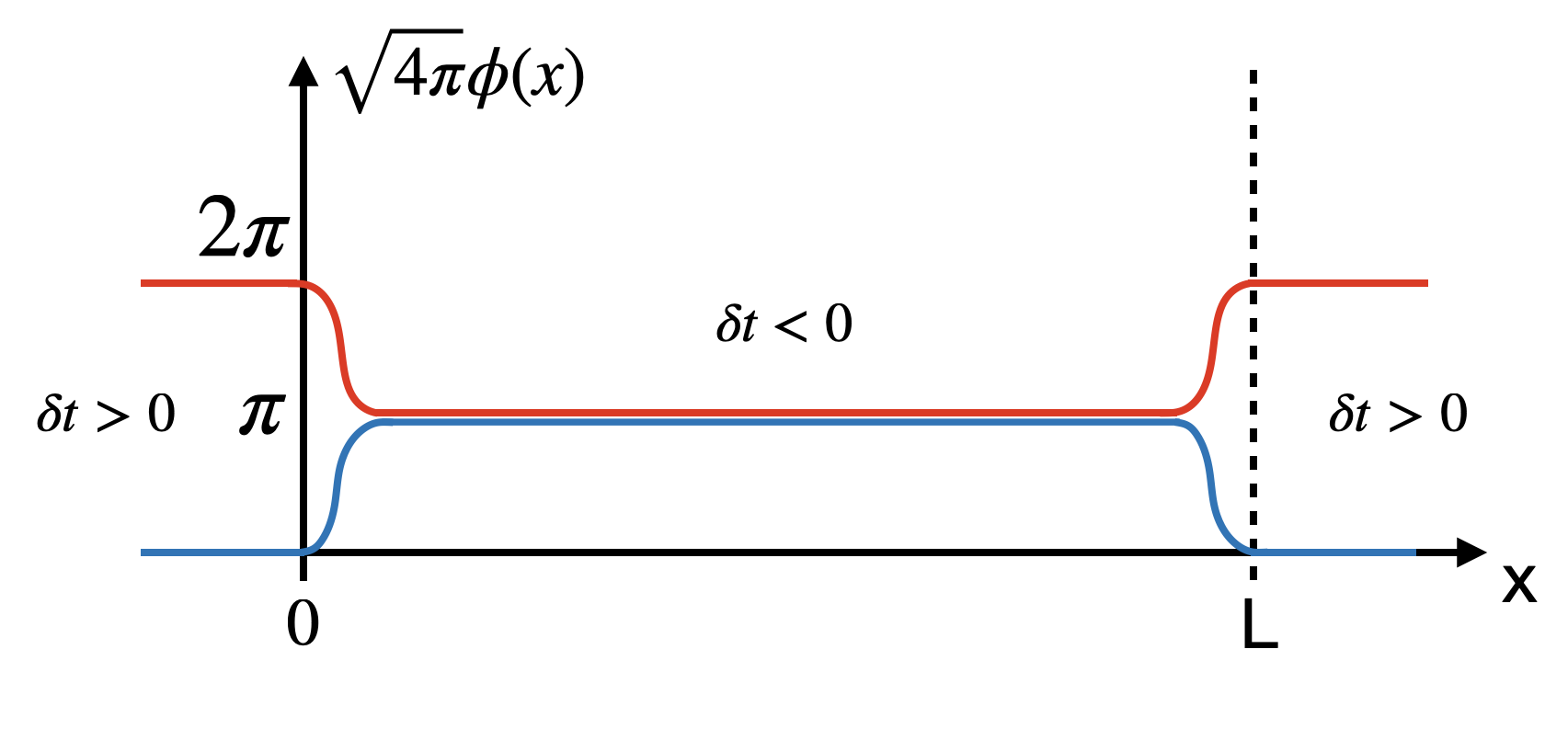}
\caption{The quasiclassical bosonic ground state for an interface between topologically distinct phases with $\delta t>0$ and $\delta t<0$ of a non-interacting SSH chain (\ref{SSH_bosonised}).  The kinks in the bosonic field $\phi(x)$ at the boundaries correspond to the edge states.  }
\label{SSH_bosons_fig}
\end{figure}

  The gapless edge states are manifested as kinks in bosonic fields at the boundary.  To see that we recall the fact that on the quasiclassical level, the bosonic fields in the bulk are fixed to the minima of the potential energy (\ref{SSH_bosonic_GS}) such that:  
\begin{align}
\label{bulk_phi_SSH}
\sqrt{4\pi} \phi(x)  = \begin{cases}
\pi \mod 2\pi,  x>0 \\ 
0 \mod 2\pi,  x<0
\end{cases} 
\end{align}
In anticipation of our upcoming discussion of finite systems with two edges let us consider a finite region with $\delta t<0$ sandwiched between regions with $\delta t>0$.  The possible field configurations are illustrated in Fig.  \ref{SSH_bosons_fig}.  
The kinks in bosonic fields are associated with the jump of fermionic density at the boundary.   There are four degenerate kinks (two at each edge),  that connect different ground states in the regions with $\delta t>0$ and $\delta t<0$.   The electric charge of the edge states can be computed as follows: 
\begin{align}
\label{edgestate_charge}
Q= \frac{e}{\sqrt{\pi} }\int\limits_{x_0-\epsilon}^{x_0+\epsilon} \partial_x \phi (x) dx =\pm \frac{e}{2}, 
\end{align}
where $x_0=0$ or $L$.  This fractional charge was one of the key points of the original SSH paper,  however the analogue in terms of the edge states is much more mundane.   The electric charges of the two states differ by $e$,  therefore the edge state has a charge of $e$. 

The many-body ground state of a finite system with the size $L$ is fourfold degenerate,  as each edge state can be either empty or occupied.  This is consistent with a fermionic description of the edge states in the SSH model.

\subsection{Edge states at a physical edge}\label{strong_imp_subsec}
Let us now we consider a physical edge where we will demonstrate that the physics is identical to the phase boundary discussed above.  

We will model the physical edge by putting a single (infinitely) strong impurity at the boundary while the bulk is in the topological phase.   Let us start with a model of an impurity placed at the lattice site $j=0$: 
\begin{align}
\label{strong_impurity}
V_{\text{imp}}= V_0 \sum_j \delta_{j0} c^{\dagger}_jc_j +\text{h.c}.
\end{align}
The limit of a strong impurity implies we wish to consider $V_0 \rightarrow \infty$.  Now let us bosonize this term and focus on the non-linear part that fixes the value of the bosonic field $\phi$ at the impurity.  
\begin{align}
\label{scattering_operator}
V_{\text{imp}}= \frac{V_0}{\pi a}  \sin[\sqrt{4\pi} \phi(x=0)- \text{sgn} [V_0] \delta], \hspace{0.3cm} \delta = \pi/2.   
\end{align}
Here we added an additional phase $\delta$ to the bosonic field at the boundary.  For a weak impurity $\delta \rightarrow 0$,  however it is not surprising that an additional phase shift should occur in the strong impurity limit.   We hypothesize that $\delta \rightarrow \pi/2 $ as $V_0 \rightarrow \infty$.  This is empirical but we will show that it works for all models that we consider.  We interpret this phase as a forward scattering phase for electrons with quadratic dispersion that scatter on the delta potential in the limit of the infinite strength $V_0 \rightarrow \infty$.  

As $V_0 \rightarrow \infty$,   the impurity term $V_{\text{imp}}$ has a quasiclassical minima at $ \sqrt{4\pi} \phi = 0 \mod 2\pi$.  This matches the bulk phase when $\delta t>0$ (\ref{SSH_bosonic_GS}).  We can thus consider $\delta t >0$ as the trivial phase of the SSH model where nothing happens at the edges while $\delta t <0$ is the topological phase exhibiting zero-energy edge modes and a field profile  identical to the one already shown in Fig.  \ref{SSH_bosons_fig}.

\subsection{Edge states in an interacting SSH model}\label{subsec:edge_states_SSH_int}
So far we have described the non-interacting SSH model in the bosonic language and reproduced the known results for the edge states.  It is now easy to add interactions. 

Specifically we consider an interacting SSH chain with corresponding bosonized Hamiltonian given by (\ref{SSH_int_H}).  We will first focus on the case when the interactions are small compared to the single particle gap.  In that case the umklapp can be neglected and the only effect of interaction is to change $K$ that enters the kinetic part of the Hamiltonian.  This has no effect on the topological properties of the solitons and the picture remains as in Fig.  \ref{SSH_bosons_fig}. The only effect will be on the exact shape of the solitons, this will be discussed in Sections \ref{open_boundaries} and \ref{loc_lengh_int}.

In case when interactions are repulsive $g>0$ and strong,  we showed in Section \ref{SSH_interacting}  that there is a phase transition to the charge density wave phase (CDW).  Let us now consider the properties of this phase in the presence of boundaries.   

The bosonic field in CDW phase is fixed  quasiclassically in the bulk at one of the potential minima $\phi(x) \approx \pm \pi/2 \mod 2\pi$,  as is discussed in Appendix \ref{Ising_app}.  At the boundary,  the bosonic field interpolates between the bulk values  and the value $\phi=0 \mod 2\pi$ that is fixed by impurity.

\begin{figure}
\includegraphics[scale=0.25]{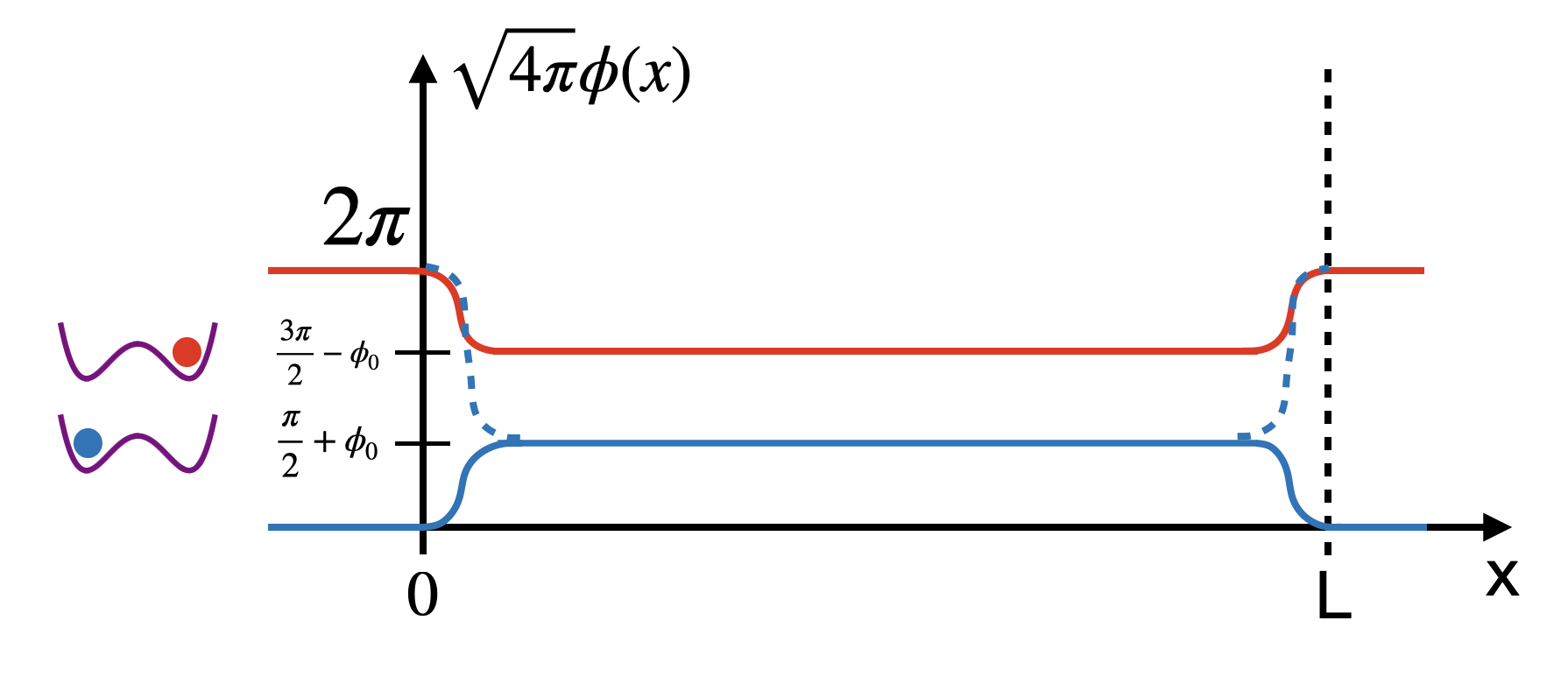}
\caption{ Possible ground state configurations of the interacting SSH model in the CDW phase (solid lines).  The system spontaneously breaks $\mathbb{Z}_2$ symmetry in the bulk by choosing one of the minima shown in red or blue.   For a given bulk state,  there is a unique kink structure at the edges with lowest energy.   Such states differ from the topological edge states in non-interacting SSH model,  that have an edge degeneracy.  }
\label{bosons_nontop}
\end{figure}

The two possible ground state field configurations in a finite system are illustrated in Fig.  \ref{bosons_nontop}.    They are degenerate,  however this degeneracy is related to the spontaneously broken $\mathbb{Z}_2$ symmetry in the bulk,  and not to the topological properties of the model.  In particular, in a case of non-interacting SSH chain the bulk ground state is unique and degeneracy occurs due to two bosonic kinks that interpolate between bulk and boundary values.  Such states differ by a unit charge,  that implies that  one can add or remove an electron from the edge without energy cost.  In the CDW phase,  however,  adding a particle at the edge,   i.e.  adding a kink of $\pi$ magnitude,  requires finite energy as it corresponds to the excitation of the model.

\subsection{Estimation of the localization length in the interacting SSH model without umklapp} \label{open_boundaries}
While we have explained that the presence of the edge states in the SSH model is unaffected by weak interactions,  the profile of the edge  states and in particular their localization length will depend on the interaction strength.  This was previously considered in \cite{Giamarchi2023} which used a more precise treatment of open boundary conditions within bosonisation  \cite{Fabrizio1995,Mattsson1997,Cazalilla2002}.  

This gave rise to complicated non-linear  Euler-Lagrange equations for the decaying bosonic mode that the authors solved to compute the localization length as a function of the Luttinger parameter.   In particular,  they found that the localization length is non-monotonic as a function of the interaction strength.  
We demonstrate here that we get the same features by using simple boundary conditions that we have previously discussed. 

With this boundary conditions the edge state corresponds to a half-soliton in the sine-Gordon model,  see \cite{Mirlin2018}.   Therefore the localization length can be estimated as the half-width of the sine-Gordon soliton,  which is given by $l =v/\Delta$,  where $v$ is the renormalized Fermi velocity $v=v_F/K$ and $\Delta$ is the energy gap,  that can be estimated via scaling analysis.  For $K>1/2$ where the umklapp term can be neglected,  the RG equation for the dimerisation $\delta t$ reads: 
\begin{align}
\frac{d (\delta t)}{dl} = (2-K)\delta t(l)
\end{align}
The RG scale  $l_*$ associated with the energy gap is determined by the condition $\delta t(l_*)/\Lambda_0=1$,  where $\Lambda_0$ is a non-universal high-energy cutoff.   Thus for the gap we obtain $\Delta =\Lambda_0 e^{-l_*}=\Lambda_0 (\delta t/\Lambda_0)^{1/(2-K)}$.   Therefore as we increase $K$ the gap $\Delta$ decreases,  this is related to the fact that the operator $V^{\text{SSH}}_{\text{gap}}$ becomes less relevant.  For the localization length of the edge state we obtain thus: 
\begin{align}
\label{loc_length}
l_{\text{loc}}= \frac{v_F}{K} \frac{1}{\Lambda_0}\left(\frac{\Lambda_0}{\delta t}\right)^{\frac{1}{2-K}}
\end{align}
As $K$ increases $1/\Delta$ increases as well,  but the renormalized Fermi velocity decreases.  Therefore the function $l_{\text{loc}}$ is non-monotonic,  and changes  slope at $K \approx 1$ when $\delta t/\Lambda_0 \approx  e^{(1-\delta K)/\delta K}$,  where $\delta K =K-1,  \delta K \ll 1$ for a fixed cutoff $\Lambda_0$,  in full agreement with  \cite{Giamarchi2023}.  Note that at $K=2$ the localization length diverges due to the decreasing of the gap, which was also discussed in \cite{Giamarchi2023}.  Thus we showed that by treating the edge states as half-solitons in the sine-Gordon model and using the scaling analysis one gets qualitatively the same results for the localization length as with the open boundary bosonization approach.   

The discussion so far has been for completely general interaction.  We will now illustrate this for concrete lattice models in the next Section where we also include possible effects of the umklapp term. 

\begin{figure}[t]
\begin{subfigure}{0.5\textwidth}
\includegraphics[scale=0.26]{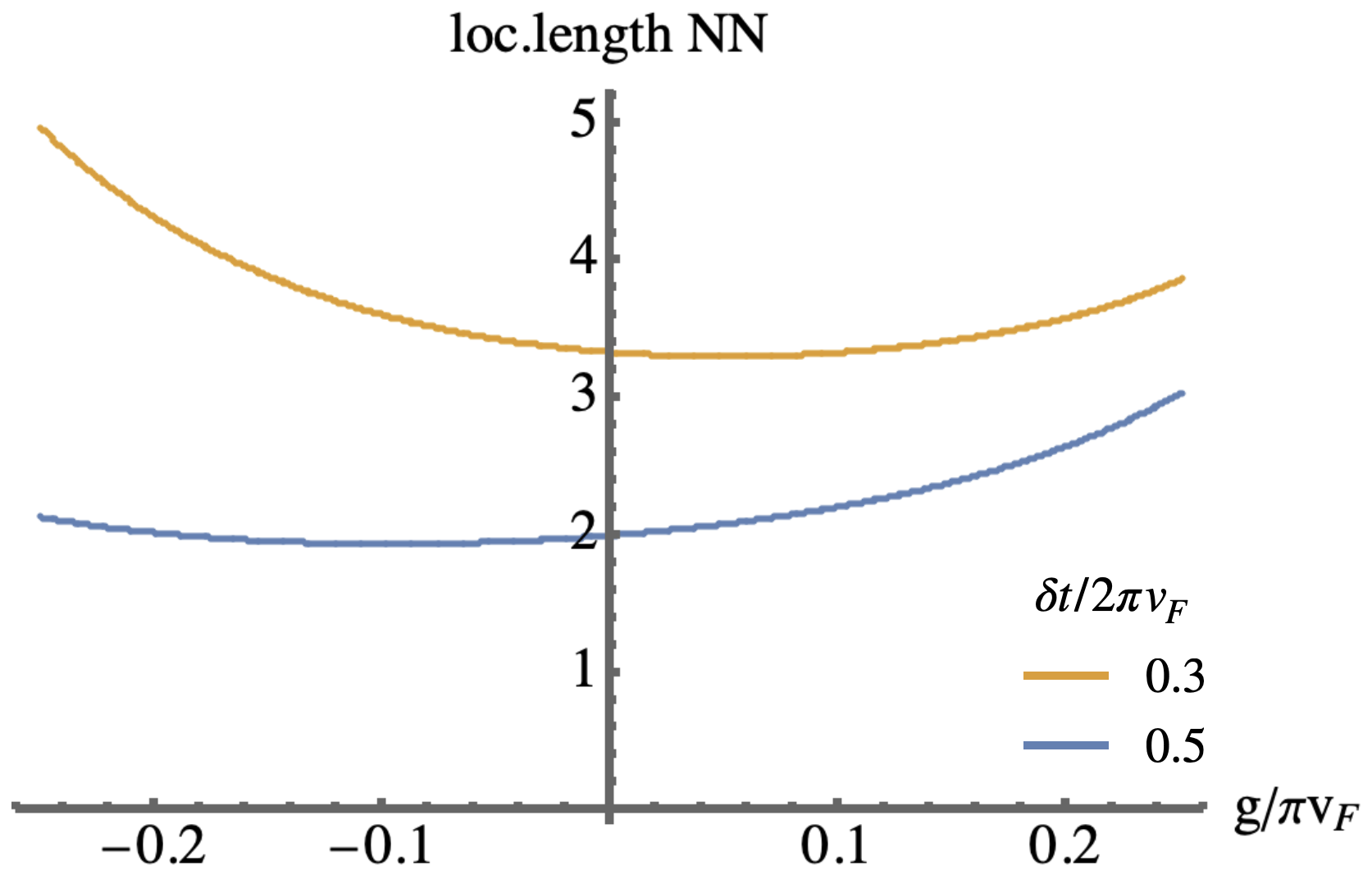}
\caption{}
\label{Loc_length_NN}
\end{subfigure}
\vspace{0.5cm}
\begin{subfigure}{0.5\textwidth}
\includegraphics[scale=0.27]{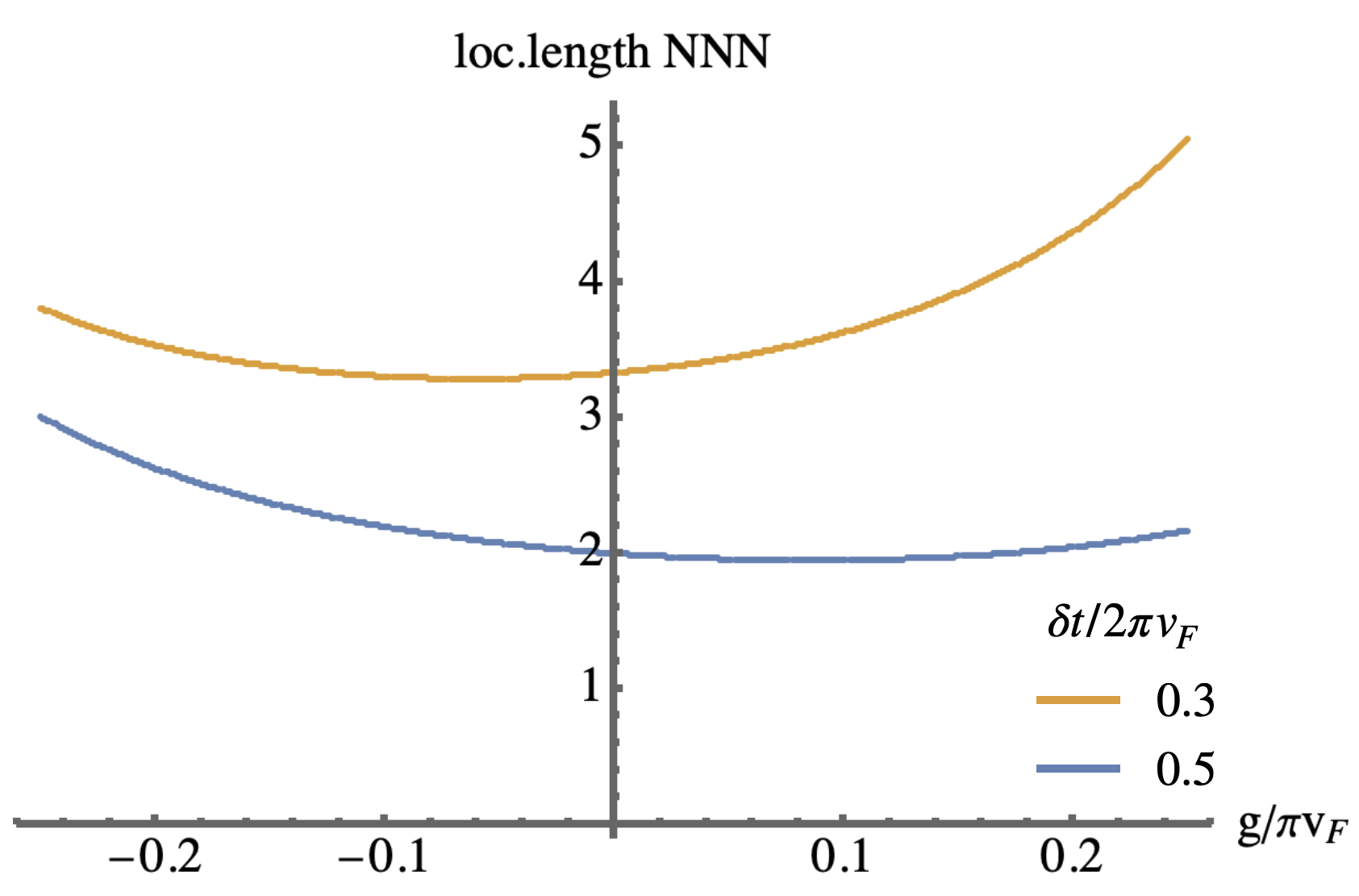}
\caption{}
\label{Loc_length_NNN}
\end{subfigure}
\caption{The localization length of the edge states in interacting SSH model as a function of the interaction strength $g=a_0V_0$,  where $V_0$ is the interaction strength in a lattice model.   The localization length is measured in the units $v_F/ \Lambda_0$.  a) Localization length in case of nearest-neighbor interactions (\ref{int_NN})  b) The localization length for next-nearest neighbor interactions (\ref{int_NNN}).  }
\end{figure}
\subsection{Localization length for nearest neighbor and next nearest neighbor interactions}\label{loc_lengh_int}

We consider here two possible  interactions one can add to the SSH model: nearest neighbor interactions (NN) and next nearest neighbor interactions (NNN).    The first couples atoms of different sublattices,  while the latter couples atoms belonging to the same sublattice.  It is known for the non-interacting SSH model,  see e.g.  \cite{Matveeva2023} that the edge states are localized on only one of the sublattices.  Thus one might ask if there is any difference between 
the weak NN interactions that couple the edge states to the bulk compared to weak NNN interactions which directly affect the wavefunction of the edge states.

We start with the nearest-neighbor interaction: 
\begin{align}
\label{int_NN}
H_{\text{int,NN}}= V_0  \sum_j (n_j-1/2) (n_{j+1}-1/2).   
\end{align}
In this model the Luttinger parameter $K$  and backscattering amplitude $g$ are not independent and are related by  $K =  1 - 2g /(\pi v_F)$ in the limit  $g \ll E_F$,  here $g= a_0 V_0$ as follows from  Section \ref{SSH_interacting}.  To compute the localization length one needs to evaluate the gap first.   To do that we need to solve the RG equations that include also the umklapp term around $K=1$ and find a length scale where one of the coupling constants in (\ref{SSH_int_H}) umklapp $g/\Lambda_0$ or dimerization $\delta t/\Lambda_0$ reaches 1.  The RG equations can be derived via Operator Product Expansion outlined in \cite{Senechal2004,francesco2012conformal} and in the second order yield:
\begin{align}
\label{RG_equations}
& \frac{d (\widetilde{\delta t})}{dl} = (2-K) \widetilde{\delta t} -  \pi \widetilde{\delta t}(l)\widetilde{g}(l) \nonumber \\
&\frac{d K}{dl}  = - 64\pi^2\widetilde{g}^2  \nonumber \\ 
&\frac{d \tilde{g} }{dl} =(2-4K)\widetilde{g}(l)
\end{align}
where $\widetilde{\delta t}= -\delta t/ 2v_F\pi a_0, \widetilde{g} = g/4v_F(\pi a_0)^2$.  Note that $\widetilde{\delta t}$ is renormalized by the umklapp term such that $g>0$ effectively decreases $\widetilde{\delta t}(l)$  and thus decreases the gap.  This is consistent with the results of the quasiclassical analysis of the interacting SSH model in the  
Sec. \ref{SSH_interacting} where we demonstrated that the system goes through the phase transition if $g>0$ and $\widetilde{g} \simeq |\widetilde{\delta t}|$.  By numerically solving the RG equations (\ref{RG_equations}) for the energy gap we compute the localization length as a function of $g$ via $l_{\text{loc}}=v_F/(K\Delta)$.  The results for $\widetilde{g} \ll |\widetilde{\delta t}|$ are shown in the Fig.  \ref{Loc_length_NN}.  The umklapp term does not change the non-monotonic behavior of the localization length.   The results are also consistent with the fact that the umklapp repulsive interaction decreases the energy gap.   

Now let us consider the next-nearest neighbor interactions: 
\begin{align}
\label{int_NNN}
H_{\text{int,NNN}}= V_0  \sum_j (n_j-1/2) (n_{j+2}-1/2).   
\end{align}
The Luttinger parameter and umklapp scattering amplitude of the bosonized version of this model  (\ref{SSH_int_H}) are given by $K =  1 + 2g /(\pi v_F)$ and $\tilde{g} = -g=-a_0 V_0$ correspondingly.  The localization length for NNN interactions is plotted in Fig.  \ref{Loc_length_NNN}.  

Therefore the only difference between the NN interaction and NNN interactions is the signs of the coupling constants that enter in the bosonic models,  namely repulsive interactions become attractive if one replaces NNN interactions with NN and vice versa.

\subsection{Symmetry that protects the edge modes}\label{symmetry_protectionSSH}

The non-interacting SSH model falls into the BDI class that has time-reversal $T$,  particle-hole $P$ and chiral $C$ symmetries.   The winding number does not change if one breaks $T$ and $P$ but preserves $C$.    Furthermore,  if $T$  and $C$  are broken but $P$ remains then the model falls into the $D$ topological class characterized by a $\mathbb{Z}_2$ invariant (this is more naturally discussed in the context of Kitaev chain,  but the physics is identical).  As the $\mathbb{Z}$ and $\mathbb{Z}_2$ invariants are indistinguishable for a model with topological index $0$ and $1$,  the edge modes in the non-interacting SSH model will remain protected so long as either $C$ or $P$ symmetries are present.  We now show this remains true in the weakly interacting case.  
 
To do that we derive in Appendix \ref{bosonized_symmetries} how the bosonic fields transform under the anti-unitary symmetry operations of a single SSH chain.   The results are shown in Table \ref{CT_bosonic_singlechain_main}.  The important result is that models with chiral symmetry $C$ or particle-hole symmetry $P$ should be symmetric under a change of sign of $\phi$. 

In our calculations we show that the value of $\phi$ is fixed by a gap opening operator (\ref{bulk_phi_SSH}) and in the topological phase it is given by $\sqrt{4\pi}\phi=\pm \pi$.   For topological insulators  the  allowed Hamiltonians conserve the number of particles, 
and therefore the perturbations consistent with the particle conservation   depend on  $\phi$ and are independent of  $\theta$.  
There are two kinds of possible perturbation terms,  
those that shift the position  of the minima and those that do not. 
Indeed,  a small  perturbation of the form $V_n\sin [n \sqrt{4\pi}\phi]$ shifts a minima away from  $\sqrt{4\pi}\phi=\pm \pi$ by $\delta \phi \propto V_n a_0/\delta t$ where $\delta t/a_0$ corresponds to the amplitude of the bulk operator (\ref{SSH_bosonised}).  Thus such perturbation removes the degeneracy.  
However,  it breaks the chiral symmetry and particle-hole symmetry, and therefore  is forbidden as long as the system retains at least one of these symmetries. 

Thus the allowed terms should be even functions of $\phi$,  such functions also need to be $2\pi$ periodic in $\sqrt{4\pi}\phi$,  so a generic allowed perturbation has a form:
\begin{align}
V =\sum_{n=2}^\infty A_n \cos(\sqrt{4\pi} n \phi).
\end{align}   
However if the perturbation is small,  $A_n \ll \delta t/a_0$,  such terms  do not shift the minima of the potential energy in the  topological phase.  
Therefore as long as the chiral symmetry is preserved the degeneracy of the edge states in the SSH model is protected.  

\begin{table}
\begin{tabular}{c|ccc}
\toprule
\multicolumn{1}{c}{} & \multicolumn{3}{c}{\textbf{Symmetry $S$}}  \\
\cmidrule(rl){2-4} 
\textbf{Fields } & {$T$} & {$P$} & {$C$} \\
\hline
$S^{-1} \phi S$ & $\phi$& $-\phi$ & $-\phi$\\
\hline
$S^{-1}\theta S$ & $-\theta$ & $\theta$& $-\theta$\\
\bottomrule
\end{tabular}
\caption{Symmetry transformation of bosonic fields $\phi$ and $\theta$.   $T$ denotes time-reversal symmetry with $T^2=+1$,   $P$ is the particle hole symmetry with $P^2=+1$ and $C= T \cdot P$ is the chiral symmetry.  We note that the cases $P^2=-1$ and $T^2=-1$ are not realisable for a single chain.  }
\label{CT_bosonic_singlechain_main}
\end{table}

\section{Capacitively coupled interacting SSH chains}\label{Sec:uncoupledSSH} 
Before we consider a generic coupled chain model of a $\mathbb{Z}$ topological insulator,  we will focus on the limit of a two-chain model coupled only by interactions (capacitive coupling).   We consider two cases: (i) both chains are identical (either topological or trivial),  and (ii) one chain is topological and the other is trivial.   

Specifically we will consider Hamiltonians of the form: 
\begin{align}
\label{SSH_capacit}
H = \sum_{i,n} (t c^{\dagger}_{n,i} c_{n+1,i} + \delta t_i (-1)^n c^{\dagger}_{n,i} c_{n+1,i} + \text{h.c.} ) + \nonumber \\ 
+ U \sum_n (c^{\dagger}_{n,1} c_{n,1}-1/2) (c^{\dagger}_{n,2} c_{n,2}-1/2), 
\end{align}
where $\delta t_1 = \pm \delta t_2$.   Bosonization of this model yields a sum of three parts:  
 \begin{align}
\label{2chains_ident_int}
H=H_c+H_s + V_{\text{gap}}.  
\end{align}
The first two terms $H_c$ and $H_s$ are the conventional charge and spin separated Hamiltonians \cite{Giamarchi_book,gogolin2004bosonization},   given by:
\begin{align}
\label{Hcs}
H_{c,s}= \frac{v_{c,s}}{2} \left[K_{c,s}\Pi_{s,c}^2 + K^{-1}_{c,s}(\partial_x \phi_{s,c})^2 \right] + \nonumber \\+\frac{g_{s,c}}{\pi a^2} \cos[\sqrt{8\pi}\phi_{s,c}], 
\end{align}
where we defined spin and charge fields as $\phi_c = (\phi_1 + \phi_2)/\sqrt{2},  \phi_s = (\phi_1 - \phi_2)/\sqrt{2}$.   The parameters can be related to the original lattice values as: $g_c=-g_s=-U$,  $K_{c,s} \simeq 1+g_{c,s}/(\pi v_F)$.   

 The single particle term $V_{\text{gap}}$ couples the charge and spin sectors and originates from the dimerization $\delta t_i$ in the lattice model.   The bosonized form of this depends on the relative values of $\delta t$ on each chain for the case $\delta t_1= \delta t_2$ we obtain: 
\begin{align}
\label{Vgap1}
V_{\text{gap,1}}= -\frac{2\delta t}{\pi a_0}\cos (\sqrt{2\pi } \phi_c) \cos(\sqrt{2\pi } \phi_s), 
\end{align}
and in the other case when $\delta t_1 = -\delta t_2$ we get: 
\begin{align}
\label{Vgap2}
V_{\text{gap,2}}= \frac{2\delta t}{\pi a_0}\sin (\sqrt{2\pi} \phi_c) \sin(\sqrt{2\pi} \phi_s).
\end{align}
To describe the topological properties of the model we require an edge.  To do that we place a strong impurity at the edge $x=0$  as we did for a single chain (\ref{scattering_operator}).  The bosonized form of the impurity Hamiltonian  is given by: 
\begin{align}
\label{impurity_bosonised_2chains}
H_{\text{imp}}(0)= V_0 \cos (\sqrt{2\pi} \phi_c(0)) \cdot \cos(\sqrt{2\pi} \phi_s(0)).
\end{align}
We will now analyze the edge states of such models.  
\subsection{Identical SSH chains}\label{subsec:identical} 
 In the topological phase $\delta t<0$,  there are two edge states for the non-interacting model.  Each of these states can be either empty or occupied,   which leads to a four-fold degeneracy  per edge of the many-body ground state.  
 
\begin{figure*}[!t]
\begin{subfigure}{.4\linewidth}
\includegraphics[scale=0.31]{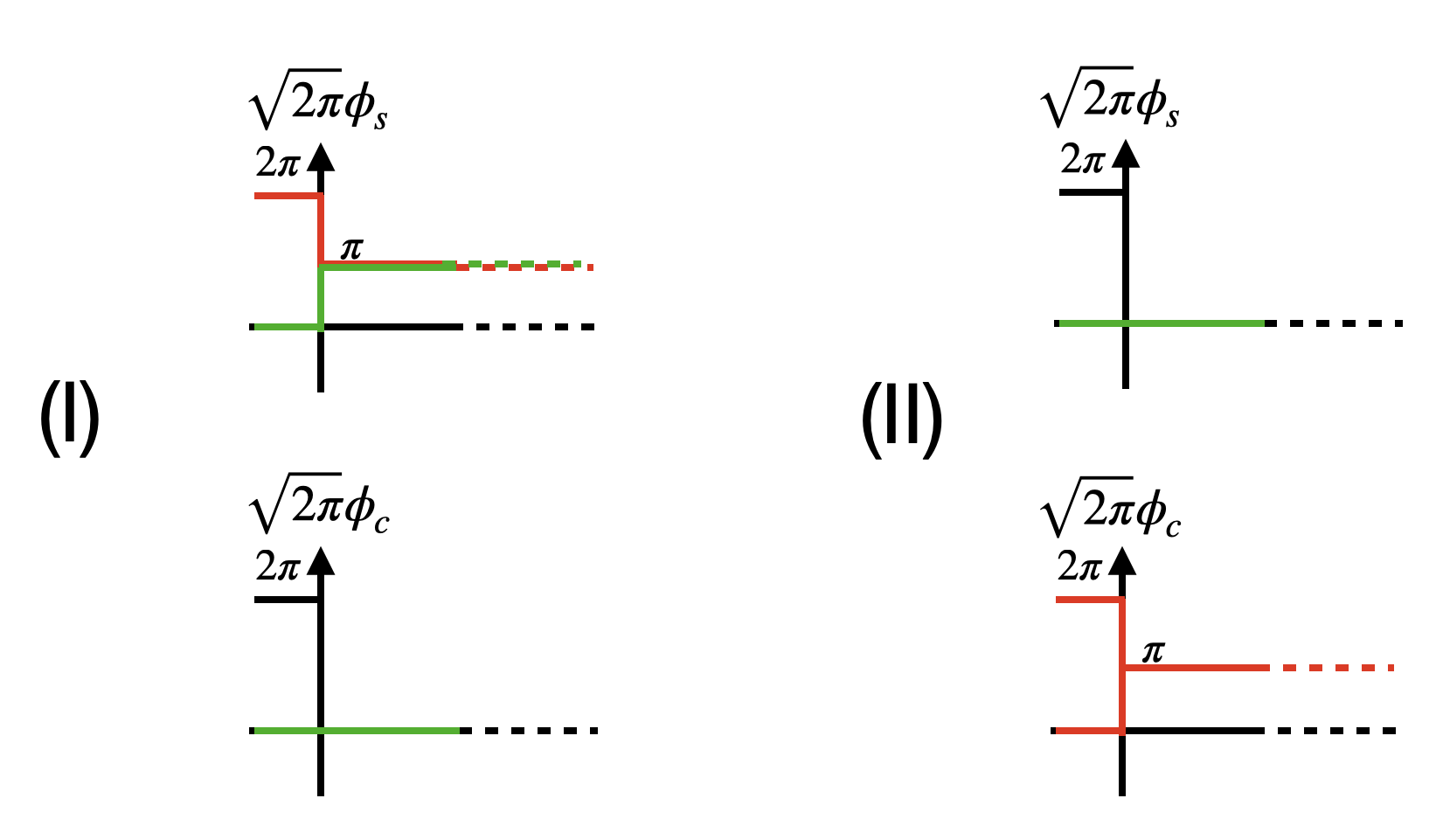}
\caption{}
\end{subfigure}
\hspace{2cm}
\begin{subfigure}{.4\linewidth}
\includegraphics[scale=0.25]{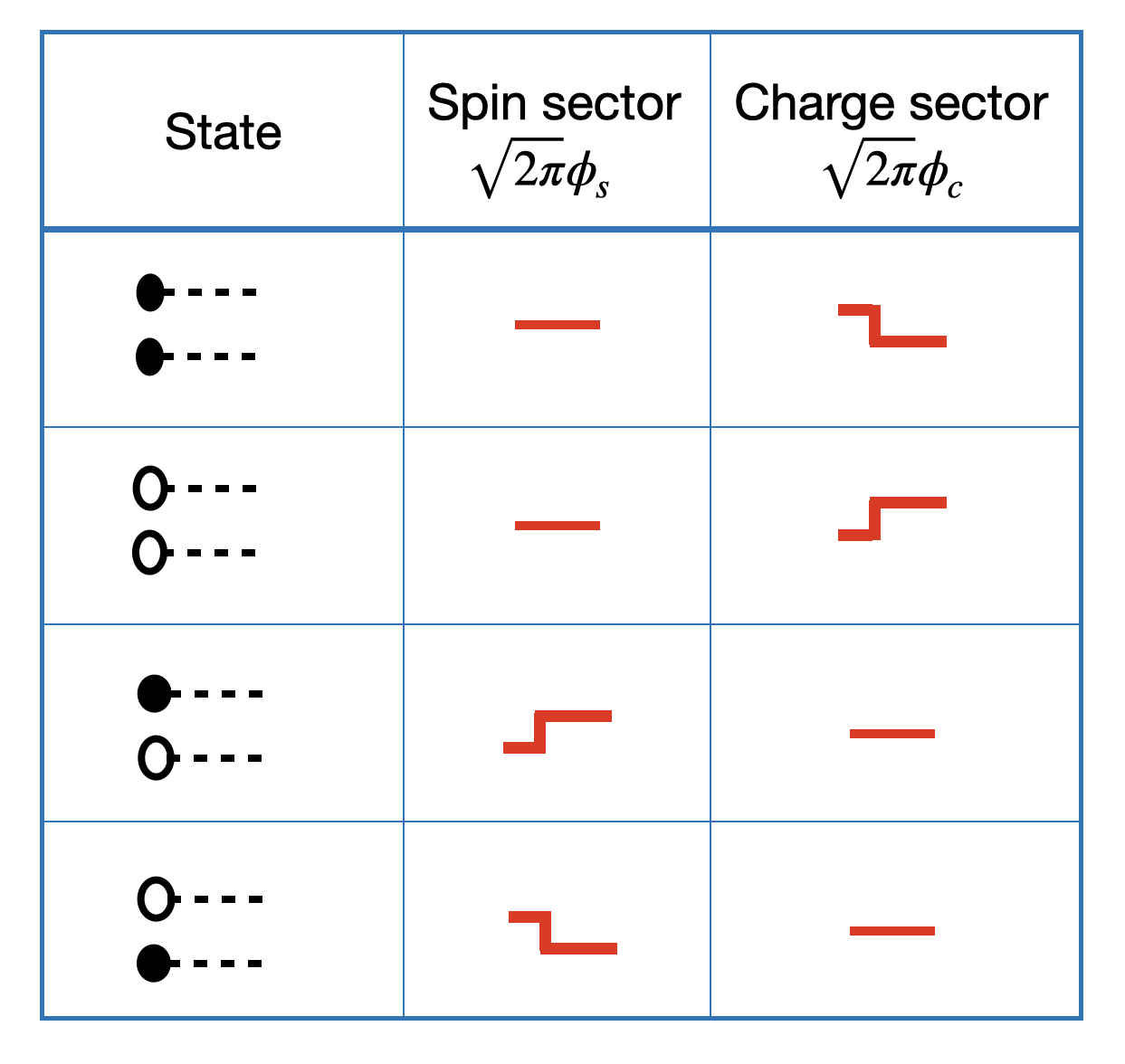}
\caption{}
\end{subfigure}
\caption{a) Profiles of the charge and spin fields for two identical SSH chains  in topologically non-trivial phase with $\delta t<0$ b) Field configurations for fourfould degenerate many-body states.  The column "State" illustrates the fermionic occupation for each of the states in real space.  Full circle corresponds to the occupied edge state and an empty circle corresponds to an empty state. }  
\label{Topological_2chains_ident}
\end{figure*}

We now reproduce these result in bosonic language.  Without interactions the spin and charge sectors are identical,  $K_c=K_s=1$,  and $g_c=g_s=0$.  That means that we only need to consider minima of the potentials $V_{\text{gap,1}}$ (\ref{Vgap1}) and $H_{\text{imp}}$ (\ref{impurity_bosonised_2chains}).  We see for the trivial phase $\delta t>0$ these operators are identical thus nothing happens at the edge,  as expected.  In the topological phase the operators are of  opposite sign and therefore have minima in different locations.  Similar to the single chain one thus expects kinks at the edge.   We find four degenerate kinks which are illustrated in  Fig. \ref{Topological_2chains_ident}.   

\begin{figure*}[t]
\begin{subfigure}{0.9\linewidth}
\includegraphics[scale=0.4]{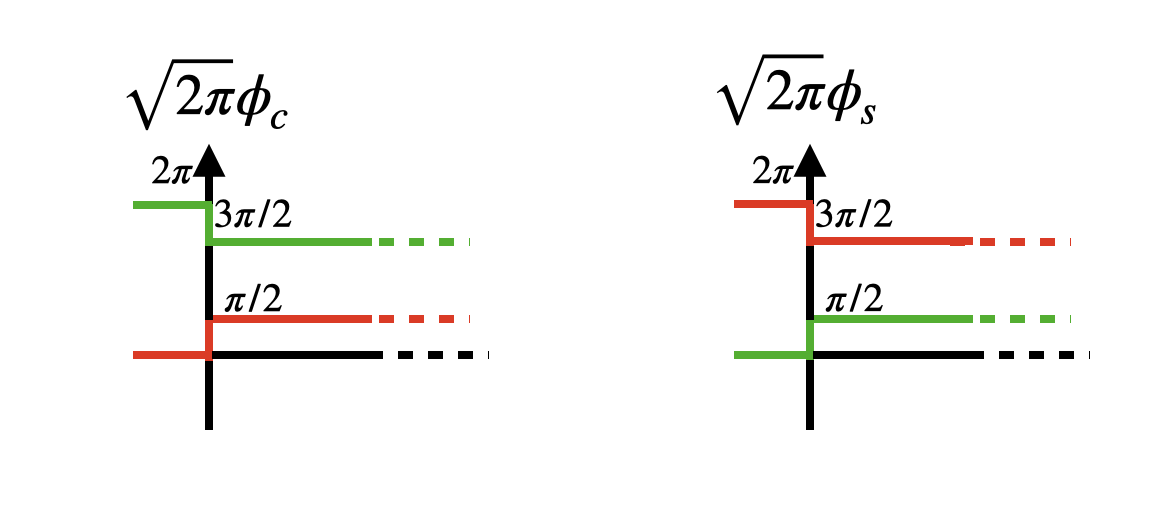}
\end{subfigure}
\caption{Profiles of the charge and spin fields for inequivalent SSH chains with $\delta t<0$.  Kinks of bosonic fields at the boundaries have the same magnitude and correspond to the edge states.  The ground state at half-filling is two-fold degenerate,  the  two degenerate states are marked by green/red colors.  }
\label{Topological_2chains_ineq}
\end{figure*}

 Let us now add interactions to the model.  As known for the Hubbard model \cite{gogolin2004bosonization},  if $U>0$ the spin sector is gapless and charge sector is gapped as the  $g_c$ cosine term (umklapp scattering) is relevant.  It fixes the bosonic fields at the values $\sqrt{2\pi K_c} \phi_c = 0 \mod \pi$ which is compatible with the non-interacting ground state illustrated in the Fig.  \ref{Topological_2chains_ident}.  Note that the model is symmetric with respect to $U \rightarrow -U$ and $\phi_c \rightarrow \phi_s$,  so the same conclusions hold for attractive interactions.  

Even though the quasiclassical ground states are compatible,  interaction reduce the ground state degeneracy because $K_{c}  \neq K_s$  in the bulk,  therefore the magnitude of kinks is different for the spin and the charge sector.    In particular,  for repulsive interactions $K_c<K_s$,  and thus the two kinks in the spin sector have the lower energy than those in the charge sector.   For attractive interactions it is the other way around. 

Thus  the fourfold degeneracy of the ground state is reduced to twofold.  This has a simple physical interpretation: due to electrostatic interactions the two states with two electrons localized near the edge have different energy than the two states with a single electron.  

As the bosonic form of the Hamiltonian is independent of the microscopic details of the interaction, such a degeneracy reduction  occurs also in the case of next-nearest neighbor Hubbard interaction,  that couples the atoms from different sublattices on different chains.   The edge states in a non-interacting model however are localized on the same sublattice \cite{Matveeva2023},  so such degeneracy breaking is less obvious.  It may be attributed to second order processes that are generated in the interacting model and effectively couple atoms from the same sublattice. 

Note that the bulk ground state of this Hubbard SSH model was studied earlier in  \cite{Tsuchiizu2004,Benthien2006,Ejima2016,Ejima2018}.  These works focus on the extended Hubbard model,  that includes  nearest-neighbor interaction on top of  Hubbard term.   It was demonstrated that when the nearest-neighbor interaction is sufficently strong compared to the Hubbard term,  the sign of coupling constant in charge sector changes and  the system undergoes Ising transition to the phase characterized by CDW order parameter with broken $\mathbb{Z}_2$ symmetry $\phi_c \rightarrow -\phi_c$ similar to the case of the Ising transition in a  single interacting SSH chain that we considered earlier in Subsection \ref{SSH_interacting}.   It was also shown that for stronger interactions,  the system goes through a tricritical point where the phase transition changes universality class and becomes first order \cite{Ejima2016,Ejima2018}.   None of these strong coupling phases exhibit topological properties, so we will not discuss them further.    

\subsection{Chains in opposite phases}\label{subsec:inequivalent}
Now we will focus on the model of inequivalent SSH chains,  when one of the chains is in topological phase,  and another one is trivial. This model is described by the Hamiltonian (\ref{2chains_ident_int}),  with the dimerization term given by (\ref{Vgap2}).   The ground state of the non-interacting model is illustrated in the Fig.   \ref{Topological_2chains_ineq} and is two-fold degenerate.  

Weak interactions do not break the degeneracy of the edge states as the magnitudes of the kinks corresponding to the two edge states in the Fig.  \ref{Topological_2chains_ineq} remain equal.  

However the ground state of the non-interacting model is incompatible with the values of fields that minimize the interacting terms.  Namely,  in repulsive Hubbard model the cosine in charge sector fixes bosonic field to $\sqrt{2\pi} \phi_c = 0 \mod \pi$ (for attractive Hubbard model $\phi_c \rightarrow \phi_s$) while  the single particle term (\ref{Vgap2}) has minima at $\sqrt{2\pi} \phi_c = \pi/2 \mod  \pi $.  So for strong interactions,   there is a phase transition in charge or spin sectors depending on  whether one deals with repulsion or attraction.  The corresponding order parameters in the strong coupling phases are spin density wave (SDW) $\propto \cos(\sqrt{2\pi} \phi_c) \sin(\sqrt{2\pi} \phi_s)$   in the repulsive case or CDW in case of attraction.    

\begin{table}
\begin{tabular}{*{3}{|p{2.5cm}}|}
      \multicolumn{3}{c}{Many-body ground state degeneracy\rule[-1em]{0pt}{2em} }\\
      \hline
      Winding number $\nu$ &  Non-interacting & Interacting \\
      \hline
      1 & 2& 2\\
      2 & 4 & 2\\
      3 & 8 & 2\\
      4 & 16 & 1\\
      \hline
\end{tabular}
\caption{\small{Many-body ground state degeneracy depending on the winding number (or number of SSH chains in topological phase) in the non-interacting and weakly interacting chain models.   The degeneracy for the non-interacting model is given by  $2^{\nu} $}}
\label{manybody_degeneracy}
\end{table}
\subsection{Inequivalent chains}
Finally,  let us analyze the case when both gap opening terms (\ref{Vgap1}) and (\ref{Vgap2}) are present.   Such model corresponds to  a model of two inequivalent chains,  that can be written as:  
\begin{align}
\label{Vgap_inequiv}
& V_{\text{gap}} =  -\frac{(\delta t_1 +\delta t_2)}{\pi a_0}\cos(\sqrt{2\pi } \phi_c) \cos(\sqrt{2\pi} \phi_s) - \nonumber \\
& -\frac{(\delta t_2 - \delta t_1)}{\pi a_0}\sin(\sqrt{2\pi} \phi_c) \sin(\sqrt{2\pi} \phi_s)= \nonumber  \\ 
& = - \frac{\delta t_1}{\pi a_0} \cos(\sqrt{4\pi} \phi_1(x)) -\frac{\delta t_2}{\pi a_0}   \cos(\sqrt{4\pi} \phi_2(x)). 
\end{align}
As long as $\delta t_2$ and $\delta t_1$ have the same sign the ground state of the model is determined by the minima of the term $\propto (\delta t_1 +\delta t_2)$ that describes the identical chains,  whereas when $\delta t_2$ and $\delta t_1$ have the opposite sign,  the ground state of the model correspond to the minima of $(\delta t_1 - \delta t_2)$ term.    This is not obvious in the charge/spin basis,  but becomes so if one relates $\phi_c$ and $\phi_s$ to the original $\phi_1$ and $\phi_2$ basis as shown in the final line of (\ref{Vgap_inequiv}).

\subsection{Symmetries that protect the degenerate edge states}
\label{inequivalent}
Above we found that in the presence of weak interactions the many-body degeneracy of the ground state for two capacitatively coupled SSH chains is reduced,  and for 
two inequivalent chains it remains unchanged.  
This degeneracy reduction is absolutely general and 
can be computed solely from the symmetry arguments.
The original argument was given in the framework of Majorana
 fermions \cite{FidkowskiKitaev2011}. 
 This results was reproduced by various alternative methods \cite{Meidan2014,Queiroz2016,MorimotoMudry2015,Song2017}.
It was also shown that other fermionic models,  with no apparent relation to Majorana chains can be mapped  on this model \cite{Verresen2017}.  There are  two symmetries that are relevant in the classification:
chiral symmetry and the fermionic parity.   
If chiral symmetry is broken,  the interacting model is trivial.
If chiral symmetry is present, but parity is not the classification is $\mathbb{Z}_2$. 
If both are preserved,  the non-interacting classification $\mathbb{Z}$ is reduced to $\mathbb{Z}_4$.   In Table \ref{manybody_degeneracy} we review these results. 

Here we will focus on two-chain case and we will prove by using bosonization that chiral symmetry protects the edge state degeneracy in $\mathbb{Z}$ topological classes.  To do that we need to establish  the action of symmetries in terms of bosonic fields.    The fermionic representation of symmetries of two-chain model was constructed in  \cite{Matveeva2023},  and the action of these symmetries on bosonic fields is derived in Appendix \ref{bosonized_symmetries} and is shown in Table \ref{charge_spin_symm_main}.   Note,  that there are two chiral symmetry operators,  $C_1$ and $C_2$.  We will focus here only on $C_1$ as we will see in the next section,   $C_2$ does not protect the degeneracy of the edge states.     
 
Note that the bosonic model that describes identical SSH chains (\ref{Vgap1}) is invariant under all of the symmetries discussed above.  Thus it falls into multiple universality classes.  Such ambiguity is related to the presence of additional unitary symmetries that the uncoupled model has.  In \cite{Matveeva2023} we showed that inter-chain coupling breaks some of these symmetries and puts a model into certain symmetry class of the tenfold classification.  We study such models in the next section in more detail.  Here we will focus on the role of chiral symmetry only.  

\begin{table}
\begin{tabular}{c|cccccc}
\toprule
\multicolumn{1}{c}{} & \multicolumn{6}{c}{\textbf{Symmetry action $ \tilde{\phi}_j \equiv S^{-1} \phi_j S$}}  \\
\cmidrule(rl){2-7} 
\textbf{ } & {$T_+$} & {$T_-$} & {$P_+$} & {$P_-$} & {$C_1$}& {$C_2$}  \\
\hline
$ \tilde{\phi}_c$ & $\phi_c$& ($\phi_c +\sqrt{\frac{\pi}{2}}$) & $-\phi_c$ &$- (\phi_c + \sqrt{\frac{\pi}{2}})$ & $-\phi_c$ & $-(\phi_c +\sqrt{\frac{\pi}{2}})$ \\
\hline
$ \tilde{\phi}_s$ & $-\phi_s$ & $-(\phi_s+\sqrt{\frac{\pi}{2}})$ &  $\phi_s$& $(-\phi_s +\sqrt{\frac{\pi}{2}})$ & $-\phi_s$& $-(\phi_s+\sqrt{\frac{\pi}{2}})$\\
\bottomrule
\end{tabular}
\caption{Action of symmetry operators on bosonic fields in two-chain model.  $T_{\pm}$ denotes time-reversal symmetry:  $T^2_{\pm}=\pm 1$,   and $P_{\pm}$ denotes particle-hole symmetry: $P^2_{\pm}=\pm 1$.  Chiral symmetry is given by their product: $C_1=P_{\pm} T_{\pm}$,  $C_2=P_{\pm} T_{\mp}$.}
\label{charge_spin_symm_main}
\end{table}

As follows from Table \ref{charge_spin_symm_main}  the allowed by chiral symmetry $C_1$ perturbations are even functions of the bosonic fields: 
\begin{align}
V = \sum_{n,m=0}^\infty A_{n,m} \cos(\sqrt{2\pi} m \phi_c) \cos(\sqrt{2\pi} n \phi_s)
+\\  +B_{n,m}\sin(\sqrt{2\pi} n \phi_c) \sin(\sqrt{2\pi}m \phi_s)\,.
\end{align}
Provided  $A$ and $B$  are small,  this perturbation  does not shift the position of the minima of the potential energy  the topological phase, 
see the discussion of a single SSH chain.
However any perturbation that breaks chiral symmetry,  namely $\alpha \cos(\sqrt{2\pi} n \phi_c)\sin(\sqrt{2\pi} m \phi_s)$ or  $\beta \cos(\sqrt{2\pi} n \phi_s)\sin(\sqrt{2\pi} m \phi_c)$  shift the position of minima of potential energy in the topological phases, determined by the terms in (\ref{Vgap1}) and (\ref{Vgap2}).  Therefore  it removes the degeneracy.  

There is an interesting observation that we would like to mention here without going into detail.  Above we found that the fourfold degeneracy of the edge states  in capacitively coupled chains is reduced in the presence of  interactions.   Surprisingly,  the remaining twofold degeneracy of the ground state may be protected by another set of symmetries rather than just chiral.  For example,  
in case of attractive interactions the ground state corresponds to two degenerate bosonic kinks in charge sector.  Such kinks are not splitted by chiral symmetry breaking perturbation $\alpha \cos(\sqrt{2\pi} n \phi_c)\sin(\sqrt{2\pi} m \phi_s)$,  that affects position of kinks only in spin sector.  However,  such perturbation preserves the particle-hole symmetry $P_+$ as follows from the Table \ref{charge_spin_symm_main},  so this symmetry may potentially serve as a protective one.   Even though a set of protective symmetries depends on the sign of interactions,  this still might have interesting experimental implications,  as usually experiments deal with interactions of a certain sign.

\section{Properties of non-interacting coupled chain model}\label{sec:coupled_chains}
Above we mentioned that model of two identical capacitively coupled SSH chains has additional symmetries and falls into multiple chiral topological classes.  Such ambiguity can be resolved by adding inter-chain coupling.  Symmetries of the coupling terms put a model into certain topological class.  A coupled model can be invariant under one of two possible chiral symmetry operators,  and as it turns out \cite{Matveeva2023} the winding number of a weakly coupled system depends only on a type of chiral symmetry and does not change if other symmetries are broken.  Let us demonstrate this on bosonic language.

\subsection{Two-chain model of topological insulators }\label{Sub:coupled_nonint}
First of all,   let us introduce a fermionic model of coupled chains: 
\begin{figure}{}
\begin{subfigure}{1\linewidth}
\caption{}
\includegraphics[scale=0.22]{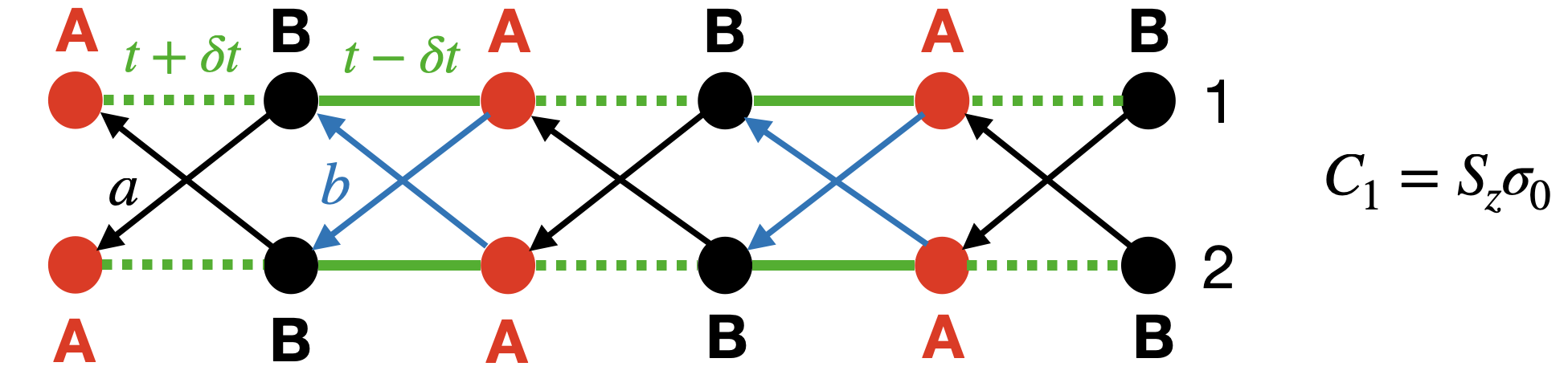}
\label{Coupling_chains_a}
\end{subfigure}\\
\begin{subfigure}{1\linewidth}
\caption{}
\includegraphics[scale=0.22]{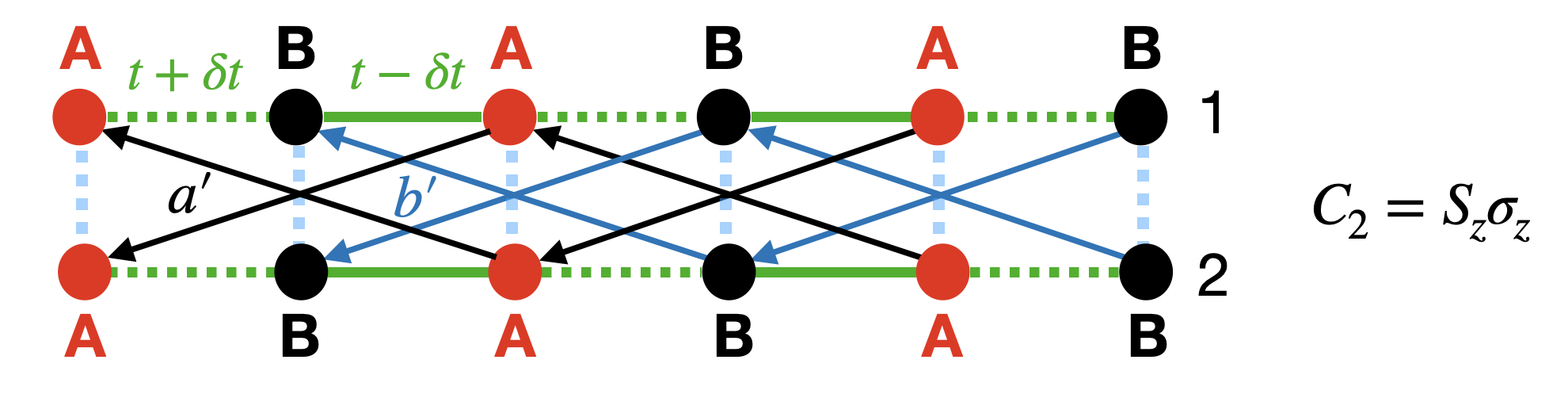}
\label{Coupling_chains_b}
\end{subfigure}
\caption{Coupled chain model of the topological classes a) with chiral symmetry $C_1$ that describe classes  BDI/CII or AIII and b) with chiral symmetry $C_2$ that describes classes CI, DIII or AIII.  Chiral symmetry $C_1$ allows coupling between $A$ and $B$ atoms only,  while $C_2$ allows inter-chain coupling only between atoms of the same type. }
\label{Coupling_chains}
\end{figure}

\begin{equation}
 \hat{H}_{\text{TI}} = \hat{H}_0+ \hat{V}_i,
 \label{H1}
 \end{equation}
 where $\hat{H}_0$ describes the Hamiltonian of two uncoupled identical SSH chains.  There are two possible coupling terms  $\hat{V}_i$ that are invariant under chiral operators $C_1$ or $C_2$.   The two models are illustrated in Fig.  \ref{Coupling_chains}.  

Coupling term $\hat{V}_1$  that describes model with chiral symmetry  $C_1$  is given by: 
 \begin{align}
 \label{BDI_CII}
 \hat{V}_1 &=  \sum_n a \left( c^\dagger_{A,1,n} c_{B,2,n} + c^\dagger_{A,2,n} c_{B,1,n} \right)  \nonumber \\
 & + b  \left( c^\dagger_{B,1,n} c_{A,2,n+1} 
 + c^\dagger_{B,2,n} c_{A,1,n+1}    \right) + \text{h.c.}
 \end{align} 
For a generic complex hopping amplitudes $a$ and $b$ this model (\ref{H1}) belongs to the topological class AIII,   if the coupling amplitudes are real the model belongs to topological class BDI and  in case when the parameters $a$ and $b$ are imaginary,   the model belongs to the class CII.  The winding number of the model with this type of coupling is given by a sum of winding number of uncoupled chains.  

The coupling invariant with respect to the second chiral operator $C_2$ is given by: 
  \begin{align}
 \label{DIII_CI}
 \hat{V}_2 &= \sum_n a' \left ( c^\dagger_{A,1,n} c_{A,2,n+1} 
 + c^\dagger_{A,2,n} c_{A,1,n+1}    \right) + \nonumber \\ 
 &  + b' \left( c^\dagger_{B,1,n} c_{B,2,n+1} + c^\dagger_{B,2,n} c_{B,1,n+1} \right) + \text{h.c.}
 \end{align}
Imaginary coupling amplitudes describe class DIII,  and  real correspond to the class CI.   The winding number of such model in case of weak coupling is given by a difference of winding numbers of two chains and thus is zero.

\subsection{Model with chiral symmetry $C_1$}\label{Sub:bos_coupled}
Now let us discuss the bosonic description of the model (\ref{H1}) with a coupling term that preserves chiral symmetry $C_1$.  Instead of directly bosonizing the model (\ref{H1}) we can write all possible gap opening terms that preserve symmetries of a given class by using action of symmetries on bosonic fields,  that we derived in Appendix \ref{bosonized_symmetries}. 
Model compatible with chiral symmetry $C_1$ is given by:
\begin{align}
\label{C1_bosonic}
& V_{\text{C}_1} = -\frac{2(\delta t + \Im(a) +\Im(b))}{\pi a_0} \cos(\sqrt{2\pi} \phi_c)  \cos(\sqrt{2\pi} \phi_s) + \nonumber \\
& + \frac{2(\Re(a)-\Re(b))}{\pi a_0}  \sin(\sqrt{2\pi} \phi_c)  \cos(\sqrt{2\pi} \theta_s).
\end{align}
This corresponds to the model illustrated in Fig.  \ref{Coupling_chains_a}. and consists of the interchain hopping terms (\ref{BDI_CII}),   as well as the dimerisation term (\ref{Vgap1}).

If $a$ and $b$ are real the model belongs to the topological class BDI,  and model of CII class can be obtained by setting $ \Re(a) = \Re(b)=0$ as such terms break time-reversal symmetry $T_-$.  

In order to see that weak coupling does not change the degeneracy of the edge states we need to diagonalize inter-chain hopping terms:  $ \sin(\sqrt{2\pi} \phi_c)  \cos(\sqrt{2\pi} \theta_s)  \rightarrow \sin(\sqrt{2\pi} \phi_c)  \sin(\sqrt{2\pi} \phi_s)$,  see Appendix \ref{BDI_bosonization_app} for more details.  Such term coincides with the term that we studied earlier in Section \ref{inequivalent}.  It describes chains in the opposite topological phases,  and if this term is small we showed that it does not affect the ground state of the identical chains. 

Note that here we discarded the uniform part of inter-chain hopping that may modify the gapless part of the spectrum.   In particular it may split the position of Fermi points of different bands,   or change their Fermi velocities,  as we demonstrate in Appendix \ref{BDI_bosonization_app} on the example of BDI model.  Such effects occur in various ladder models  \cite{Starykh2008,Carr2013,Starykh2007,Moroz1999}. 
In our model they may lead to a phase transition to SDW phase for strong interactions,  as we discuss in Appendix \ref{sec:coupled_chains_int}.  
However,  as we show in Appendices  \ref{BDI_bosonization_app}  and \ref{sec:coupled_chains_int}  renormalization of the gapless spectrum  does not affect topology of a model for weak interactions.   In Appendix  \ref{BDI_bosonization_app} we explicitly bosonize the model (\ref{H1}) in case of BDI class and demonstrate that  we reproduce the gap opening terms that we wrote using symmetry arguments (\ref{C1_bosonic}).

\begin{figure*}[t!]
\begin{subfigure}{.4\linewidth}
\includegraphics[scale=0.22]{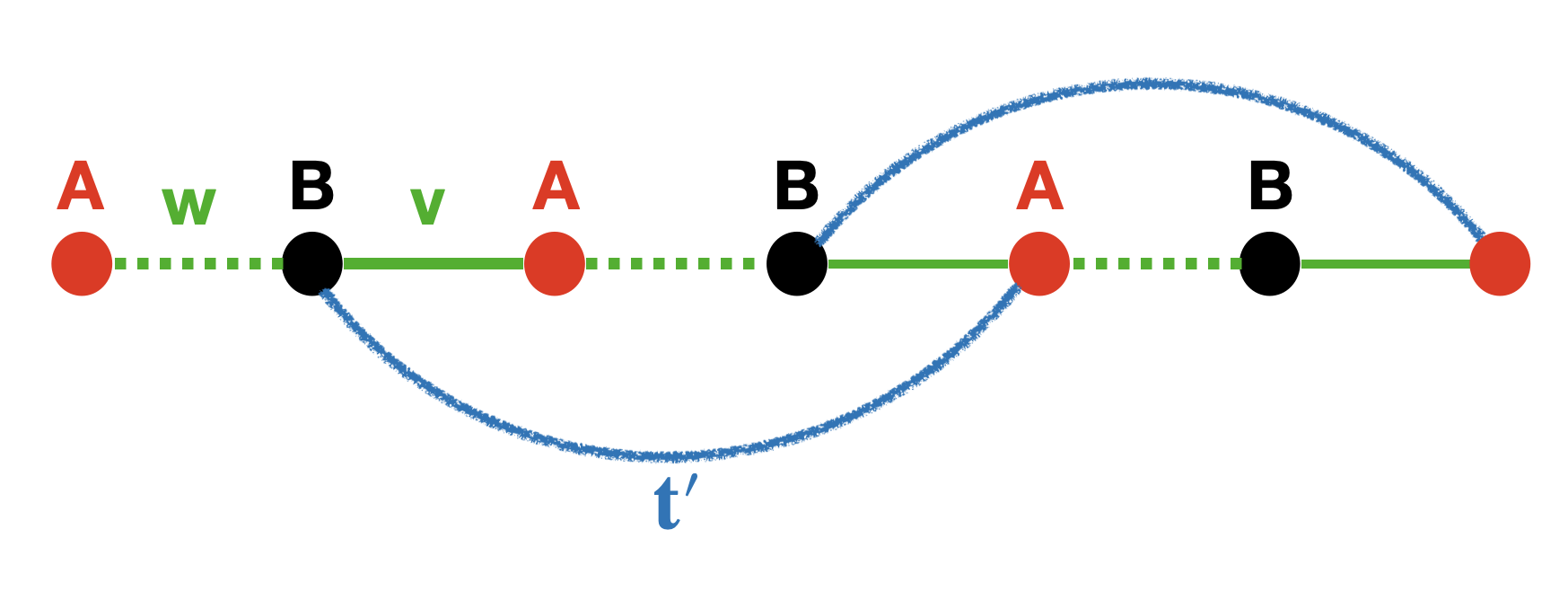}
\caption{ }
\label{Extended_SSH_pic}
\end{subfigure}
\hspace{2cm}
\begin{subfigure}{.4\linewidth}
\includegraphics[scale=0.20]{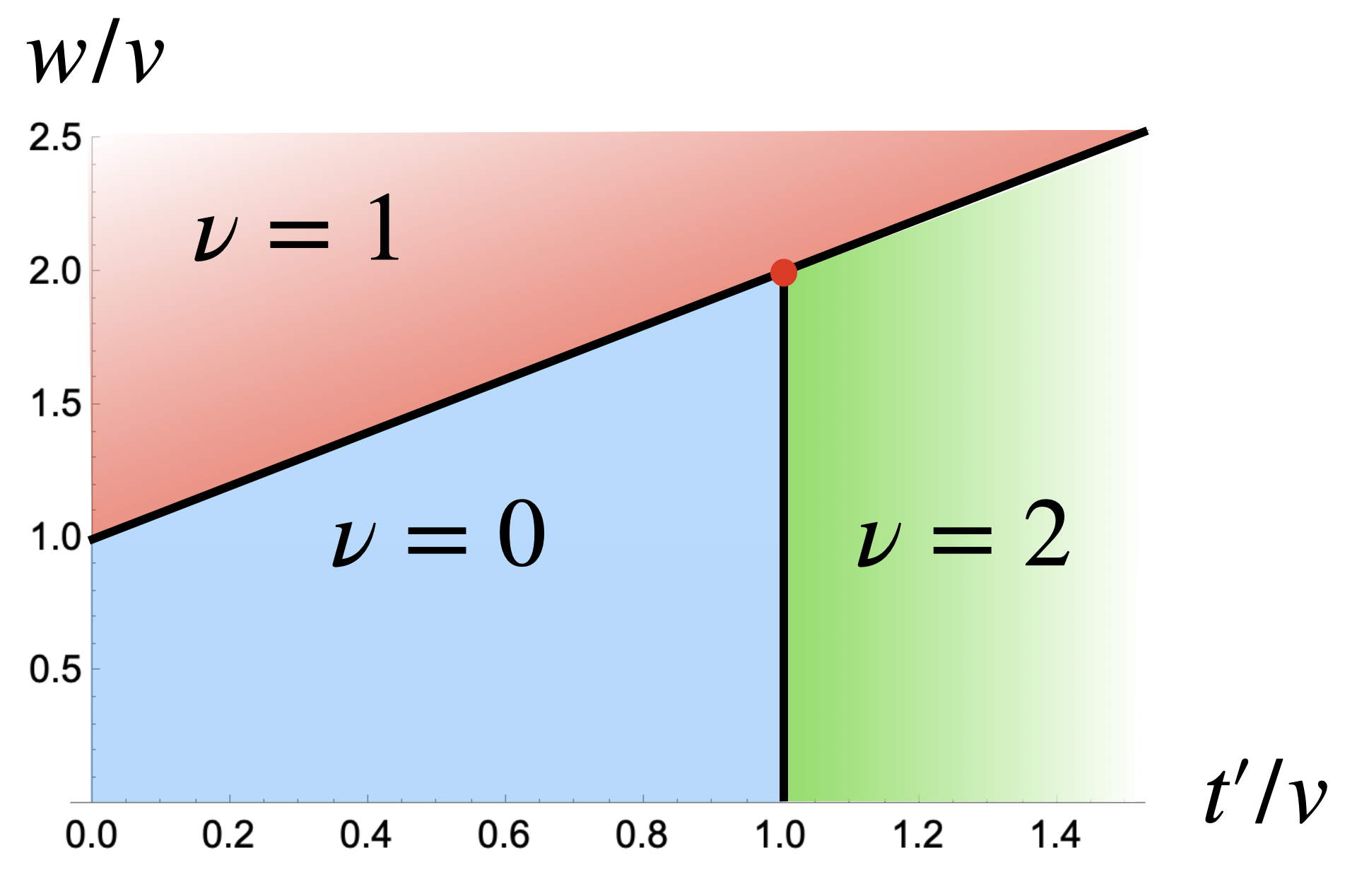}
\caption{}
\label{Extended_SSH_phases}
\end{subfigure}
\caption{a) Illustration of the hopping structure in the Extended SSH model (\ref{ESSH}).  b) Phase diagram of the Extended SSH model characterized by the winding number $\nu$. }
\end{figure*}

\subsection{Model with chiral symmetry $C_2$}
The model invariant under the chiral symmetry operator $C_2$  has zero winding number in the presence of weak inter-chain hopping terms.  That means that such hopping terms split degeneracy of the edge states if there are no additional symmetries.    Let us reproduce this result using bosonization.  A generic model invariant under chiral symmetry $C_2$ has a form: 
\begin{align}
\label{C2_bosonic}
& V_{\text{C}_2} = -\frac{2 (\delta t + \Im(a) +\Im(b))}{\pi a_0}  \cos(\sqrt{2\pi} \phi_c)  \cos(\sqrt{2\pi} \phi_s) + \nonumber \\
& +  \frac{2(\Re(a')-\Re(b'))}{\pi a_0}    \cos(\sqrt{2\pi} \phi_c)  \cos(\sqrt{2\pi} \theta_s).  
\end{align}
Such model is illustrated in Fig.   \ref{Coupling_chains_b}. The model of class CI corresponds to the case when $a$ and $b$ are real and for class DIII we need to set $\Re(a')= \Re(b')=0$.   

In order to see that the inter-chain hopping breaks the degeneracy of the edge states we need to go to the basis when inter-chain hopping is diagonal.  In this basis $\cos(\sqrt{2\pi} \phi_c) \cos(\sqrt{2\pi} \theta_s) \rightarrow \cos(\sqrt{2\pi} \phi_c) \sin(\sqrt{2\pi} \phi_s)$.  Such term shifts the position of minima in spin sector in uncoupled model,  and thus breaks the degeneracy of the edge states.  For a model with repulsive interactions  however,  with the ground state described by kinks in charge sector,  such inter-chain terms do not change the ground state degeneracy (if they are smaller than interaction strength).  Thus,  similar to the case of $C_1$ chiral symmetry,     the set of effective protective symmetries for the interacting model may be different from symmetries that protect the degeneracy of a non-interacting model.  

 Note that  such  degeneracy breaking terms  are prohibited if time-reversal symmetry $T_-$ is present.  This reflects the fact that DIII class, that has this symmetry,  has non-trivial $\mathbb{Z}_2$ classification.  The degeneracy of the edge states in that case is protected both by chiral  and  time-reversal symmetries.   In particular,   Kramers' theorem guarantees the twofold degeneracy of the single-particle states and chiral symmetry pins them at zero energy.   Therefore,  no weak perturbations can remove the degeneracy of the edge states.

\section{Extended SSH model}\label{Sec:extendedSSH}
Earlier we studied a bosonic theory of two SSH chains.  We found that the number of bosonic fields is related to the number of bands in the spectrum that become gapless at a critical point that separates two phases with different topological indices.   In this section we show that it is actually also holds for a single band model with higher winding number.  We consider a simple example of such a model that represents an extension of a single SSH chain with possible winding numbers $\nu =0,1,2$.  We also prove that a model at the transition line between two phases that differ by winding number $\Delta \nu = n$ is equivalent in bosonic language to at least $n$ chains.  

\subsection{Phase diagram of Extended SSH model}
Consider an extended SSH model that describes the SSH chain with next nearest neighbour hopping:  
\begin{align}
\label{ESSH}
& H_{\text{ESSH}} = v \sum_n c^{\dagger}_{A,n} c_{B,n} + w \sum_n c^{\dagger}_{B,n} c_{A, n+1} + \nonumber \\ &+t' \sum_n c^{\dagger}_{B,n} c_{A,n+2} +\text{h.c.} 
\end{align}
Note that properties of such a model at the critical lines were studied in \cite{Verresen2018}.   The new hopping term $t'$ preserves chiral symmetry,  as it couples different sublattices.  If hopping amplitudes are complex,  the model breaks time-reversal symmetry and thus belongs to the topological class AIII that has $\mathbb{Z}$ classification.   
\begin{figure*}[!t]
\begin{subfigure}{.4\linewidth}
\includegraphics[scale=0.23]{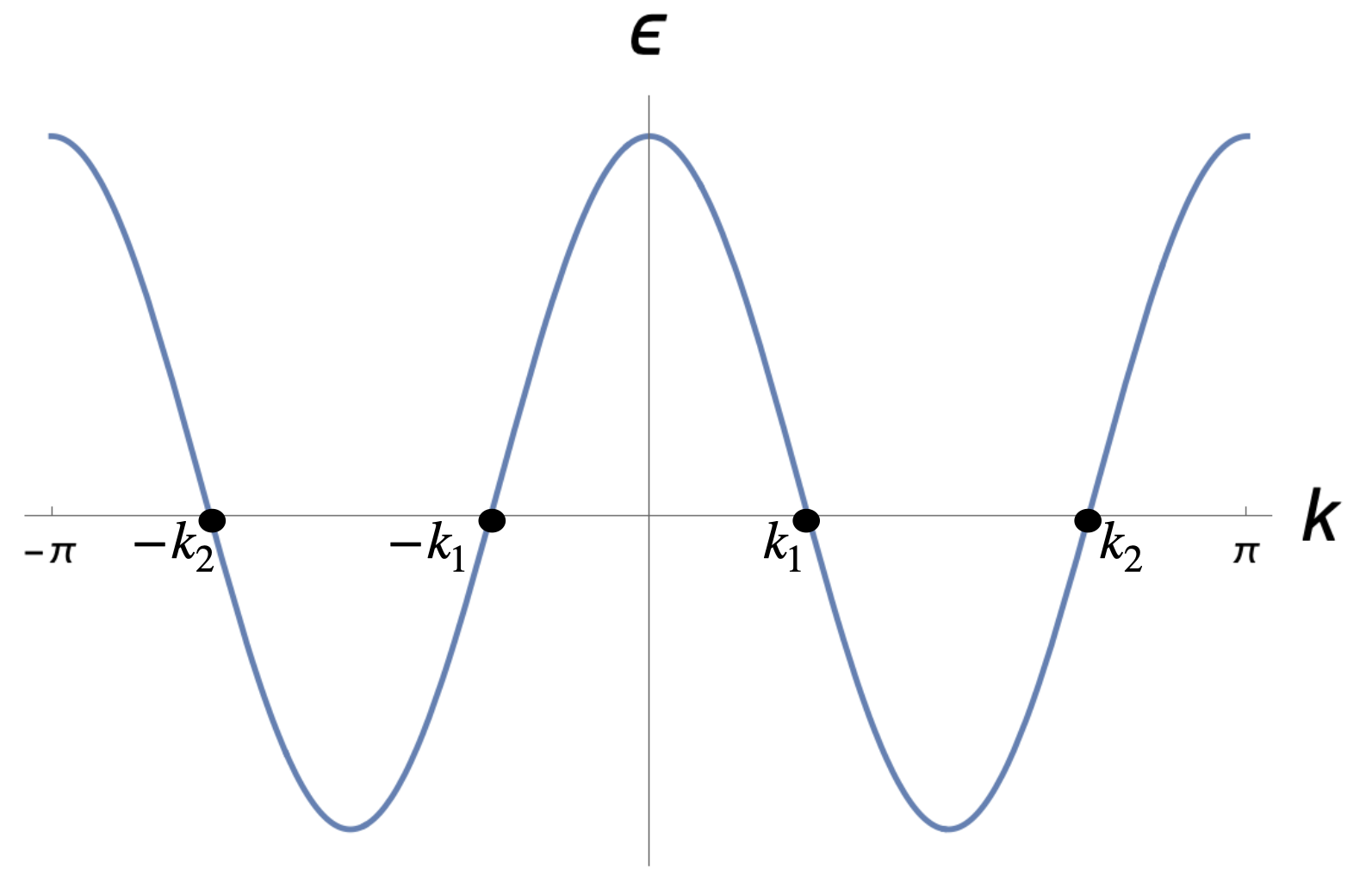}
\caption{}
\label{Extended_SSH_Spectrum02}
\end{subfigure}
\hspace{2cm}
\begin{subfigure}{.4\linewidth}
\includegraphics[scale=0.26]{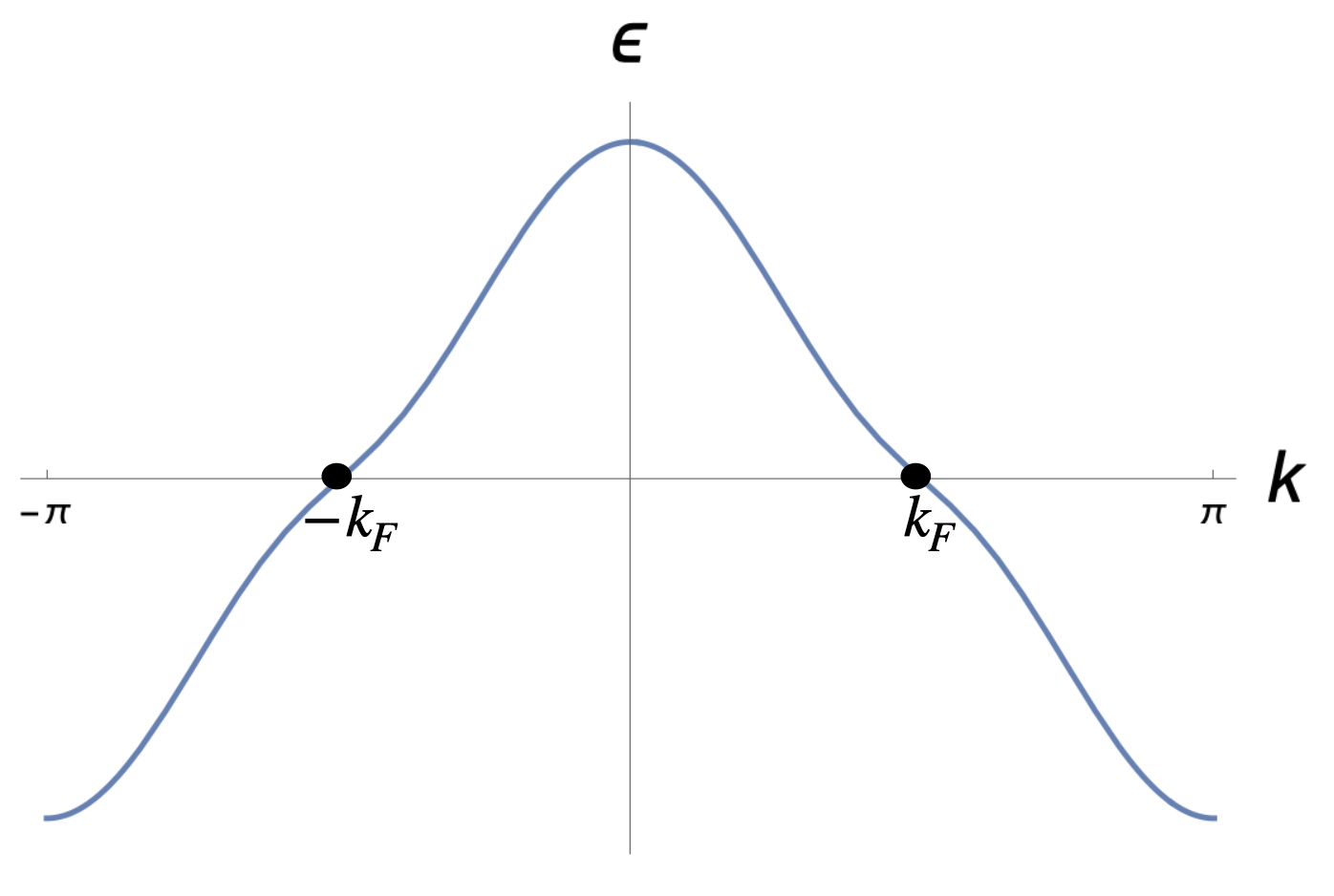}
\caption{}
\label{Extended_SSH_Spectrum01}
\end{subfigure}
\caption{a) The spectrum of the Extended SSH model at the transition line $t'=v$ between the phases $\nu=0$ and $\nu=2$.  There are two inequivalent Fermi points given by $k_{1,2} = \pm \arccos[w/2t']$ b) Spectrum at the transition line $w/v = t'/v+1$  between the phases with $\nu=1$  and $\nu=0,2$.  It crosses the Fermi level at $ \pm k_F$, where $k_F= \pi/2$. } 
\label{Extended_SSH_Spectrum}
\end{figure*}
The hopping structure of the model is illustrated in Fig.  \ref{Extended_SSH_pic}.  Let us overview its topological properties.  For illustration, we will focus on the case when all parameters are positive $w, v, t'>0$, which covers all possible values of the winding number.   One can write the Hamiltonian in $k$- space in the reduced Brillouin zone in block off-diagonal form:
\begin{align}
\label{Extended_SSH_blocks}
\hat{H}_{\text{ESSH}}= \sum_k c^{\dagger}_k h_{\text{ESSH}} c_k,   \\
h_{\text{ESSH}} = \begin{pmatrix}
0 & \Delta(k) \\ 
\Delta^*(k) &0
\end{pmatrix},  \nonumber
\end{align}
where $\Delta(k) = v+ w e^{ik}+ t' e^{2ik}$.  One can compute the winding number of the model via (\ref{phase_detq}) and obtain the phase diagram illustrated in Fig.  \ref{Extended_SSH_phases}.  The phase boundaries are determined by the equations: 
\begin{align}
\begin{cases}
\text{Between $\nu=1$ and $\nu=2$:} \hspace{0.4cm} w/v = t'/v+1,  t'/v<1  \\
\text{Between $\nu=1$ and $\nu=0$:} \hspace{0.4cm} w/v = t'/v+1, t'/v>1  \\
\text{Between $\nu=0$ and $\nu=2$:} \hspace{0.4cm} t'/v=1  \nonumber
\end{cases}
\end{align}
Next, we will focus on the transition lines and discuss the bosonic theories that characterize the model in the vicinity of those lines.  
\subsection{Bosonic description of Extended SSH model}
To bosonize the model we need to find gap closing points in the momentum space and linearize the theory around them.  For each of the critical lines we get: 
\begin{align}
\begin{cases}
\text{Between $\nu=1$ and $\nu=2,0$:} \hspace{0.5cm} k=\pi \\
\text{Between $\nu=0$ and $\nu=2$:} \hspace{0.3cm} k_{1,2}= \pm \arccos[w/2t'] \nonumber
\end{cases}
\end{align}
The spectrum for each of the three cases is illustrated in Fig.  \ref{Extended_SSH_Spectrum} in the extended Brillouin zone.  In the case when the winding number changes by 1 across the transition the spectrum crosses the Fermi level $\epsilon=0$ at $\pm k_F=\pi/2$ similar to the case of a single SSH chain.  Thus the corresponding low-energy model is described by a single Luttinger Liquid.  To demonstrate its relation to the SSH chain,  let us slightly move away from the transition line:  $v \simeq w-t +\delta v$.  We linearize the Hamiltonian in the vicinity of gap-closing points  and obtain: 
\begin{align}
h_{\text{ESSH}}(k)  \approx \delta v \sigma_x -(2t-w) k \sigma_y,  
\end{align}
where the Pauli matrices $\sigma_{x,y}$ act in the sublattice basis.   This Hamiltonian coincides with the linearized Hamiltonian of the SSH model in the vicinity of the gap closing point $w=v$. 
The linearized Hamiltonian can be written in the basis of left- and right-moving modes where $\sigma_y \rightarrow \sigma_z, \sigma_x \rightarrow \sigma_x$  and thus $\delta v$ term can be interpreted as the mass term in that basis,  as it hybridizes left and right movers.  By bosonizing this term,  and by taking into account the fact that in the original lattice model,  different sublattices are separated by an odd number of lattice sites,  that results in $\pi/2$ relative phase between left and right moving fermions as follows from  (\ref{RL_fermions}),  we reproduce the bosonic Hamiltonian of SSH model  (\ref{SSH_bosonised}).

One can use similar arguments in the case of the transition between $\nu=0$ and $\nu=2$ where the spectrum has two inequivalent Fermi points.  One can linearize the Hamiltonian around the two Fermi points and obtain a model that describes two capacitively coupled SSH chains,  that were studied in Section \ref{Sec:uncoupledSSH}.

\subsection{Winding number and number of Fermi points}
Let us show that at the transition line where the winding number changes by $n$ the spectrum has at least $n$ distinct Fermi points,  and thus the corresponding bosonic low-energy model is equivalent to the theory of $n$ (or more) coupled chains.  This statement follows from the geometric interpretation of the winding number.  Consider the $N$ band model with chiral symmetry,  so in the sublattice basis the Hamiltonian can be written in block off-diagonal form,  where the block is given by some $N \times N$ matrix $\Delta(k)$.   One can write its determinant as $\det \Delta(k) = r(k) e^{i\phi(k)}$,  where the absolute value of the determinant is given by \cite{Matveeva2023} $r(k) = \prod_j \epsilon_j(k)$,  where $\epsilon_j(k)$ -- is the energy band labeled by j.  So  $\det \Delta(k_0)=0$ at a gap closing point $k_0$,  where one or few of the bands become gapless.  The winding number is given by the number of times the complex function $\ln \det \Delta(k)$ changes the branch as $k$ runs across the Brillouin zone and essentially is given by the number of times the function winds around the point $\det \Delta(k)=0$.  Thus if one deforms the Hamiltonian such that $\det \Delta(k)$ crosses 0 at $n$ points in the  Brillouin zone the winding number changes by $n$.

\section{Discussion and conclusion}

In our work,  we developed a bosonization framework that allows to study the topological properties of interacting one-dimensional fermionic models.  Specifically,  in our paper,  we focused on the topological properties of interacting topological insulators built from SSH chains.   We studied properties of the edge states in finite models with boundary conditions imposed by a strong impurity or in the interface between trivial and topological phases.   The topological edge states are manifested as degenerate kinks in bosonic fields at the boundary that appear when the values of the bosonic fields in the bulk and at the boundary are incompatible.   We applied this idea to a single SSH chain,  capacitively coupled SSH chains,  coupled chain model of topological insulators in different symmetry classes,  and the SSH model with longer range hopping.   We found the following properties of these models: \\

(a)  We compared different approaches to the boundary conditions in a bosonic model on the example of a single SSH chain.    In particular,  we considered boundary conditions imposed by a strong impurity and at the boundary between trivial and topological phases.   We showed that they are equivalent if an additional scattering phase is added to the impurity potential.   Furthermore,  we computed the localization length in the interacting SSH model with such boundary conditions and found it qualitatively similar to results obtained via open boundary bosonization.    \\

(b) We studied ground state properties of  two capacitively coupled  SSH chains,  that can be identical or in different topological phases.  We demonstrated that the degeneracy of the topological edge states in the topological phases is determined by a number of degenerate bosonic kinks at the boundary in charge and spin sectors.   By constructing the action of anti-unitary symmetry operators on bosonic language,  we proved that degeneracy of the edge states is protected by chiral symmetry,  similar to a non-interacting model.    We showed that weak interactions can partially split the degeneracy of the non-interacting ground state.  Interestingly,  the remaining degenerate subspace of the edge states can be protected by a different set of symmetries rather than chiral,   depending on the sign of interactions.     \\

(c) Next we added inter-chain hopping to the two-chain model to break additional unitary symmetries of uncoupled chains.  Symmetries of a coupling put a model to a certain non-interacting symmetry class.  As we showed for the non-interacting model in \cite{Matveeva2023} there are two possible chiral symmetry operators for such a model,  and the winding number for a weakly coupled system is determined only by a type of chiral symmetry.   We reproduced these properties using bosonization by writing down a bosonic model compatible with symmetries of a certain class.      \\

(d) We proved that in the vicinity of the phase transition between the phases with topological indices that differ by $n$ any non-interacting model is equivalent to at least $n$ decoupled chains.   The number of chains is equal to the number of Fermi points at a gapless boundary between gapped phases.  Near each Fermi point a bosonic model is equivalent to the bosonic model of a single SSH chain.  We illustrated this idea with the example of an extended SSH model that includes next-nearest neighbor hopping in addition to nearest-neighbour terms of a SSH chain.  

\section{Acknowledgements}
D. G.  is supported by ISF-China 3119/19 and ISF 1355/20. P.M. acknowledges support by the Israel Council for Higher Education Quantum Science and Technology Scholarship.  We acknowledge useful discussions with Elio K\"onig,  Pavel Ostrovsky and Richard Berkovits.

\onecolumngrid
\appendix
\section*{Appendix} 
\section{Bosonisation conventions for spinless fermions}\label{bosonization_spinless}
Here we outline the bosonization conventions used for the spinless fermions. The right- and left-moving fermions can be expressed via exponents of bosonic fields  $\phi_R(x)$ and $\phi_L(x)$: 
\begin{align}
\label{RL_bosonized}
R(x) = \frac{1}{\sqrt{2\pi a}} e^{i\sqrt{4\pi} \phi_R},  \hspace{0.3cm} L(x) = \frac{1}{\sqrt{2\pi a}} e^{-i\sqrt{4\pi} \phi_L}. 
\end{align}
The bosonic fields $\phi_R(x)$ and $\phi_L(x)$ satisfy the following commutation relations: 
\begin{align}
\label{RLboson_commutators}
[\phi_R(x), \phi_L(y)] = \frac{i}{4},  \hspace{0.3cm}
[\phi_\eta (x), \phi_{\eta'}(y)]=  \frac{i}{4} \eta \delta_{\eta, \eta'} \text{sign}(x-y).
\end{align}
This allows us to introduce the following fields: 
\begin{align}
\phi(x)= \phi_R(x)+\phi_L(x),  \hspace{0.3cm}  \theta(x) = \phi_L(x) -\phi_R(x).
\end{align}
The fields $\phi(x)$ and $\Pi(x) = \partial_x \theta(x)$ are canonically conjugate,  meaning $[\phi(x), \Pi(x')]= i \delta(x-x')$.   Using these conventions,   
we obtain the following useful relations: 
\begin{align}
R^{\dagger}L = \frac{-i}{2\pi a} e^{-i\sqrt{4\pi} \phi}, \nonumber \\
R^{\dagger} R + L^{\dagger}L= \frac{1}{\sqrt{\pi}} \partial_x\phi,
\end{align}
where we also used the Campbell-Baker-Hausdorff formula for two non-commuting operators A and B: 
\begin{align}
\label{Campbell-Baker-Hausdorff}
e^Ae^B=e^{A+B}e^{\frac{1}{2}[A,B]}.
\end{align}
The density operator in terms of right- and left-movers is given by:  
\begin{align}
n_j \rightarrow R^{\dagger}(x) R(x) + L^{\dagger}(x) L(x) + e^{ik_Fx} R^{\dagger}(x) L(x) + e^{-ik_Fx}L^{\dagger}(x) R(x).  
\end{align}
\section{Bosonization conventions for fermions with spin}\label{bosonization_spinful}
Bosonization rules for spinless fermions reviewed in Appendix \ref{bosonization_spinless} can be generalized to the spinful case.   
The equation (\ref{RL_bosonized}) is generalized by introducing  Klein factors,  that ensure the correct anticommutation relations between  fermionic operators with different spin indexes: 
\begin{align}
\label{RL_bosonizedS}
R_{\sigma}(x) = \frac{\kappa_{\sigma}}{\sqrt{2\pi a}} e^{i\sqrt{4\pi} \phi_{R\sigma}},  \hspace{0.3cm} L_{\sigma}(x) = \frac{\kappa_{\sigma}}{\sqrt{2\pi a}} e^{-i\sqrt{4\pi} \phi_{L\sigma}}. 
\end{align}
The Klein factors $\kappa_{\sigma}$ are Hermitian operators and they satisfy the algebra $\{\kappa_{\sigma}, \kappa_{\sigma'}\}= 2\delta_{\sigma \sigma'}$.   The specific representation satisfying this algebra can be chosen such as $\kappa_{\uparrow}\kappa_{\downarrow} = -\kappa_{\downarrow}\kappa_{\uparrow}=i$.   The commutation relations between the bosonic fields (\ref{RLboson_commutators}) are generalized such that fields with different spin index commute: 
 \begin{align}
\label{RLboson_commutators_spin}
[\phi_{R\sigma}(x), \phi_{L\sigma'}(y)] = \frac{i}{4} \delta_{\sigma,\sigma'}\nonumber \\ 
[\phi_{\eta, \sigma} (x), \phi_{\eta'\sigma'}(y)]=  \frac{i}{4} \eta \delta_{\eta, \eta'} \delta_{\sigma,\sigma'} \text{sign}(x-y),
\end{align}
here the indices $\sigma$ and $\sigma'$ denote the spin degree of freedom.  
It is useful to introduce the charge and spin bosonic fields: 
\begin{align}
\label{charge_spin_bosons}
\phi_c = \frac{\phi_{\uparrow}+\phi_{\downarrow}}{\sqrt{2}}, \hspace{0.3cm} \phi_s = \frac{\phi_{\uparrow}-\phi_{\downarrow}}{\sqrt{2}}.
\end{align}
One can use those fields to express the non-oscillatory part of charge and spin densities:  
\begin{align}
\rho_c = \sum_{\sigma} \Psi^{\dagger}_{\sigma} \Psi_{\sigma}= \sqrt{\frac{2}{\pi}} \partial_x\phi_c, \nonumber\\ 
\rho_s = \frac{1}{2} \sum_{\sigma\sigma'} \Psi_{\sigma}^{\dagger} (\sigma_z)_{\sigma\sigma'}  \Psi_{\sigma'}= \frac{1}{\sqrt{2\pi}} \partial_x\phi_s.  
\end{align}

\section{Ising phase transition in interacting SSH chain}\label{Ising_app}
Consider the non-linear terms in the bosonized interacting SSH chain model (\ref{SSH_int_H}):  
\begin{align}
\label{VgapSSH_app}
V_{\text{gap}}= \frac{g}{2(\pi a_0)^2} \cos[2\sqrt{4\pi} \phi] -\frac{\delta t }{\pi a_0} \int dx \cos[\sqrt{4\pi} \phi(x)].
\end{align}
We focus on the case $g>0$ when the minima of the two cosines are incompatible.  If interactions are weak,  i.e.  $g \ll |\delta t|$ the minima of the potential  (\ref{VgapSSH_app}) coincides with the minima in the non-interacting model,  given by: 
\begin{align}
\begin{cases}
\delta t>0:  \sqrt{4\pi} \phi(x)=0  \mod 2\pi  \\
\delta t<0: \sqrt{4\pi} \phi(x) = \pi  \mod 2\pi 
\end{cases}
\end{align}
In the limit of strong interactions $g \gg |\delta t|$ the position of minima are shifted to $\sqrt{4\pi}\phi = \pm \pi/2  \mp \phi_0$,  where $\phi_0 \approx \delta t (\pi a_0)/2g$.   The two phases can be distinguished by a local charge density wave order parameter,  which describes the oscillating $2k_F$ component of a density,  $n_{\text{osc}}(x) = (1/\pi a_0) \sin[\sqrt{4\pi}\phi(x)] $: 
\begin{align}
\begin{cases}
 n_{\text{osc}}(x) =0 ,  \hspace{0.2cm} g \ll |\delta t|  \\
n_{\text{osc}}(x) \approx  \pm 1/\pi a_0,  \hspace{0.2cm} g \gg |\delta t| 
\end{cases}
\end{align}
Therefore there is a phase transition between the phases with $\langle n_{\text{osc}}(x) \rangle=0$ and $ \langle n_{\text{osc}}(x) \rangle \neq 0$  as one tunes the ratio $\delta t/g$.   Such a phase transition is described by an effective $\phi^4$ theory that can be obtained if one expands the potential (\ref{VgapSSH_app}) around one of the minima of the non-interacting cosine: 
\begin{align}
\label{phi4_SSH}
& V_{\text{eff}}= \alpha_0 + \frac{\alpha_2}{2!} (\delta \phi)^2+ \frac{\alpha_4}{4!} (\delta\phi)^4,  \nonumber \\ 
& \alpha_0= \frac{g}{2(\pi a_0)^2}+\frac{|\delta t| }{\pi a_0},  \alpha_2= 8\left(-\frac{g}{a_0^2 \pi}+\frac{|\delta t|}{2a_0}\right), \nonumber \\  & \alpha_4= 16 \left( \frac{8g}{a_0^2} -\frac{\pi |\delta t|}{a_0} \right).
\end{align}
Here $\delta \phi$ is the deviation of $\sqrt{4\pi}\phi$ from $0$ or from $\pi$. The transition happens at the point determined by $\alpha_2=0$,  i.e.  when $|\delta t| = 2g/a_0\pi$.  Such phase transition is known in the context of quantum Ising chain model \cite{Delfino1996}.  

\section{Symmetries in bosonic language}\label{bosonized_symmetries}
\subsection{Single chain limit}
Let us derive the action of  chiral symmetry onto the bosonic fields in the case of a single chain.  In the many-body space chiral symmetry acts as \cite{Ludwig2016}: 
 \begin{align}
 \label{chiral_manybody_lattice}
\hat{C}^{-1} c_j \hat{C} = (-1)^{j} c^{\dagger}_j.  
 \end{align}
Note,  that in the single-particle description this operator corresponds to the unitary matrix $- \sigma_z$ in the space of sublattices $A$ and $B$.  This is consistent with the symmetry implemented in the single particle Fock space, 
where  single-particle Hamiltonian (\ref{SSH_tight_binding_k}) anticommutes with this matrix,
$\{h_{SSH},\sigma_z\}=0\,$.
The chiral symmetry acting in the many-particle Fock space is an antiunitary operator,  i.e.  $\hat{C}^{-1} i \hat{C} = -i$.  
To figure out the action of the chiral symmetry onto the bosonic fields we first rewrite (\ref{chiral_manybody_lattice}) in the continuous limit:
\begin{align}
\label{chiral_manybody_cont}
C^{-1}\hat{\Psi}(x) C =(-1)^{x/a_0} \hat{\Psi}^{\dagger}(x). 
\end{align}
By using the right- and left- decomposition of the fermionic operator (\ref{RL_fermions}) we obtain: 
\begin{align}
\label{chiral_action1_cont}
C^{-1}\hat{\Psi}(x)C = e^{i\pi x} \left(e^{-ik_Fx} R^{\dagger}(x)+ e^{ik_F x} L^{\dagger}(x) \right).
\end{align}
On the other hand: 
\begin{align}
\label{chiral_action2_cont}
C^{-1}\hat{\Psi}(x)C = \left(e^{-ik_Fx} C^{-1} R(x) C+ e^{ik_F x} C^{-1}L(x) C \right).
\end{align}
Now we take into account that we focus on a half-filled model $k_F=\pi/2a_0$.   By comparing (\ref{chiral_action1_cont}) and (\ref{chiral_action2_cont}) we obtain the following transformation rules: 
\begin{align}
\begin{cases}
C^{-1} R(x) C = L^{\dagger}(x)  \\
C^{-1}L(x) C = R^{\dagger}(x).
\end{cases}
\end{align}
We can rewrite them in terms of bosonic fields using (\ref{RL_bosonized}) and obtain: 
\begin{align}
\begin{cases}
 C^{-1} \phi_R(x) C =-  \phi_{L}(x)\\
 C^{-1} \phi_L(x) C =-  \phi_{R}(x). 
 \end{cases}
\end{align}
The time-reversal symmetry for a single chain does not change the fermionic operator $\hat{\Psi}(x)$,  if we use the definition from \cite{Ludwig2016},  i.e. $T^{-1} \hat{\Psi}(x) T=\hat{\Psi}(x)$,  however it is an anti-unitary symmetry,  so for a single chain is coincides with complex conjugation  $T=\mathcal{K}$.  Thus we obtain: 
\begin{align}
\begin{cases}
T^{-1} R(x) T = L(x)  \\
T^{-1}L(x) T = R(x).
\end{cases}
\end{align}
By using (\ref{RL_bosonized}) we rewrite this via bosonic fields as: 
\begin{align}
\begin{cases}
 T^{-1} \phi_L(x) T = \phi_{R}(x)\\
 T^{-1} \phi_R(x) T =  \phi_{L}(x). 
 \end{cases}
\end{align}
Finally we represent the action of time-reversal and chiral symmetries in terms of the conjugated bosonic fields $\phi(x)$ and $\theta(x)$: 
\begin{align}
\label{CT_bosonic_singlechain}
C:
\begin{cases}
\phi(x) \rightarrow -\phi(x) \\
\theta(x) \rightarrow \theta(x) 
\end{cases}
\hspace{0.5cm}T: 
\begin{cases}
\phi(x) \rightarrow \phi(x) \\
\theta(x) \rightarrow -\theta(x)
\end{cases}
\end{align}
Note,  that the action of the particle-hole symmetry is given by $P=C \cdot T^{-1}$. 
\subsection{Symmetries of two chains}
In \cite{Matveeva2023} we constructed the following set of symmeties  for the two chain model: 
\begin{equation}
\label{time-reversal}
\text{Time-reversal}: \\
\begin{cases}
 T^2=+1 : \hspace{0.2cm} T_+= S_0 \sigma_x  \mathcal{K},  \\
 T^2=-1 : \hspace{0.2cm} T_-=  iS_0 \sigma_y\mathcal{K}.
\end{cases} 
\text{Particle-hole}:  \hspace{0.1cm} \\
\begin{cases}
 P^2=+1 : \hspace{0.3cm} P_+= i S_z \sigma_x \mathcal{K} \\
P^2=-1 : \hspace{0.3cm} P_-= -iS_z \sigma_y\mathcal{K}
\end{cases}
\end{equation}
And for chiral symmetry,  which is the product of time-reversal and particle-hole symmetries we get:   
\begin{equation}
\text{Chiral}:\\
\begin{cases}
C_1 = P_+ T_+ = P_- T_- = S_z\sigma_0\\
C_2 = P_- T_+ = P_+ T_- = S_z\sigma_z ,
\end{cases}
\end{equation}
where Pauli matrices $S_i$ act on sublattice degrees of freedom and $\sigma_i$ describe chain degree of freedom.   Let us derive bosonic representation of these symmetries. 
We start with chiral symmetry that can be represented by two operators $C_1=S_z\sigma_0$ and $C_2=S_z\sigma_z$ in a single-particle space.     Let us  start with the right-left decomposition of the fermionic operators: 
\begin{align}
\label{PsiPM}
\hat{\Psi}_{\sigma}(x) = e^{ik_{F}x} R_{\sigma}(x)+ e^{-ik_{F}x} L_{\sigma}(x).
\end{align}
The chiral symmetry $C_1$  acts on the operators $\hat{\Psi}_{\sigma}$ as follows: 
\begin{align}
\label{chiral_manybody_contS}
C_1^{-1}\hat{\Psi}_{\sigma}(x) C_1 =(-1)^{x/a_0} \hat{\Psi}_{\pm}^{\dagger}(x). 
\end{align}
By using (\ref{PsiPM}) one obtains: 
\begin{align}
C^{-1}_1\hat{\Psi}_{\sigma}(x)C_1 = e^{i\pi x} \left(e^{-ik_{F}x} R^{\dagger}_{\sigma}(x)+ e^{ik_{F}x} L^{\dagger}_{\sigma}(x) \right).
\end{align}
That implies the following action of the chiral symmetry onto the fermionic operators $\hat{\Psi}_{\sigma}$: 
\begin{align}
C_1^{-1}\hat{\Psi}_{\sigma}(x)C_1 = \left(e^{-ik_{F}x} C_1^{-1} L_{\sigma}(x) C_1+ e^{ik_F x } C_1^{-1}R_{\sigma}(x) C_1 \right).
\end{align}
By comparing the two expressions we obtain the following transformation rules in terms of chiral fermions:
\begin{align}
C_1^{-1} R_{\sigma}(x) C_1 = L^{\dagger}_{\sigma}(x) \nonumber \\
C_1^{-1} L_{\sigma}(x) C_1 = R^{\dagger}_{\sigma}(x).
\end{align}
In order to rewrite that in terms of bosonic fields,  one needs to use (\ref{RL_bosonizedS}) and take into account the action of complex conjugation onto the Klein factors.  By using the identity $\eta_1\eta_2=i$,  and therefore $\mathcal{K} \eta_1\eta_2\mathcal{K}=-i$,  we may assume that  $\mathcal{K} \eta_{1} \mathcal{K} =\eta_2$ and $\mathcal{K} \eta_{1} \mathcal{K} =\eta_2$.  That yields:
\begin{align}
\label{RL_ boson_C1}
 C^{-1} _1\phi_{R,\sigma}(x) C_1 =-  \phi_{L,\sigma}(x) \mp \pi/(2 \sqrt{4\pi}) \nonumber  \\ 
 C^{-1} _1\phi_{L,\sigma}(x) C_1 =-  \phi_{R,\sigma}(x) \pm \pi/(2 \sqrt{4\pi}).
\end{align}
That corresponds to the following transformation of the bosonic fields: 
\begin{align}
\label{boson_C1}
 C^{-1} _1\phi_{\sigma}(x) C_1 =-  \phi_{\sigma}(x)  \\ 
C^{-1} _1 \theta_{\sigma}(x) C_1 = \theta_{\sigma}(x) \mp \sqrt{\pi}/2.
\end{align}
In terms of charge and spin degrees of freedom,  we obtain: 
\begin{align}
\label{charge_spin_C1}
C_1:
\begin{cases}
\phi_c \rightarrow -\phi_c \\
\phi_s \rightarrow -\phi_s 
\end{cases}
\text{For the conjugated fields:} \hspace{0.5cm}
\begin{cases}
\theta_c \rightarrow \theta_c \\
\theta_s \rightarrow \theta_s + \sqrt{\pi/2}.   
\end{cases}
\end{align}
The action of the second chiral symmetry operator $C_2$ can be obtained in a similar way,  taking into account that it acts also on a chain degree of freedom,  it particular it multiplies the fermionic operator on the second chain by $-1$.  Thus we get: 

\begin{align}
\label{charge_spin_C2}
C_2:
\begin{cases}
\phi_c \rightarrow -(\phi_c +\sqrt{\pi/2}) \\
\phi_s \rightarrow -(\phi_s +\sqrt{\pi/2})
\end{cases}
\text{For the conjugated fields:} \hspace{0.5cm}
\begin{cases}
\theta_c \rightarrow  \theta_c \\
\theta_s \rightarrow  \theta_s +\sqrt{\pi/2}
\end{cases}
\end{align}

Now let us focus on the time-reversal symmetry.  The time-reversal symmetry for a two-chain model is represented by two operators with properties $T^2_+=+1 $ and $T^2_-=-1$ (\ref{time-reversal}).   Let us consider the action of $T_+$ first,  the derivation for $T_-$ can be obtained in a similar way.  We use that the operator of time reversal symmetry combines a spin flip expressed by $\sigma_x$ and the complex conjugation and obtain:     
\begin{align}
\label{TR}
T^{-1}_{+} R_1(x) T_{+}  = L_{2}(x)  \\ 
T^{-1}_{+}  R_2(x)T_{+} =L_{1}(x)
\end{align}
Next, we bosonize the fermionic operators by using  (\ref{RL_bosonizedS}) and take into account the action of complex conjugation onto the Klein factors,  and obtain:  
\begin{align}
\begin{cases}
 \tilde{\phi}_{R,1}  =  \phi_{L,2}   \\
 \tilde{\phi}_{R,2}  =  \phi_{L,1}  \\
 \tilde{\phi}_{L,1}  = \phi_{R,2} \\
 \tilde{\phi}_{L,2}  = \phi_{R,1}, 
\end{cases}
\end{align}
where $\tilde{\phi}_{R/L,\sigma} \equiv T^{-1}_+ \phi_{R/L,\sigma} T_+$.
In terms of charge and spin degrees of freedom,  we get: 
\begin{align}
T^2_+=1:
\begin{cases}
\phi_c \rightarrow \phi_c \\
\phi_s \rightarrow -\phi_s 
\end{cases}
\begin{cases}
\theta_c \rightarrow -\theta_c \\
\theta_s \rightarrow \theta_s 
\end{cases}
\end{align}
Similarly,  we can obtain the action of the second time-reversal symmetry operator $T_-$:
\begin{align}
T^2_-=-1:
\begin{cases}
\phi_c \rightarrow \phi_c + \sqrt{\pi/2} \\
\phi_s \rightarrow -\phi_s - \sqrt{\pi/2}
\end{cases}
\begin{cases}
\theta_c \rightarrow -\theta_c \\
\theta_s \rightarrow \theta_s 
\end{cases}
\end{align}
The action of particle-hole symmetry operators is given by $P_+ =C_1 T_+^{-1}$ and $P_- =C_1 T_-^{-1}$: 
\begin{align}
P^2_+=1:
\begin{cases}
\phi_c \rightarrow -\phi_c \\
\phi_s \rightarrow \phi_s 
\end{cases}
\begin{cases}
\theta_c \rightarrow - \theta_c \\
\theta_s \rightarrow  \theta_s +\sqrt{\pi/2}
\end{cases} \\ 
P^2_-=-1:
\begin{cases}
\phi_c \rightarrow -(\phi_c + \sqrt{\pi/2}) \\
\phi_s \rightarrow - \phi_s + \sqrt{\pi/2}
\end{cases}
 \hspace{0.5cm}
\begin{cases}
\theta_c \rightarrow -\theta_c \\
\theta_s \rightarrow  \theta_s+\sqrt{\pi/2}
\end{cases}
\end{align}

\section{Bosonisation of non-interacting topological insulator in class BDI}\label{BDI_bosonization_app} 
\subsection{Two band limit}
\begin{figure}
\begin{subfigure}{.3\linewidth}
\includegraphics[width=6cm]{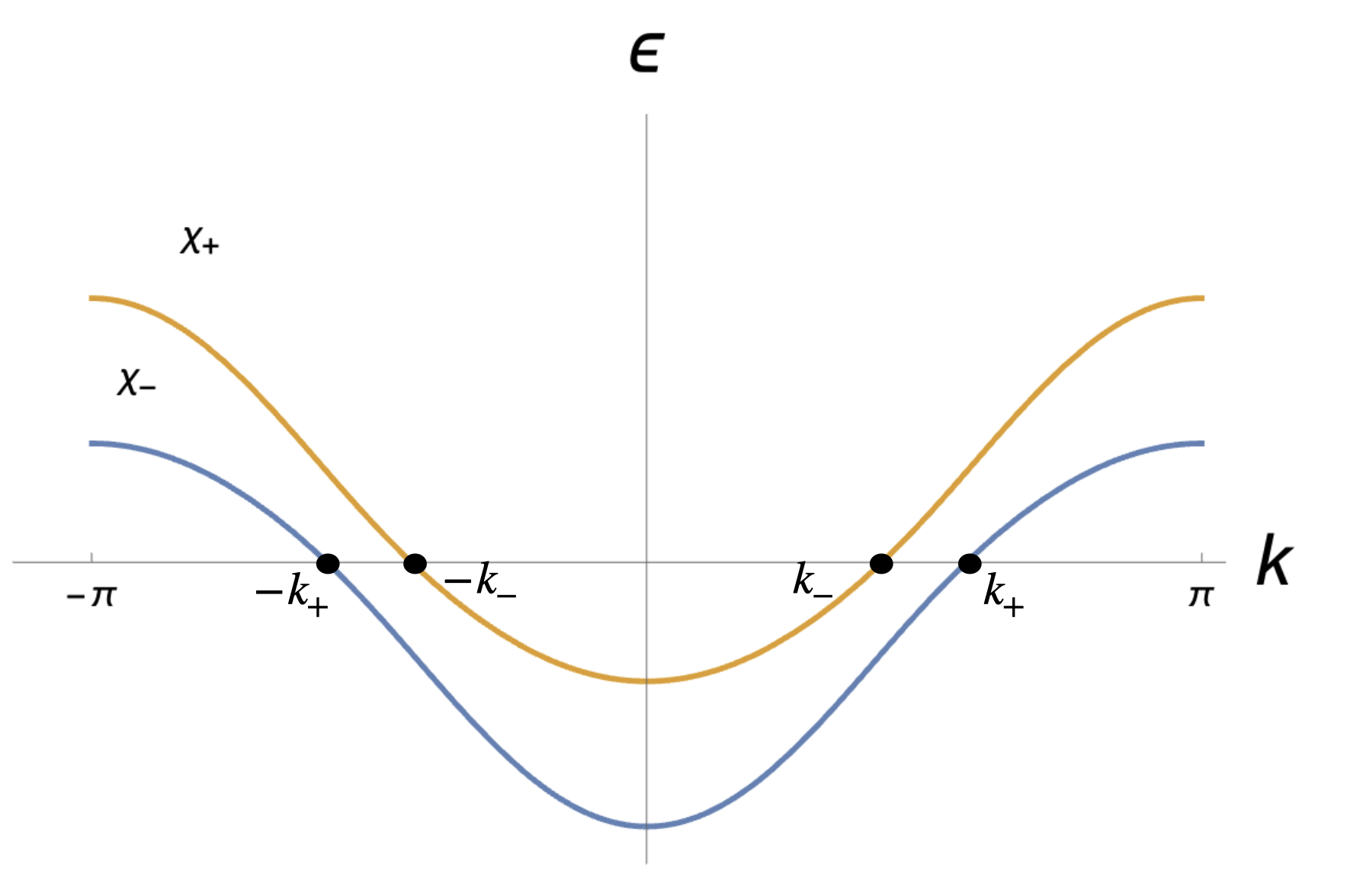}
\caption{}
\label{gapless_spectrum_app_a}
\end{subfigure}
\hspace{3cm}
\begin{subfigure}{.3\linewidth}
\includegraphics[width=6cm]{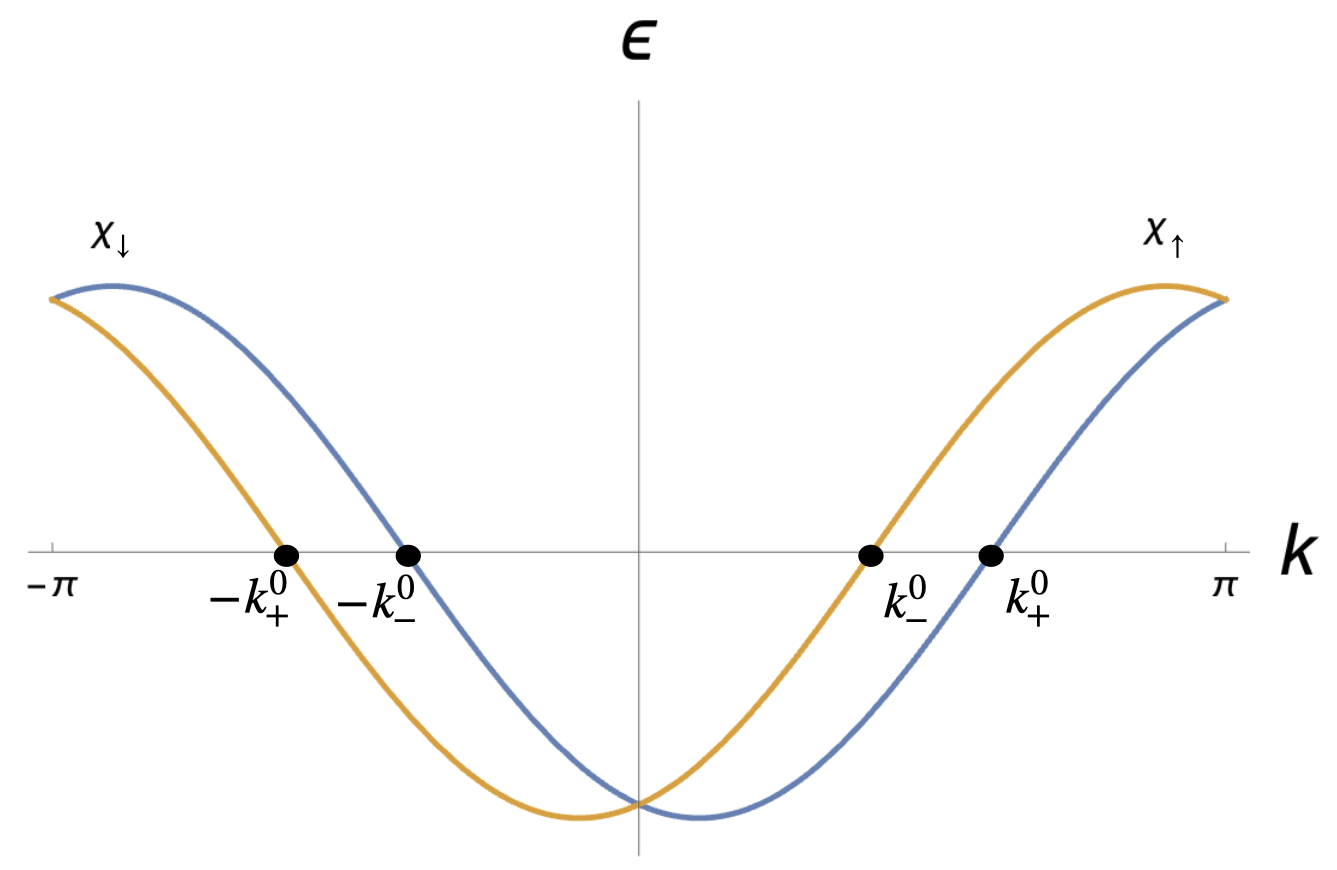}
\caption{}
\label{gapless_spectrum_app_b}
\end{subfigure}
\caption{a)Gapless part of the spectrum of the BDI model (\ref{gapless_fermions_BDI}) for generic parameters $t<0, t_z>0,  t_x \neq 0$,  the difference between the Fermi momenta is given by $\delta k= k_+-k_-= 2\arctan(t_z/\sqrt{t^2-t_x^2})$ b)Gapless part of the spectrum of the BDI model (\ref{gapless_fermions_BDI}) for the case  $t_x=0$,  $t<0, t_z>0$. The difference between the Fermi momenta $\delta k^0= k^0_+-k^0_-=2\arctan(t_z/t)$. }
\end{figure}

Here we bosonize the model of the topological insulator in the class BDI.  We start with the coupled model from the main text (\ref{H1}): 
\begin{equation}
\label{H1_app}
 \hat{H}_1 = \hat{H}_0+ \hat{V}_1,
 \end{equation}
 where $\hat{H}_0$ describes the Hamiltonian of two uncoupled SSH chains. Here we will focus on a more general model,  compatible with symmetries of BDI class.  In such model the uncoupled SSH chains have complex hopping parameters $w_i$ and $v_i$,  related by complex conjugation.  They can be represented as $w_1 = w_2^*=|w| e^{i\phi_0}, v_1= v_2^*= |v| e^{i\phi_0}$.  The inter-chain coupling term $\hat{V}_1$ is given by: 
 \begin{align}
 \label{BDI_app}
 \hat{V}_1 &=  \sum_n a \left( c^\dagger_{A,1,n} c_{B,2,n} + c^\dagger_{A,2,n} c_{B,1,n} \right) + b  \left( c^\dagger_{B,1,n} c_{A,2,n+1} 
 + c^\dagger_{B,2,n} c_{A,1,n+1}    \right) + \text{h.c.}
 \end{align} 
The parameters $a$ and $b$ are real.  Next we separate the gapless part of the Hamiltonian and the terms that open up a gap.  
The gapless part of the Hamiltonian is written  as:
\begin{align}
\label{gapless_fermions_BDI}
H_0= \sum_n t c_n^{\dagger}\sigma_0 c_{n+1}+ t_x c_n^{\dagger}  \sigma_x c_{n+1} +i t_z c_n^{\dagger}  \sigma_z c_{n+1} +  \text{h.c.},
\end{align}
here $n$ -- is the site index,  and the operators $c_n$ are written in the chain basis,  $\hat{c}^{\text{T}}_n=\{c_{n,1}, c_{n,2}\}$,   the parameters $t=(|w|+|v|)\cos[\phi_0]/2$,  $t_z = (|w|+|v|)\sin[\phi_0]/2$,  $t_x= (a+b)/2$.  
The gap opening part is given by:  
\begin{align}
\label{BDI_gapped_app}
V_{\text{BDI}}= \sum_n (-1)^n \left[ \delta  t \hat{c}_n^{\dagger}\sigma_0 \hat{c}_{n+1}+i\delta t_z \hat{c}_n^{\dagger}\sigma_z \hat{c}_{n+1} + \delta t_x \hat{c}_n^{\dagger} \sigma_x \hat{c}_{n+1}  \right] + \text{h.c.}
\end{align}
here $\delta t=(|w|-|v|)\cos[\phi_0]/2$,  $\delta t_z = (|v|-|w|)\sin[\phi_0]/2$,  $\delta t_x= (a-b)/2$.   
We first bosonize the gapless part of the model. In order to do that,  we write  (\ref{gapless_fermions_BDI}) in $k$- space:
\begin{align}
\label{gapless_fermions_CII_kspace}
H_0= 2  \sum_k  c_k^{\dagger} h_k c_{k},  \hspace{0.3cm} h_k=\cos[ka_0]\left(t\sigma_0+t_x \sigma_x\right) +t_z \sin[k a_0] \sigma_z.
\end{align}
The spectrum of this Hamiltonian is given by: 
\begin{align}
\epsilon_{\pm}= 2t\cos[k a_0] \pm  \sqrt{ 2(t^2_x +t_z^2+ (t_x^2-t_z^2)\cos[k a_0])}.
\end{align}
It is plotted in Fig.  \ref{gapless_spectrum_app_a}.   It represents two bands separated in momentum space at half filling by $\delta k= 2\arctan(t_z/\sqrt{t^2-t_x^2})$ (in the units of $1/a_0$).  The Fermi velocity for the both bands is the same and is given by (in the leading order of $t_{x,z}/t$)  by $v_F= 2ta_0(1-t_z^2/t^2)$. 
First note that the model (\ref{gapless_fermions_BDI}) in the case $t_x=0$ has additional time-reversal symmetry $T_-^{2}=-1$,   and the spectrum in this case is illustrated in the Fig.  \ref{gapless_spectrum_app_b}.  The two eigenstates with the opposite spin $\chi_{\uparrow}$ and $\chi_{\downarrow}$,  marked by yellow and blue correspondingly,  are shifted in $k$ space by $ \delta k= k^0_+-k^0_-$.   
For the case $t_x \neq 0$ the bands corresponding two eigenstates $\chi_{\pm}$,  see the Fig.  \ref{gapless_spectrum_app_a} cross the Fermi level at the points $k_{\pm}$ correspondingly.   If we start with the picture $t_x \neq 0$ and continuously tune $t_x=0$ we arrive to the case of Fig.  \ref{gapless_spectrum_app_b} for $t_x=0$.  The eigenstates  $\chi_{\pm}$ also evolve continiously,  but they do not coincide with the spin up- and down- states $\chi_{\uparrow}, \chi_{\downarrow}$ at $t_x=0$.  In particular,  they are related as follows: 
\begin{align}
\begin{cases}
\chi_{\downarrow}(k)= \chi^0_{-}(k) \theta(k)+\chi^0_{+}(k) \theta(-k), \\
\chi_{\uparrow}(k)= \chi^0_{+}(k) \theta(k)+\chi^0_{-}(k) \theta(-k)
\end{cases}
\end{align}
here by $\chi^0_{\pm}(k)$ we denoted the eigenstates $\chi_{\pm}$ at $t_x=0$.  We can rewrite the effective low-energy operators in the real space,  projected onto the left- moving and right-moving modes. 
\begin{align}
\label{plus_minus}
\begin{cases}
\Psi_{\downarrow}(x)=R_{-}(x)+L_{+}(x) \\ 
\Psi_{\uparrow}(x)=R_{+}(x)+L_{-}(x), 
\end{cases}
\end{align}
where we included the oscillating prefactors $e^{i k^0_{\pm} x}$ into the definition of the left- and right-moving operators.
We need this expression in order to write down a bosonic model that continiously interpolates between the two cases $t_x=0$ and $t_x \neq 0$.  Now consider a generic case $t_x \neq 0$,  that generates the term that hybridizes the $R_{\pm}$ and $L_{\pm}$ operators and opens a gap at $k=0$.  We substitute (\ref{plus_minus}) to (\ref{gapless_fermions_BDI}) and obtain the Hamiltonian that decomposes to direct sum of the terms that describe right- and left- moving particles: $H_0 =H_R \oplus H_L$.   We can diagonalize the left-moving part and right-moving part independently and obtain the following relation between $R_{\pm}/L_{\pm}$ and the operators $\tilde{R}_{\pm}/\tilde{L}_{\pm}$ in the basis where $H_0$ is diagonal: 
\begin{align}
\label{rotatedRL}
\begin{cases}
R_{+}(x)=\cos\frac{\gamma}{2} \tilde{R}_{+} -\sin\frac{\gamma}{2} \tilde{R}_{-}  \\
R_{-}(x)=\cos\frac{\gamma}{2} \tilde{R}_{-} -\sin\frac{\gamma}{2} \tilde{R}_{+} 
\end{cases},\hspace{0.3cm}
\begin{cases}
L_{+}(x)=\cos\frac{\gamma}{2} \tilde{L}_{+} -\sin\frac{\gamma}{2} \tilde{L}_{-} \\
L_{-}(x)=\cos\frac{\gamma}{2} \tilde{L}_{-} -\sin\frac{\gamma}{2} \tilde{L}_{+} ,
\end{cases}
\end{align}
where the parameter $\gamma$ is given by: 
\begin{align}
\label{gamma_deltak}
\gamma= \arctan \left[\frac{t_x }{t_z } \cot \frac{\delta k}{2} \right] \simeq \frac{t_xt}{t_z^2}.
\end{align}
For the operators $\tilde{R}_{\pm}/\tilde{L}_{\pm}$ we apply the standart bosonization procedure: 
\begin{align}
\label{LR_bosonisation}
\tilde{R}_{\pm}(x) = \frac{\eta_{\pm}}{\sqrt{2\pi a}}e^{i\sqrt{4\pi} \phi_{R,\pm}(x) + ik_{\pm} x}, \hspace{0.3cm}
\tilde{L}_{\pm}(x) = \frac{\eta_{\pm}}{\sqrt{2\pi a}}e^{-i\sqrt{4\pi} \phi_{L,\pm}(x) - ik_{\pm} x},
\end{align}
where $k_{\pm} =k_F \pm \delta k/2$ and $\eta_{\pm}$ are Klein factors,  that were defined in the  Appendix \ref{bosonization_spinful}. 
We introduce canonically conjugated bosonic fields $\Pi_{\pm}(x)$ and $\phi_{\pm}(x)$ for each spin species:  
\begin{align}
\label{canonical}
\Pi_{\pm}(x)=\sqrt{\pi}(\rho_{L, \pm}-\rho_{R, \pm}),  \hspace{0.3cm}
\phi_{\pm}(x)=\phi_{L, \pm}+\phi_{R, \pm}, 
\end{align}
where $\rho_{R/L, \pm}= (1/\sqrt{\pi}) \partial_x \phi_{R/L, \pm}$.  One introduces spin and charge degrees of freedom: 
\begin{align}
\label{spin_charge_bosons}
\phi_c=\frac{\phi_{+}+\phi_{-}}{\sqrt{2}},\hspace{0.3cm}
\phi_s=\frac{\phi_{+}-\phi_{-}}{\sqrt{2}}. 
\end{align}
So for the gapless part of the BDI model we obtain the following Luttinger Liquid Hamiltonian: 
\begin{align}
H_{\text{LL}}= \sum_{j=c,s} \frac{v_F}{2} \left[\Pi^2_j +(\partial_x\phi_j)^2\right].
\end{align}
After we substitute (\ref{plus_minus}), (\ref{rotatedRL}), (\ref{LR_bosonisation}) to the gapped part (\ref{BDI_gapped_app}),  we obtain: 
\begin{align}
\label{gapped_bosonised}
H_{\text{gap}} = -\alpha \sin (\sqrt{2\pi} \phi_c) \cdot \sin(\sqrt{2\pi} \theta_s) +\beta \cos( \sqrt{2\pi}\phi_c) \cdot \cos (\delta k\cdot x +\sqrt{2\pi} \phi_s),
\end{align}
where $\alpha,  \beta$ are given by: 
\begin{align}
\label{parameters_BDI}
\alpha= \frac{4}{\pi a_0} \left(\delta t \cos\frac{\delta k}{2} -\cos \gamma \sin \frac{\delta k}{2} \delta t_z + \delta t_x  \sin \gamma \cos \frac{\delta k}{2} \right),\hspace{0.3cm}
\beta =  \frac{4}{\pi a_0}  \delta t_x.
\end{align}
Next we can bring (\ref{gapped_bosonised}) to the more convenient form if we introduce a new set of commuting fields as follows:
 \begin{align}
 \label{new_fields}
 \begin{cases}
 \phi_1 =-(\phi_{R, -}+\phi_{L,+}),  \\
 \phi_2 = -\sqrt{\pi}/2-(\phi_{R, +}+\phi_{L,-}) 
 \end{cases}
 \begin{cases}
 \theta_1 =-(\phi_{L, +}-\phi_{L,-}),  \\
 \theta_2 = -(\phi_{L, -}-\phi_{R,+}) 
 \end{cases}
 \end{align}
For bosons in charge and spin sector of freedom it implies the following transformation: 
\begin{align}
\label{new_charge_spin}
\tilde{\phi}_c = \phi_c -\frac{\sqrt{\pi}}{2\sqrt{2}}, \hspace{0.3cm}
\tilde{\phi}_s=\frac{\sqrt{\pi}}{2\sqrt{2}}-\theta_s,  \hspace{0.2cm}\tilde{\theta}_s = -\phi_s.  
\end{align}
In terms of these degrees of freedom the Hamiltonian (\ref{gapped_bosonised}) takes the following form: 
\begin{align}
\label{gapped_bosonised_tilde}
V_{\text{BDI}} = -\alpha \cos (\sqrt{2\pi} \tilde{\phi}_c) \cdot \cos(\sqrt{2\pi} \tilde{\phi_s}) +\beta \sin(\sqrt{2\pi}\tilde{\phi_c}) \cdot \sin (\delta k\cdot x -\sqrt{2\pi} \tilde{\theta}_s). 
\end{align}
We will further omit $\sim$ for convenience.  Note,  that the inter-chain hopping (\ref{BDI_app}) is not the most general one for BDI class.  It contains terms $\propto \sigma_x$ in chain basis.  One can also add $\sigma_y$ terms to this model.  In the bosonic language such terms correspond to $ \propto \sin(\sqrt{2\pi}\tilde{\phi_c}) \cdot \cos (\delta k\cdot x -\sqrt{2\pi} \tilde{\theta}_s)$.   

At the energies $\epsilon > v_F \cdot \delta k $ the $\theta_s$ term in (\ref{gapped_bosonised_tilde}) oscillates slowly,   and thus we obtain the same bosonic model that we wrote solely out of symmetry arguments in the main text,  see (\ref{C1_bosonic}).    The  $\theta_s$ term can be analyzed in a basis where $\delta t_x$ term in (\ref{BDI_gapped_app}) is diagonal,  i.e.  $\sigma_x \rightarrow \sigma_z$.   If we also set $t_x=0$ (so there is no need to use a band basis,  as the Hamiltonian becomes diagonal in chain space),   we eventually get a Hamiltonian of uncoupled inequivalent chains,  that we studied in Section \ref{inequivalent} of the main text.  

Also notice that,  in the limit $\delta k=0$ or $t_z=0$ but $t_x \neq 0$ in (\ref{gapless_fermions_BDI}),   the Fermi velocities of the two bands differ by $ \delta v \propto a_0 t_x$.  This generates a term that mixes charge and spin sectors in the bosonic Hamiltonian \cite{Kainaris2015}.  Such a term might change the interacting phase diagram,   as we discuss in Appendix  \ref{sec:coupled_chains_int},  as it changes the RG flow of marginally irrelevant backscattering.  However it does not quantitatively affect the flow of single-particle gap opening terms  (\ref{gapped_bosonised_tilde}),   so for the weakly interacting model it can be neglected.  

\subsection{Single band limit}
\begin{figure}
\centering
\captionsetup{justification=centering}
\begin{subfigure}{0.4\linewidth}
\includegraphics[scale=0.35]{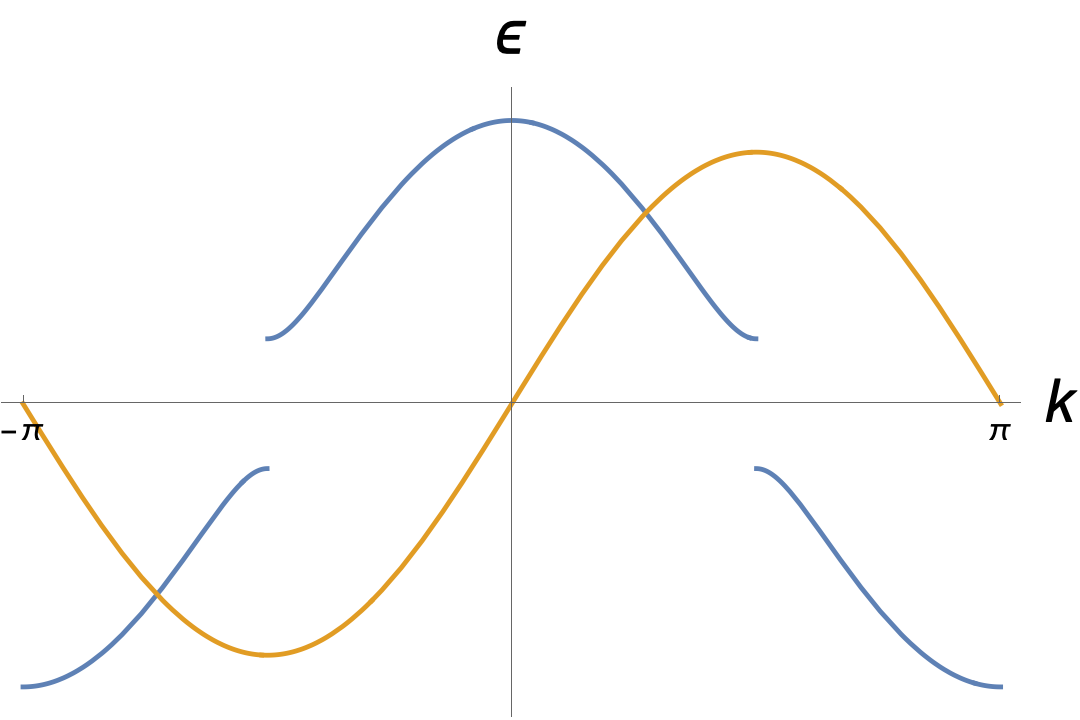}
\end{subfigure}
\caption{Spectrum of BDI model in a single band limit}
\label{1chain}
\end{figure}

Above we focused on the case of weakly coupled chains,  at the gapless point such model is equivalent to two Luttinger liquids.   However,   for strong inter-chain hopping there is also a critical point where the model is equivalent to a single chain.  In particular,  if we assume that $w,v$ are real and set $w - v = a - b$ (or equivalently $w - v = -(a - b)$).  In that limit the model (\ref{H1_app}) is equivalent to decoupled chains -- one is gapless and the second one is gapped with a gap $\Delta \propto |a-b|$.    Spectrum of such model is illustrated schematically in Fig.  \ref{1chain}.   Thus near the half-filling  this model is equivalent to a single SSH chain.  

\section{Interacting BDI model}\label{sec:coupled_chains_int}

\subsection{Interactions in single band limit}
As we discussed earlier in the single band limit the BDI model (\ref{H1_app}) described by a single bosonic mode.   We will focus on the case of weak interactions -- smaller than a gap of one of the bands $\Delta \propto |a-b|$.  Thus, for interactions that do not couple two bands,  the model is equivalent to a single interacting SSH chain discussed in Section \ref{Section_InteractingSSH}.   
The  interaction,  that couples two bands can be treated perturbatively:  $H_{\text{int}}= U (\hat{n}_{1,j}-1/2)( \hat{n}_{2,j}-1/2) \rightarrow U \langle (\hat{n}_{1,j}-1/2) \rangle (\hat{n}_{2,j}-1/2)$,  where by $\langle .\rangle$ we denote the average over the non-interacting ground state.   In the gapped phase one can neglect fluctuations of density on site and get $\langle n_{1,j} \rangle = 1/2 $.   Thus $\langle H_{\text{int}} \rangle =0$ in this limit,  therefore in the first order of perturbation theory one can neglect the inter-band interactions and only focus on the interactions within bands.

\subsection{Interactions in the two band limit}
Now let us discuss the interacting model in the two-band case.  
Note that the coupled chain model breaks SU(2) symmetry in chain space.  In the low-energy description,  this generates scattering processes in the Hubbard interaction term that do not conserve the chain ("spin") degree of freedom.  The processes involve forward-  and back-scattering of the fermions with the opposite spin,  that enter the low-energy model with the oscillating amplitude $\propto e^{i \delta k x}$.   These terms can be neglected at the low energies $\epsilon < v_F \cdot \delta k$,  as they oscillate fast on those scales,  and they need to be taken into account if $\epsilon > v_F \cdot \delta k$.   This results in two distinct low-energy bosonic models at the two energy scales.   Let us study their properties.  

\subsubsection{Model at energies $\epsilon < v_F \cdot \delta k $ }
In this limit, the model one can neglect the oscillating single-particle term in (\ref{gapped_bosonised_tilde}) as well as spin non-conserving processes in the interaction.  That effectively restores SU(2) symmetry and the model is reduced to the model of two identical capacitively coupled SSH chains in the presence of Hubbard interaction studied before.  The only difference compared to the case of capacitively chains is that the coupling constants get renormalized $g_c=-g_s= -U\cos^2 \gamma$,  where the angle parameter $\gamma$ is related to the inter-chain coupling parameters (\ref{gamma_deltak}).  

\subsubsection{Model at energies $\epsilon > v_F \cdot \delta k $ }
Now consider the interacting model at the energies $\epsilon > v_F \cdot \delta k $ so the spin non-conserving processes need to be taken into account.  In that case, we obtain:
\begin{align}
\label{BDI_interactions_brokenSU2}
H=H_c+H_s +H_{cs},  \nonumber \\
H_{cs} = g_{cs} [\partial_x \phi_c \partial_x \phi_s + \Pi_c \Pi_s],
\end{align}
where the charge and spin parts of the Hamiltonian $H_{c,s}$ are given by (\ref{Hcs}).  
The parameters of the model are given by: 
\begin{align}
K_{c}  \simeq 1+ \frac{g_{c}}{\pi v_F},  K_{s}  \simeq 1- \frac{g_{s}}{\pi v_F} \nonumber \\
g_c=-g_s= -U\cos^2 \gamma,  g_{cs}= -\frac{2U}{\pi} \sin [\gamma]. 
\end{align}
Here we used a different basis from that used before in Appendix \ref{BDI_bosonization_app},  see   \footnote{Namely,  we rotated the band left-moving and right-moving fermionic operators (\ref{LR_bosonisation}) such that $\sigma_z \rightarrow \sigma_x$,  we also shifted the bosonic fields $\sqrt{2\pi}\phi_{s,c} \rightarrow \sqrt{2\pi}\phi_{s,c} + \pi/2$.  With this shift the single-particle terms take the form studied earlier in the Section \ref{Sec:uncoupledSSH}}.    In this basis,  $H_{cs}$ takes a simple quadratic form given in (\ref{BDI_interactions_brokenSU2}).  
Note that the term $H_{cs}$ does not open a gap,  but generates the difference between Fermi velocities for electrons with the opposite band index.  Such types of terms have been studied earlier in one-dimensional systems with spin-orbit coupling \cite{Starykh2008, Kainaris2015}.   It was demonstrated that due to the presence of such terms the backscattering in the spin sector becomes relevant for sufficiently strong spin-orbit coupling.   From the results of \cite{Kainaris2015} it follows that for our model (\ref{BDI_interactions_brokenSU2}) it implies that $\tan^2 \gamma >(a \pi v_F/U)$.  The gap opening single-particle terms in the new basis take the following form: 
\begin{align}
\label{gap_BDI_int}
V_{\text{BDI}} = \alpha \cos (\sqrt{2\pi } \phi_c) \cos(\sqrt{2\pi} \phi_s)- \nonumber \\ - \beta \sin(\sqrt{2\pi}\phi_c)\sin (\sqrt{2\pi} \phi_s).  
\end{align}
Therefore depending on the ratio $|\alpha|/|\beta|$ the model is reduced either to capacitively coupled identical SSH chains or SSH chains in the opposite topological phases.   The only difference from the interacting models considered before in Section \ref{Sec:uncoupledSSH} is that one needs to take into account the relevant backscattering in the spin sector in (\ref{BDI_interactions_brokenSU2}) for repulsive interactions.  The backscattering term fixes the bosonic field $\sqrt{2\pi} \phi_s= \pi /2 \mod \pi$ and therefore it is incompatible with the single-particle term in (\ref{gap_BDI_int})  that describes two identical chains.   Thus there is an Ising phase transition to the SDW phase if interactions are sufficiently strong.   

\bibliography{Manuscript_v1}

\begin{thebibliography}{76}%
\makeatletter
\providecommand \@ifxundefined [1]{%
 \@ifx{#1\undefined}
}%
\providecommand \@ifnum [1]{%
 \ifnum #1\expandafter \@firstoftwo
 \else \expandafter \@secondoftwo
 \fi
}%
\providecommand \@ifx [1]{%
 \ifx #1\expandafter \@firstoftwo
 \else \expandafter \@secondoftwo
 \fi
}%
\providecommand \natexlab [1]{#1}%
\providecommand \enquote  [1]{``#1''}%
\providecommand \bibnamefont  [1]{#1}%
\providecommand \bibfnamefont [1]{#1}%
\providecommand \citenamefont [1]{#1}%
\providecommand \href@noop [0]{\@secondoftwo}%
\providecommand \href [0]{\begingroup \@sanitize@url \@href}%
\providecommand \@href[1]{\@@startlink{#1}\@@href}%
\providecommand \@@href[1]{\endgroup#1\@@endlink}%
\providecommand \@sanitize@url [0]{\catcode `\\12\catcode `\$12\catcode
  `\&12\catcode `\#12\catcode `\^12\catcode `\_12\catcode `\%12\relax}%
\providecommand \@@startlink[1]{}%
\providecommand \@@endlink[0]{}%
\providecommand \url  [0]{\begingroup\@sanitize@url \@url }%
\providecommand \@url [1]{\endgroup\@href {#1}{\urlprefix }}%
\providecommand \urlprefix  [0]{URL }%
\providecommand \Eprint [0]{\href }%
\providecommand \doibase [0]{https://doi.org/}%
\providecommand \selectlanguage [0]{\@gobble}%
\providecommand \bibinfo  [0]{\@secondoftwo}%
\providecommand \bibfield  [0]{\@secondoftwo}%
\providecommand \translation [1]{[#1]}%
\providecommand \BibitemOpen [0]{}%
\providecommand \bibitemStop [0]{}%
\providecommand \bibitemNoStop [0]{.\EOS\space}%
\providecommand \EOS [0]{\spacefactor3000\relax}%
\providecommand \BibitemShut  [1]{\csname bibitem#1\endcsname}%
\let\auto@bib@innerbib\@empty
\bibitem [{\citenamefont {Landau}\ \emph {et~al.}(1996)\citenamefont {Landau},
  \citenamefont {Landau}, \citenamefont {M.},\ and\ \citenamefont
  {P.}}]{landau_statistics}%
  \BibitemOpen
  \bibfield  {author} {\bibinfo {author} {\bibfnamefont {L.~D.}\ \bibnamefont
  {Landau}}, \bibinfo {author} {\bibfnamefont {L.~D.}\ \bibnamefont {Landau}},
  \bibinfo {author} {\bibfnamefont {L.~E.}\ \bibnamefont {M.}},\ and\ \bibinfo
  {author} {\bibfnamefont {P.~L.}\ \bibnamefont {P.}},\ }\href@noop {} {\emph
  {\bibinfo {title} {Statistical physics}}}\ (\bibinfo  {publisher}
  {Butterworth-Heinemann c1980},\ \bibinfo {year} {1996})\BibitemShut {NoStop}%
\bibitem [{\citenamefont {Thouless}\ \emph {et~al.}(1982)\citenamefont
  {Thouless}, \citenamefont {Kohmoto}, \citenamefont {Nightingale},\ and\
  \citenamefont {den Nijs}}]{Thouless1982}%
  \BibitemOpen
  \bibfield  {author} {\bibinfo {author} {\bibfnamefont {D.~J.}\ \bibnamefont
  {Thouless}}, \bibinfo {author} {\bibfnamefont {M.}~\bibnamefont {Kohmoto}},
  \bibinfo {author} {\bibfnamefont {M.~P.}\ \bibnamefont {Nightingale}},\ and\
  \bibinfo {author} {\bibfnamefont {M.}~\bibnamefont {den Nijs}},\ }\bibfield
  {title} {\bibinfo {title} {Quantized hall conductance in a two-dimensional
  periodic potential},\ }\href {https://doi.org/10.1103/PhysRevLett.49.405}
  {\bibfield  {journal} {\bibinfo  {journal} {Phys. Rev. Lett.}\ }\textbf
  {\bibinfo {volume} {49}},\ \bibinfo {pages} {405} (\bibinfo {year}
  {1982})}\BibitemShut {NoStop}%
\bibitem [{\citenamefont {Haldane}(1988)}]{Haldane1988}%
  \BibitemOpen
  \bibfield  {author} {\bibinfo {author} {\bibfnamefont {F.~D.~M.}\
  \bibnamefont {Haldane}},\ }\bibfield  {title} {\bibinfo {title} {Model for a
  quantum hall effect without landau levels: Condensed-matter realization of
  the "parity anomaly"},\ }\href {https://doi.org/10.1103/PhysRevLett.61.2015}
  {\bibfield  {journal} {\bibinfo  {journal} {Phys. Rev. Lett.}\ }\textbf
  {\bibinfo {volume} {61}},\ \bibinfo {pages} {2015} (\bibinfo {year}
  {1988})}\BibitemShut {NoStop}%
\bibitem [{\citenamefont {Kane}\ and\ \citenamefont
  {Mele}(2005)}]{KaneMele2005}%
  \BibitemOpen
  \bibfield  {author} {\bibinfo {author} {\bibfnamefont {C.~L.}\ \bibnamefont
  {Kane}}\ and\ \bibinfo {author} {\bibfnamefont {E.~J.}\ \bibnamefont
  {Mele}},\ }\bibfield  {title} {\bibinfo {title} {$\mathbb{Z}_{2}$ topological
  order and the quantum spin hall effect},\ }\href
  {https://doi.org/10.1103/PhysRevLett.95.146802} {\bibfield  {journal}
  {\bibinfo  {journal} {Phys. Rev. Lett.}\ }\textbf {\bibinfo {volume} {95}},\
  \bibinfo {pages} {146802} (\bibinfo {year} {2005})}\BibitemShut {NoStop}%
\bibitem [{\citenamefont {Fu}\ \emph {et~al.}(2007)\citenamefont {Fu},
  \citenamefont {Kane},\ and\ \citenamefont {Mele}}]{Fu2007}%
  \BibitemOpen
  \bibfield  {author} {\bibinfo {author} {\bibfnamefont {L.}~\bibnamefont
  {Fu}}, \bibinfo {author} {\bibfnamefont {C.~L.}\ \bibnamefont {Kane}},\ and\
  \bibinfo {author} {\bibfnamefont {E.~J.}\ \bibnamefont {Mele}},\ }\bibfield
  {title} {\bibinfo {title} {Topological insulators in three dimensions},\
  }\href {https://doi.org/10.1103/PhysRevLett.98.106803} {\bibfield  {journal}
  {\bibinfo  {journal} {Phys. Rev. Lett.}\ }\textbf {\bibinfo {volume} {98}},\
  \bibinfo {pages} {106803} (\bibinfo {year} {2007})}\BibitemShut {NoStop}%
\bibitem [{\citenamefont {Qi}\ \emph {et~al.}(2008)\citenamefont {Qi},
  \citenamefont {Hughes},\ and\ \citenamefont {Zhang}}]{Qi2008}%
  \BibitemOpen
  \bibfield  {author} {\bibinfo {author} {\bibfnamefont {X.-L.}\ \bibnamefont
  {Qi}}, \bibinfo {author} {\bibfnamefont {T.~L.}\ \bibnamefont {Hughes}},\
  and\ \bibinfo {author} {\bibfnamefont {S.-C.}\ \bibnamefont {Zhang}},\
  }\bibfield  {title} {\bibinfo {title} {Topological field theory of
  time-reversal invariant insulators},\ }\href
  {https://doi.org/10.1103/PhysRevB.78.195424} {\bibfield  {journal} {\bibinfo
  {journal} {Phys. Rev. B}\ }\textbf {\bibinfo {volume} {78}},\ \bibinfo
  {pages} {195424} (\bibinfo {year} {2008})}\BibitemShut {NoStop}%
\bibitem [{\citenamefont {Roy}(2009)}]{Roy2009}%
  \BibitemOpen
  \bibfield  {author} {\bibinfo {author} {\bibfnamefont {R.}~\bibnamefont
  {Roy}},\ }\bibfield  {title} {\bibinfo {title} {Topological phases and the
  quantum spin hall effect in three dimensions},\ }\href
  {https://doi.org/10.1103/PhysRevB.79.195322} {\bibfield  {journal} {\bibinfo
  {journal} {Phys. Rev. B}\ }\textbf {\bibinfo {volume} {79}},\ \bibinfo
  {pages} {195322} (\bibinfo {year} {2009})}\BibitemShut {NoStop}%
\bibitem [{\citenamefont {Hsieh}\ \emph {et~al.}(2009)\citenamefont {Hsieh},
  \citenamefont {Xia}, \citenamefont {Qian}, \citenamefont {Wray},
  \citenamefont {Dil}, \citenamefont {Meier}, \citenamefont {Osterwalder},
  \citenamefont {Patthey}, \citenamefont {Checkelsky}, \citenamefont {Ong},
  \citenamefont {Fedorov}, \citenamefont {Lin}, \citenamefont {Bansil},
  \citenamefont {Grauer}, \citenamefont {Hor}, \citenamefont {Cava},\ and\
  \citenamefont {Hasan}}]{Hsieh2009}%
  \BibitemOpen
  \bibfield  {author} {\bibinfo {author} {\bibfnamefont {D.}~\bibnamefont
  {Hsieh}}, \bibinfo {author} {\bibfnamefont {Y.}~\bibnamefont {Xia}}, \bibinfo
  {author} {\bibfnamefont {D.}~\bibnamefont {Qian}}, \bibinfo {author}
  {\bibfnamefont {L.}~\bibnamefont {Wray}}, \bibinfo {author} {\bibfnamefont
  {J.~H.}\ \bibnamefont {Dil}}, \bibinfo {author} {\bibfnamefont
  {F.}~\bibnamefont {Meier}}, \bibinfo {author} {\bibfnamefont
  {J.}~\bibnamefont {Osterwalder}}, \bibinfo {author} {\bibfnamefont
  {L.}~\bibnamefont {Patthey}}, \bibinfo {author} {\bibfnamefont {J.~G.}\
  \bibnamefont {Checkelsky}}, \bibinfo {author} {\bibfnamefont {N.~P.}\
  \bibnamefont {Ong}}, \bibinfo {author} {\bibfnamefont {A.~V.}\ \bibnamefont
  {Fedorov}}, \bibinfo {author} {\bibfnamefont {H.}~\bibnamefont {Lin}},
  \bibinfo {author} {\bibfnamefont {A.}~\bibnamefont {Bansil}}, \bibinfo
  {author} {\bibfnamefont {D.}~\bibnamefont {Grauer}}, \bibinfo {author}
  {\bibfnamefont {Y.~S.}\ \bibnamefont {Hor}}, \bibinfo {author} {\bibfnamefont
  {R.~J.}\ \bibnamefont {Cava}},\ and\ \bibinfo {author} {\bibfnamefont
  {M.~Z.}\ \bibnamefont {Hasan}},\ }\bibfield  {title} {\bibinfo {title} {A
  tunable topological insulator in the spin helical dirac transport regime},\
  }\href {https://doi.org/10.1038/nature08234} {\bibfield  {journal} {\bibinfo
  {journal} {Nature}\ }\textbf {\bibinfo {volume} {460}},\ \bibinfo {pages}
  {1101} (\bibinfo {year} {2009})}\BibitemShut {NoStop}%
\bibitem [{\citenamefont {Hsieh}\ \emph {et~al.}(2008)\citenamefont {Hsieh},
  \citenamefont {Qian}, \citenamefont {Wray}, \citenamefont {Xia},
  \citenamefont {Hor}, \citenamefont {Cava},\ and\ \citenamefont
  {Hasan}}]{Hsieh2008}%
  \BibitemOpen
  \bibfield  {author} {\bibinfo {author} {\bibfnamefont {D.}~\bibnamefont
  {Hsieh}}, \bibinfo {author} {\bibfnamefont {D.}~\bibnamefont {Qian}},
  \bibinfo {author} {\bibfnamefont {L.}~\bibnamefont {Wray}}, \bibinfo {author}
  {\bibfnamefont {Y.}~\bibnamefont {Xia}}, \bibinfo {author} {\bibfnamefont
  {Y.~S.}\ \bibnamefont {Hor}}, \bibinfo {author} {\bibfnamefont {R.~J.}\
  \bibnamefont {Cava}},\ and\ \bibinfo {author} {\bibfnamefont {M.~Z.}\
  \bibnamefont {Hasan}},\ }\bibfield  {title} {\bibinfo {title} {A topological
  dirac insulator in a quantum spin hall phase},\ }\href
  {https://doi.org/10.1038/nature06843} {\bibfield  {journal} {\bibinfo
  {journal} {Nature}\ }\textbf {\bibinfo {volume} {452}},\ \bibinfo {pages}
  {970} (\bibinfo {year} {2008})}\BibitemShut {NoStop}%
\bibitem [{\citenamefont {Xia}\ \emph {et~al.}(2009)\citenamefont {Xia},
  \citenamefont {Qian}, \citenamefont {Hsieh}, \citenamefont {Wray},
  \citenamefont {Pal}, \citenamefont {Lin}, \citenamefont {Bansil},
  \citenamefont {Grauer}, \citenamefont {Hor}, \citenamefont {Cava},\ and\
  \citenamefont {Hasan}}]{Xia2009}%
  \BibitemOpen
  \bibfield  {author} {\bibinfo {author} {\bibfnamefont {Y.}~\bibnamefont
  {Xia}}, \bibinfo {author} {\bibfnamefont {D.}~\bibnamefont {Qian}}, \bibinfo
  {author} {\bibfnamefont {D.}~\bibnamefont {Hsieh}}, \bibinfo {author}
  {\bibfnamefont {L.}~\bibnamefont {Wray}}, \bibinfo {author} {\bibfnamefont
  {A.}~\bibnamefont {Pal}}, \bibinfo {author} {\bibfnamefont {H.}~\bibnamefont
  {Lin}}, \bibinfo {author} {\bibfnamefont {A.}~\bibnamefont {Bansil}},
  \bibinfo {author} {\bibfnamefont {D.}~\bibnamefont {Grauer}}, \bibinfo
  {author} {\bibfnamefont {Y.~S.}\ \bibnamefont {Hor}}, \bibinfo {author}
  {\bibfnamefont {R.~J.}\ \bibnamefont {Cava}},\ and\ \bibinfo {author}
  {\bibfnamefont {M.~Z.}\ \bibnamefont {Hasan}},\ }\bibfield  {title} {\bibinfo
  {title} {Observation of a large-gap topological-insulator class with a single
  dirac cone on the surface},\ }\href {https://doi.org/10.1038/nphys1274}
  {\bibfield  {journal} {\bibinfo  {journal} {Nature Physics}\ }\textbf
  {\bibinfo {volume} {5}},\ \bibinfo {pages} {398} (\bibinfo {year}
  {2009})}\BibitemShut {NoStop}%
\bibitem [{\citenamefont {Zhang}\ \emph
  {et~al.}(2009{\natexlab{a}})\citenamefont {Zhang}, \citenamefont {Cheng},
  \citenamefont {Chen}, \citenamefont {Jia}, \citenamefont {Ma}, \citenamefont
  {He}, \citenamefont {Wang}, \citenamefont {Zhang}, \citenamefont {Dai},
  \citenamefont {Fang}, \citenamefont {Xie},\ and\ \citenamefont
  {Xue}}]{Zhang2009}%
  \BibitemOpen
  \bibfield  {author} {\bibinfo {author} {\bibfnamefont {T.}~\bibnamefont
  {Zhang}}, \bibinfo {author} {\bibfnamefont {P.}~\bibnamefont {Cheng}},
  \bibinfo {author} {\bibfnamefont {X.}~\bibnamefont {Chen}}, \bibinfo {author}
  {\bibfnamefont {J.-F.}\ \bibnamefont {Jia}}, \bibinfo {author} {\bibfnamefont
  {X.}~\bibnamefont {Ma}}, \bibinfo {author} {\bibfnamefont {K.}~\bibnamefont
  {He}}, \bibinfo {author} {\bibfnamefont {L.}~\bibnamefont {Wang}}, \bibinfo
  {author} {\bibfnamefont {H.}~\bibnamefont {Zhang}}, \bibinfo {author}
  {\bibfnamefont {X.}~\bibnamefont {Dai}}, \bibinfo {author} {\bibfnamefont
  {Z.}~\bibnamefont {Fang}}, \bibinfo {author} {\bibfnamefont {X.}~\bibnamefont
  {Xie}},\ and\ \bibinfo {author} {\bibfnamefont {Q.-K.}\ \bibnamefont {Xue}},\
  }\bibfield  {title} {\bibinfo {title} {Experimental demonstration of
  topological surface states protected by time-reversal symmetry},\ }\href
  {https://doi.org/10.1103/PhysRevLett.103.266803} {\bibfield  {journal}
  {\bibinfo  {journal} {Phys. Rev. Lett.}\ }\textbf {\bibinfo {volume} {103}},\
  \bibinfo {pages} {266803} (\bibinfo {year} {2009}{\natexlab{a}})}\BibitemShut
  {NoStop}%
\bibitem [{\citenamefont {Zhang}\ \emph
  {et~al.}(2009{\natexlab{b}})\citenamefont {Zhang}, \citenamefont {Liu},
  \citenamefont {Qi}, \citenamefont {Dai}, \citenamefont {Fang},\ and\
  \citenamefont {Zhang}}]{ZhangHaijun2009}%
  \BibitemOpen
  \bibfield  {author} {\bibinfo {author} {\bibfnamefont {H.}~\bibnamefont
  {Zhang}}, \bibinfo {author} {\bibfnamefont {C.-X.}\ \bibnamefont {Liu}},
  \bibinfo {author} {\bibfnamefont {X.-L.}\ \bibnamefont {Qi}}, \bibinfo
  {author} {\bibfnamefont {X.}~\bibnamefont {Dai}}, \bibinfo {author}
  {\bibfnamefont {Z.}~\bibnamefont {Fang}},\ and\ \bibinfo {author}
  {\bibfnamefont {S.-C.}\ \bibnamefont {Zhang}},\ }\bibfield  {title} {\bibinfo
  {title} {Topological insulators in $\text{Bi}_2\text{Se}_3$,
  $\text{Bi}_2\text{Te}_3$ and $\text{Sb}_2\text{Te}_3$ with a single dirac
  cone on the surface},\ }\href {https://doi.org/10.1038/nphys1270} {\bibfield
  {journal} {\bibinfo  {journal} {Nature Physics}\ }\textbf {\bibinfo {volume}
  {5}},\ \bibinfo {pages} {438} (\bibinfo {year}
  {2009}{\natexlab{b}})}\BibitemShut {NoStop}%
\bibitem [{\citenamefont {Chen}\ \emph {et~al.}(2009)\citenamefont {Chen},
  \citenamefont {Analytis}, \citenamefont {Chu}, \citenamefont {Liu},
  \citenamefont {Mo}, \citenamefont {Qi}, \citenamefont {Zhang}, \citenamefont
  {Lu}, \citenamefont {Dai}, \citenamefont {Fang}, \citenamefont {Zhang},
  \citenamefont {Fisher}, \citenamefont {Hussain},\ and\ \citenamefont
  {Shen}}]{Chen2009}%
  \BibitemOpen
  \bibfield  {author} {\bibinfo {author} {\bibfnamefont {Y.~L.}\ \bibnamefont
  {Chen}}, \bibinfo {author} {\bibfnamefont {J.~G.}\ \bibnamefont {Analytis}},
  \bibinfo {author} {\bibfnamefont {J.-H.}\ \bibnamefont {Chu}}, \bibinfo
  {author} {\bibfnamefont {Z.~K.}\ \bibnamefont {Liu}}, \bibinfo {author}
  {\bibfnamefont {S.-K.}\ \bibnamefont {Mo}}, \bibinfo {author} {\bibfnamefont
  {X.~L.}\ \bibnamefont {Qi}}, \bibinfo {author} {\bibfnamefont {H.~J.}\
  \bibnamefont {Zhang}}, \bibinfo {author} {\bibfnamefont {D.~H.}\ \bibnamefont
  {Lu}}, \bibinfo {author} {\bibfnamefont {X.}~\bibnamefont {Dai}}, \bibinfo
  {author} {\bibfnamefont {Z.}~\bibnamefont {Fang}}, \bibinfo {author}
  {\bibfnamefont {S.~C.}\ \bibnamefont {Zhang}}, \bibinfo {author}
  {\bibfnamefont {I.~R.}\ \bibnamefont {Fisher}}, \bibinfo {author}
  {\bibfnamefont {Z.}~\bibnamefont {Hussain}},\ and\ \bibinfo {author}
  {\bibfnamefont {Z.-X.}\ \bibnamefont {Shen}},\ }\bibfield  {title} {\bibinfo
  {title} {Experimental realization of a three-dimensional topological
  insulator, $\text{Bi}_2\text{Te}_3$},\ }\href
  {https://doi.org/10.1126/science.1173034} {\bibfield  {journal} {\bibinfo
  {journal} {Science}\ }\textbf {\bibinfo {volume} {325}},\ \bibinfo {pages}
  {178} (\bibinfo {year} {2009})}\BibitemShut {NoStop}%
\bibitem [{\citenamefont {Su}\ \emph {et~al.}(1979)\citenamefont {Su},
  \citenamefont {Schrieffer},\ and\ \citenamefont {Heeger}}]{SSH1979}%
  \BibitemOpen
  \bibfield  {author} {\bibinfo {author} {\bibfnamefont {W.~P.}\ \bibnamefont
  {Su}}, \bibinfo {author} {\bibfnamefont {J.~R.}\ \bibnamefont {Schrieffer}},\
  and\ \bibinfo {author} {\bibfnamefont {A.~J.}\ \bibnamefont {Heeger}},\
  }\bibfield  {title} {\bibinfo {title} {Solitons in polyacetylene},\ }\href
  {https://doi.org/10.1103/PhysRevLett.42.1698} {\bibfield  {journal} {\bibinfo
   {journal} {Phys. Rev. Lett.}\ }\textbf {\bibinfo {volume} {42}},\ \bibinfo
  {pages} {1698} (\bibinfo {year} {1979})}\BibitemShut {NoStop}%
\bibitem [{\citenamefont {Ryu}\ and\ \citenamefont {Hatsugai}(2006)}]{Ryu2006}%
  \BibitemOpen
  \bibfield  {author} {\bibinfo {author} {\bibfnamefont {S.}~\bibnamefont
  {Ryu}}\ and\ \bibinfo {author} {\bibfnamefont {Y.}~\bibnamefont {Hatsugai}},\
  }\bibfield  {title} {\bibinfo {title} {Entanglement entropy and the berry
  phase in the solid state},\ }\href
  {https://doi.org/10.1103/PhysRevB.73.245115} {\bibfield  {journal} {\bibinfo
  {journal} {Phys. Rev. B}\ }\textbf {\bibinfo {volume} {73}},\ \bibinfo
  {pages} {245115} (\bibinfo {year} {2006})}\BibitemShut {NoStop}%
\bibitem [{\citenamefont {Kitaev}(2001)}]{Kitaev2001}%
  \BibitemOpen
  \bibfield  {author} {\bibinfo {author} {\bibfnamefont {A.~Y.}\ \bibnamefont
  {Kitaev}},\ }\bibfield  {title} {\bibinfo {title} {Unpaired majorana fermions
  in quantum wires},\ }\href {https://doi.org/10.1070/1063-7869/44/10S/S29}
  {\bibfield  {journal} {\bibinfo  {journal} {Physics-Uspekhi}\ }\textbf
  {\bibinfo {volume} {44}},\ \bibinfo {pages} {131} (\bibinfo {year}
  {2001})}\BibitemShut {NoStop}%
\bibitem [{\citenamefont {Zirnbauer}(1996)}]{Zirnbauer1996}%
  \BibitemOpen
  \bibfield  {author} {\bibinfo {author} {\bibfnamefont {M.~R.}\ \bibnamefont
  {Zirnbauer}},\ }\bibfield  {title} {\bibinfo {title} {{Riemannian symmetric
  superspaces and their origin in random‐matrix theory}},\ }\href
  {https://doi.org/10.1063/1.531675} {\bibfield  {journal} {\bibinfo  {journal}
  {Journal of Mathematical Physics}\ }\textbf {\bibinfo {volume} {37}},\
  \bibinfo {pages} {4986} (\bibinfo {year} {1996})}\BibitemShut {NoStop}%
\bibitem [{\citenamefont {Altland}\ and\ \citenamefont
  {Zirnbauer}(1997)}]{Altland1997}%
  \BibitemOpen
  \bibfield  {author} {\bibinfo {author} {\bibfnamefont {A.}~\bibnamefont
  {Altland}}\ and\ \bibinfo {author} {\bibfnamefont {M.~R.}\ \bibnamefont
  {Zirnbauer}},\ }\bibfield  {title} {\bibinfo {title} {Nonstandard symmetry
  classes in mesoscopic normal-superconducting hybrid structures},\ }\href
  {https://doi.org/10.1103/PhysRevB.55.1142} {\bibfield  {journal} {\bibinfo
  {journal} {Phys. Rev. B}\ }\textbf {\bibinfo {volume} {55}},\ \bibinfo
  {pages} {1142} (\bibinfo {year} {1997})}\BibitemShut {NoStop}%
\bibitem [{\citenamefont {Kitaev}(2009)}]{Kitaev2009}%
  \BibitemOpen
  \bibfield  {author} {\bibinfo {author} {\bibfnamefont {A.}~\bibnamefont
  {Kitaev}},\ }\bibfield  {title} {\bibinfo {title} {{Periodic table for
  topological insulators and superconductors}},\ }\href
  {https://doi.org/10.1063/1.3149495} {\bibfield  {journal} {\bibinfo
  {journal} {AIP Conference Proceedings}\ }\textbf {\bibinfo {volume} {1134}},\
  \bibinfo {pages} {22} (\bibinfo {year} {2009})}\BibitemShut {NoStop}%
\bibitem [{\citenamefont {Ryu}\ \emph {et~al.}(2010)\citenamefont {Ryu},
  \citenamefont {Schnyder}, \citenamefont {Furusaki},\ and\ \citenamefont
  {Ludwig}}]{Ryu2010}%
  \BibitemOpen
  \bibfield  {author} {\bibinfo {author} {\bibfnamefont {S.}~\bibnamefont
  {Ryu}}, \bibinfo {author} {\bibfnamefont {A.~P.}\ \bibnamefont {Schnyder}},
  \bibinfo {author} {\bibfnamefont {A.}~\bibnamefont {Furusaki}},\ and\
  \bibinfo {author} {\bibfnamefont {A.~W.~W.}\ \bibnamefont {Ludwig}},\
  }\bibfield  {title} {\bibinfo {title} {Topological insulators and
  superconductors: tenfold way and dimensional hierarchy},\ }\href
  {https://doi.org/10.1088/1367-2630/12/6/065010} {\bibfield  {journal}
  {\bibinfo  {journal} {New Journal of Physics}\ }\textbf {\bibinfo {volume}
  {12}},\ \bibinfo {pages} {065010} (\bibinfo {year} {2010})}\BibitemShut
  {NoStop}%
\bibitem [{\citenamefont {Tsui}\ \emph {et~al.}(1982)\citenamefont {Tsui},
  \citenamefont {Stormer},\ and\ \citenamefont {Gossard}}]{Tsui1982}%
  \BibitemOpen
  \bibfield  {author} {\bibinfo {author} {\bibfnamefont {D.~C.}\ \bibnamefont
  {Tsui}}, \bibinfo {author} {\bibfnamefont {H.~L.}\ \bibnamefont {Stormer}},\
  and\ \bibinfo {author} {\bibfnamefont {A.~C.}\ \bibnamefont {Gossard}},\
  }\bibfield  {title} {\bibinfo {title} {Two-dimensional magnetotransport in
  the extreme quantum limit},\ }\href
  {https://doi.org/10.1103/PhysRevLett.48.1559} {\bibfield  {journal} {\bibinfo
   {journal} {Phys. Rev. Lett.}\ }\textbf {\bibinfo {volume} {48}},\ \bibinfo
  {pages} {1559} (\bibinfo {year} {1982})}\BibitemShut {NoStop}%
\bibitem [{\citenamefont {Kitaev}\ and\ \citenamefont
  {Preskill}(2006)}]{Kitaev2006}%
  \BibitemOpen
  \bibfield  {author} {\bibinfo {author} {\bibfnamefont {A.}~\bibnamefont
  {Kitaev}}\ and\ \bibinfo {author} {\bibfnamefont {J.}~\bibnamefont
  {Preskill}},\ }\bibfield  {title} {\bibinfo {title} {Topological entanglement
  entropy},\ }\href {https://doi.org/10.1103/PhysRevLett.96.110404} {\bibfield
  {journal} {\bibinfo  {journal} {Phys. Rev. Lett.}\ }\textbf {\bibinfo
  {volume} {96}},\ \bibinfo {pages} {110404} (\bibinfo {year}
  {2006})}\BibitemShut {NoStop}%
\bibitem [{\citenamefont {Levin}\ and\ \citenamefont {Wen}(2006)}]{Levin2006}%
  \BibitemOpen
  \bibfield  {author} {\bibinfo {author} {\bibfnamefont {M.}~\bibnamefont
  {Levin}}\ and\ \bibinfo {author} {\bibfnamefont {X.-G.}\ \bibnamefont
  {Wen}},\ }\bibfield  {title} {\bibinfo {title} {Detecting topological order
  in a ground state wave function},\ }\href
  {https://doi.org/10.1103/PhysRevLett.96.110405} {\bibfield  {journal}
  {\bibinfo  {journal} {Phys. Rev. Lett.}\ }\textbf {\bibinfo {volume} {96}},\
  \bibinfo {pages} {110405} (\bibinfo {year} {2006})}\BibitemShut {NoStop}%
\bibitem [{\citenamefont {Wen}\ and\ \citenamefont {Zee}(1992)}]{Wen1992}%
  \BibitemOpen
  \bibfield  {author} {\bibinfo {author} {\bibfnamefont {X.~G.}\ \bibnamefont
  {Wen}}\ and\ \bibinfo {author} {\bibfnamefont {A.}~\bibnamefont {Zee}},\
  }\bibfield  {title} {\bibinfo {title} {Classification of abelian quantum hall
  states and matrix formulation of topological fluids},\ }\href
  {https://doi.org/10.1103/PhysRevB.46.2290} {\bibfield  {journal} {\bibinfo
  {journal} {Phys. Rev. B}\ }\textbf {\bibinfo {volume} {46}},\ \bibinfo
  {pages} {2290} (\bibinfo {year} {1992})}\BibitemShut {NoStop}%
\bibitem [{\citenamefont {Blok}\ and\ \citenamefont {Wen}(1990)}]{Blok1990}%
  \BibitemOpen
  \bibfield  {author} {\bibinfo {author} {\bibfnamefont {B.}~\bibnamefont
  {Blok}}\ and\ \bibinfo {author} {\bibfnamefont {X.~G.}\ \bibnamefont {Wen}},\
  }\bibfield  {title} {\bibinfo {title} {Effective theories of the fractional
  quantum hall effect at generic filling fractions},\ }\href
  {https://doi.org/10.1103/PhysRevB.42.8133} {\bibfield  {journal} {\bibinfo
  {journal} {Phys. Rev. B}\ }\textbf {\bibinfo {volume} {42}},\ \bibinfo
  {pages} {8133} (\bibinfo {year} {1990})}\BibitemShut {NoStop}%
\bibitem [{\citenamefont {Read}(1990)}]{Read1990}%
  \BibitemOpen
  \bibfield  {author} {\bibinfo {author} {\bibfnamefont {N.}~\bibnamefont
  {Read}},\ }\bibfield  {title} {\bibinfo {title} {Excitation structure of the
  hierarchy scheme in the fractional quantum hall effect},\ }\href
  {https://doi.org/10.1103/PhysRevLett.65.1502} {\bibfield  {journal} {\bibinfo
   {journal} {Phys. Rev. Lett.}\ }\textbf {\bibinfo {volume} {65}},\ \bibinfo
  {pages} {1502} (\bibinfo {year} {1990})}\BibitemShut {NoStop}%
\bibitem [{\citenamefont {Chen}\ \emph {et~al.}(2013)\citenamefont {Chen},
  \citenamefont {Gu}, \citenamefont {Liu},\ and\ \citenamefont
  {Wen}}]{Wen2013}%
  \BibitemOpen
  \bibfield  {author} {\bibinfo {author} {\bibfnamefont {X.}~\bibnamefont
  {Chen}}, \bibinfo {author} {\bibfnamefont {Z.-C.}\ \bibnamefont {Gu}},
  \bibinfo {author} {\bibfnamefont {Z.-X.}\ \bibnamefont {Liu}},\ and\ \bibinfo
  {author} {\bibfnamefont {X.-G.}\ \bibnamefont {Wen}},\ }\bibfield  {title}
  {\bibinfo {title} {Symmetry protected topological orders and the group
  cohomology of their symmetry group},\ }\href
  {https://doi.org/10.1103/PhysRevB.87.155114} {\bibfield  {journal} {\bibinfo
  {journal} {Phys. Rev. B}\ }\textbf {\bibinfo {volume} {87}},\ \bibinfo
  {pages} {155114} (\bibinfo {year} {2013})}\BibitemShut {NoStop}%
\bibitem [{\citenamefont {Gu}\ and\ \citenamefont {Wen}(2014)}]{Wen2014}%
  \BibitemOpen
  \bibfield  {author} {\bibinfo {author} {\bibfnamefont {Z.-C.}\ \bibnamefont
  {Gu}}\ and\ \bibinfo {author} {\bibfnamefont {X.-G.}\ \bibnamefont {Wen}},\
  }\bibfield  {title} {\bibinfo {title} {Symmetry-protected topological orders
  for interacting fermions: Fermionic topological nonlinear
  $\ensuremath{\sigma}$ models and a special group supercohomology theory},\
  }\href {https://doi.org/10.1103/PhysRevB.90.115141} {\bibfield  {journal}
  {\bibinfo  {journal} {Phys. Rev. B}\ }\textbf {\bibinfo {volume} {90}},\
  \bibinfo {pages} {115141} (\bibinfo {year} {2014})}\BibitemShut {NoStop}%
\bibitem [{\citenamefont {Gu}\ and\ \citenamefont {Wen}(2009)}]{Gu2009}%
  \BibitemOpen
  \bibfield  {author} {\bibinfo {author} {\bibfnamefont {Z.-C.}\ \bibnamefont
  {Gu}}\ and\ \bibinfo {author} {\bibfnamefont {X.-G.}\ \bibnamefont {Wen}},\
  }\bibfield  {title} {\bibinfo {title} {Tensor-entanglement-filtering
  renormalization approach and symmetry-protected topological order},\ }\href
  {https://doi.org/10.1103/PhysRevB.80.155131} {\bibfield  {journal} {\bibinfo
  {journal} {Phys. Rev. B}\ }\textbf {\bibinfo {volume} {80}},\ \bibinfo
  {pages} {155131} (\bibinfo {year} {2009})}\BibitemShut {NoStop}%
\bibitem [{\citenamefont {Turner}\ \emph {et~al.}(2011)\citenamefont {Turner},
  \citenamefont {Pollmann},\ and\ \citenamefont {Berg}}]{Turner2011}%
  \BibitemOpen
  \bibfield  {author} {\bibinfo {author} {\bibfnamefont {A.~M.}\ \bibnamefont
  {Turner}}, \bibinfo {author} {\bibfnamefont {F.}~\bibnamefont {Pollmann}},\
  and\ \bibinfo {author} {\bibfnamefont {E.}~\bibnamefont {Berg}},\ }\bibfield
  {title} {\bibinfo {title} {Topological phases of one-dimensional fermions: An
  entanglement point of view},\ }\href
  {https://doi.org/10.1103/PhysRevB.83.075102} {\bibfield  {journal} {\bibinfo
  {journal} {Phys. Rev. B}\ }\textbf {\bibinfo {volume} {83}},\ \bibinfo
  {pages} {075102} (\bibinfo {year} {2011})}\BibitemShut {NoStop}%
\bibitem [{\citenamefont {Santos}\ \emph {et~al.}(2019)\citenamefont {Santos},
  \citenamefont {Gutman},\ and\ \citenamefont {Carr}}]{SantosGutmanCarr2019}%
  \BibitemOpen
  \bibfield  {author} {\bibinfo {author} {\bibfnamefont {R.~A.}\ \bibnamefont
  {Santos}}, \bibinfo {author} {\bibfnamefont {D.~B.}\ \bibnamefont {Gutman}},\
  and\ \bibinfo {author} {\bibfnamefont {S.~T.}\ \bibnamefont {Carr}},\
  }\bibfield  {title} {\bibinfo {title} {Interplay between intrinsic and
  emergent topological protection on interacting helical modes},\ }\href
  {https://doi.org/10.1103/PhysRevB.99.075129} {\bibfield  {journal} {\bibinfo
  {journal} {Phys. Rev. B}\ }\textbf {\bibinfo {volume} {99}},\ \bibinfo
  {pages} {075129} (\bibinfo {year} {2019})}\BibitemShut {NoStop}%
\bibitem [{\citenamefont {Morimoto}\ \emph {et~al.}(2015)\citenamefont
  {Morimoto}, \citenamefont {Furusaki},\ and\ \citenamefont
  {Mudry}}]{MorimotoMudry2015}%
  \BibitemOpen
  \bibfield  {author} {\bibinfo {author} {\bibfnamefont {T.}~\bibnamefont
  {Morimoto}}, \bibinfo {author} {\bibfnamefont {A.}~\bibnamefont {Furusaki}},\
  and\ \bibinfo {author} {\bibfnamefont {C.}~\bibnamefont {Mudry}},\ }\bibfield
   {title} {\bibinfo {title} {Breakdown of the topological classification
  $\mathbb{Z}$ for gapped phases of noninteracting fermions by quartic
  interactions},\ }\href {https://doi.org/10.1103/PhysRevB.92.125104}
  {\bibfield  {journal} {\bibinfo  {journal} {Phys. Rev. B}\ }\textbf {\bibinfo
  {volume} {92}},\ \bibinfo {pages} {125104} (\bibinfo {year}
  {2015})}\BibitemShut {NoStop}%
\bibitem [{\citenamefont {You}\ and\ \citenamefont {Xu}(2014)}]{You2014}%
  \BibitemOpen
  \bibfield  {author} {\bibinfo {author} {\bibfnamefont {Y.-Z.}\ \bibnamefont
  {You}}\ and\ \bibinfo {author} {\bibfnamefont {C.}~\bibnamefont {Xu}},\
  }\bibfield  {title} {\bibinfo {title} {Symmetry-protected topological states
  of interacting fermions and bosons},\ }\href
  {https://doi.org/10.1103/PhysRevB.90.245120} {\bibfield  {journal} {\bibinfo
  {journal} {Phys. Rev. B}\ }\textbf {\bibinfo {volume} {90}},\ \bibinfo
  {pages} {245120} (\bibinfo {year} {2014})}\BibitemShut {NoStop}%
\bibitem [{\citenamefont {Fidkowski}\ and\ \citenamefont
  {Kitaev}(2010)}]{FidkowskiKitaev2010}%
  \BibitemOpen
  \bibfield  {author} {\bibinfo {author} {\bibfnamefont {L.}~\bibnamefont
  {Fidkowski}}\ and\ \bibinfo {author} {\bibfnamefont {A.}~\bibnamefont
  {Kitaev}},\ }\bibfield  {title} {\bibinfo {title} {Effects of interactions on
  the topological classification of free fermion systems},\ }\href
  {https://doi.org/10.1103/PhysRevB.81.134509} {\bibfield  {journal} {\bibinfo
  {journal} {Phys. Rev. B}\ }\textbf {\bibinfo {volume} {81}},\ \bibinfo
  {pages} {134509} (\bibinfo {year} {2010})}\BibitemShut {NoStop}%
\bibitem [{\citenamefont {Fidkowski}\ and\ \citenamefont
  {Kitaev}(2011)}]{FidkowskiKitaev2011}%
  \BibitemOpen
  \bibfield  {author} {\bibinfo {author} {\bibfnamefont {L.}~\bibnamefont
  {Fidkowski}}\ and\ \bibinfo {author} {\bibfnamefont {A.}~\bibnamefont
  {Kitaev}},\ }\bibfield  {title} {\bibinfo {title} {Topological phases of
  fermions in one dimension},\ }\href
  {https://doi.org/10.1103/PhysRevB.83.075103} {\bibfield  {journal} {\bibinfo
  {journal} {Phys. Rev. B}\ }\textbf {\bibinfo {volume} {83}},\ \bibinfo
  {pages} {075103} (\bibinfo {year} {2011})}\BibitemShut {NoStop}%
\bibitem [{\citenamefont {Haldane}(1983{\natexlab{a}})}]{Haldane19831}%
  \BibitemOpen
  \bibfield  {author} {\bibinfo {author} {\bibfnamefont {F.}~\bibnamefont
  {Haldane}},\ }\bibfield  {title} {\bibinfo {title} {Continuum dynamics of the
  1-d heisenberg antiferromagnet: Identification with the o(3) nonlinear sigma
  model},\ }\href
  {https://doi.org/https://doi.org/10.1016/0375-9601(83)90631-X} {\bibfield
  {journal} {\bibinfo  {journal} {Physics Letters A}\ }\textbf {\bibinfo
  {volume} {93}},\ \bibinfo {pages} {464} (\bibinfo {year}
  {1983}{\natexlab{a}})}\BibitemShut {NoStop}%
\bibitem [{\citenamefont {Haldane}(1983{\natexlab{b}})}]{Haldane19832}%
  \BibitemOpen
  \bibfield  {author} {\bibinfo {author} {\bibfnamefont {F.~D.~M.}\
  \bibnamefont {Haldane}},\ }\bibfield  {title} {\bibinfo {title} {Nonlinear
  field theory of large-spin heisenberg antiferromagnets: Semiclassically
  quantized solitons of the one-dimensional easy-axis n\'eel state},\ }\href
  {https://doi.org/10.1103/PhysRevLett.50.1153} {\bibfield  {journal} {\bibinfo
   {journal} {Phys. Rev. Lett.}\ }\textbf {\bibinfo {volume} {50}},\ \bibinfo
  {pages} {1153} (\bibinfo {year} {1983}{\natexlab{b}})}\BibitemShut {NoStop}%
\bibitem [{\citenamefont {Affleck}\ \emph {et~al.}(1987)\citenamefont
  {Affleck}, \citenamefont {Kennedy}, \citenamefont {Lieb},\ and\ \citenamefont
  {Tasaki}}]{Affleck1987}%
  \BibitemOpen
  \bibfield  {author} {\bibinfo {author} {\bibfnamefont {I.}~\bibnamefont
  {Affleck}}, \bibinfo {author} {\bibfnamefont {T.}~\bibnamefont {Kennedy}},
  \bibinfo {author} {\bibfnamefont {E.~H.}\ \bibnamefont {Lieb}},\ and\
  \bibinfo {author} {\bibfnamefont {H.}~\bibnamefont {Tasaki}},\ }\bibfield
  {title} {\bibinfo {title} {Rigorous results on valence-bond ground states in
  antiferromagnets},\ }\href {https://doi.org/10.1103/PhysRevLett.59.799}
  {\bibfield  {journal} {\bibinfo  {journal} {Phys. Rev. Lett.}\ }\textbf
  {\bibinfo {volume} {59}},\ \bibinfo {pages} {799} (\bibinfo {year}
  {1987})}\BibitemShut {NoStop}%
\bibitem [{\citenamefont {Verresen}\ \emph {et~al.}(2017)\citenamefont
  {Verresen}, \citenamefont {Moessner},\ and\ \citenamefont
  {Pollmann}}]{Verresen2017}%
  \BibitemOpen
  \bibfield  {author} {\bibinfo {author} {\bibfnamefont {R.}~\bibnamefont
  {Verresen}}, \bibinfo {author} {\bibfnamefont {R.}~\bibnamefont {Moessner}},\
  and\ \bibinfo {author} {\bibfnamefont {F.}~\bibnamefont {Pollmann}},\
  }\bibfield  {title} {\bibinfo {title} {One-dimensional symmetry protected
  topological phases and their transitions},\ }\href
  {https://doi.org/10.1103/PhysRevB.96.165124} {\bibfield  {journal} {\bibinfo
  {journal} {Phys. Rev. B}\ }\textbf {\bibinfo {volume} {96}},\ \bibinfo
  {pages} {165124} (\bibinfo {year} {2017})}\BibitemShut {NoStop}%
\bibitem [{\citenamefont {Gogolin}\ \emph {et~al.}(2004)\citenamefont
  {Gogolin}, \citenamefont {Nersesyan},\ and\ \citenamefont
  {Tsvelik}}]{gogolin2004bosonization}%
  \BibitemOpen
  \bibfield  {author} {\bibinfo {author} {\bibfnamefont {A.}~\bibnamefont
  {Gogolin}}, \bibinfo {author} {\bibfnamefont {A.}~\bibnamefont {Nersesyan}},\
  and\ \bibinfo {author} {\bibfnamefont {A.}~\bibnamefont {Tsvelik}},\ }\href
  {https://books.google.co.il/books?id=BZDfFIpCoaAC} {\emph {\bibinfo {title}
  {Bosonization and Strongly Correlated Systems}}}\ (\bibinfo  {publisher}
  {Cambridge University Press},\ \bibinfo {year} {2004})\BibitemShut {NoStop}%
\bibitem [{\citenamefont {Giamarchi}(2003)}]{Giamarchi_book}%
  \BibitemOpen
  \bibfield  {author} {\bibinfo {author} {\bibfnamefont {T.}~\bibnamefont
  {Giamarchi}},\ }\href
  {https://doi.org/10.1093/acprof:oso/9780198525004.001.0001} {\emph {\bibinfo
  {title} {{Quantum Physics in One Dimension}}}}\ (\bibinfo  {publisher}
  {Oxford University Press},\ \bibinfo {year} {2003})\BibitemShut {NoStop}%
\bibitem [{\citenamefont {Matveeva}\ \emph {et~al.}(2023)\citenamefont
  {Matveeva}, \citenamefont {Hewitt}, \citenamefont {Liu}, \citenamefont
  {Reddy}, \citenamefont {Gutman},\ and\ \citenamefont {Carr}}]{Matveeva2023}%
  \BibitemOpen
  \bibfield  {author} {\bibinfo {author} {\bibfnamefont {P.}~\bibnamefont
  {Matveeva}}, \bibinfo {author} {\bibfnamefont {T.}~\bibnamefont {Hewitt}},
  \bibinfo {author} {\bibfnamefont {D.}~\bibnamefont {Liu}}, \bibinfo {author}
  {\bibfnamefont {K.}~\bibnamefont {Reddy}}, \bibinfo {author} {\bibfnamefont
  {D.}~\bibnamefont {Gutman}},\ and\ \bibinfo {author} {\bibfnamefont {S.~T.}\
  \bibnamefont {Carr}},\ }\bibfield  {title} {\bibinfo {title} {One-dimensional
  noninteracting topological insulators with chiral symmetry},\ }\href
  {https://doi.org/10.1103/PhysRevB.107.075422} {\bibfield  {journal} {\bibinfo
   {journal} {Phys. Rev. B}\ }\textbf {\bibinfo {volume} {107}},\ \bibinfo
  {pages} {075422} (\bibinfo {year} {2023})}\BibitemShut {NoStop}%
\bibitem [{\citenamefont {Sirker}\ \emph {et~al.}(2014)\citenamefont {Sirker},
  \citenamefont {Maiti}, \citenamefont {Konstantinidis},\ and\ \citenamefont
  {Sedlmayr}}]{Sirker2014}%
  \BibitemOpen
  \bibfield  {author} {\bibinfo {author} {\bibfnamefont {J.}~\bibnamefont
  {Sirker}}, \bibinfo {author} {\bibfnamefont {M.}~\bibnamefont {Maiti}},
  \bibinfo {author} {\bibfnamefont {N.~P.}\ \bibnamefont {Konstantinidis}},\
  and\ \bibinfo {author} {\bibfnamefont {N.}~\bibnamefont {Sedlmayr}},\
  }\bibfield  {title} {\bibinfo {title} {Boundary fidelity and entanglement in
  the symmetry protected topological phase of the ssh model},\ }\href
  {https://doi.org/10.1088/1742-5468/2014/10/P10032} {\bibfield  {journal}
  {\bibinfo  {journal} {Journal of Statistical Mechanics: Theory and
  Experiment}\ }\textbf {\bibinfo {volume} {2014}},\ \bibinfo {pages} {P10032}
  (\bibinfo {year} {2014})}\BibitemShut {NoStop}%
\bibitem [{\citenamefont {Yahyavi}\ \emph {et~al.}(2018)\citenamefont
  {Yahyavi}, \citenamefont {Saleem},\ and\ \citenamefont
  {Hetényi}}]{Yahyavi_2018}%
  \BibitemOpen
  \bibfield  {author} {\bibinfo {author} {\bibfnamefont {M.}~\bibnamefont
  {Yahyavi}}, \bibinfo {author} {\bibfnamefont {L.}~\bibnamefont {Saleem}},\
  and\ \bibinfo {author} {\bibfnamefont {B.}~\bibnamefont {Hetényi}},\
  }\bibfield  {title} {\bibinfo {title} {Variational study of the interacting,
  spinless su–schrieffer–heeger model},\ }\href
  {https://doi.org/10.1088/1361-648X/aae0a4} {\bibfield  {journal} {\bibinfo
  {journal} {Journal of Physics: Condensed Matter}\ }\textbf {\bibinfo {volume}
  {30}},\ \bibinfo {pages} {445602} (\bibinfo {year} {2018})}\BibitemShut
  {NoStop}%
\bibitem [{\citenamefont {Lin}\ \emph {et~al.}(2020)\citenamefont {Lin},
  \citenamefont {Kennes}, \citenamefont {Pletyukhov}, \citenamefont {Weber},
  \citenamefont {Schoeller},\ and\ \citenamefont {Meden}}]{Lin2020}%
  \BibitemOpen
  \bibfield  {author} {\bibinfo {author} {\bibfnamefont {Y.-T.}\ \bibnamefont
  {Lin}}, \bibinfo {author} {\bibfnamefont {D.~M.}\ \bibnamefont {Kennes}},
  \bibinfo {author} {\bibfnamefont {M.}~\bibnamefont {Pletyukhov}}, \bibinfo
  {author} {\bibfnamefont {C.~S.}\ \bibnamefont {Weber}}, \bibinfo {author}
  {\bibfnamefont {H.}~\bibnamefont {Schoeller}},\ and\ \bibinfo {author}
  {\bibfnamefont {V.}~\bibnamefont {Meden}},\ }\bibfield  {title} {\bibinfo
  {title} {Interacting rice-mele model: Bulk and boundaries},\ }\href
  {https://doi.org/10.1103/PhysRevB.102.085122} {\bibfield  {journal} {\bibinfo
   {journal} {Phys. Rev. B}\ }\textbf {\bibinfo {volume} {102}},\ \bibinfo
  {pages} {085122} (\bibinfo {year} {2020})}\BibitemShut {NoStop}%
\bibitem [{\citenamefont {Marques}\ and\ \citenamefont
  {Dias}(2017)}]{Marques2017}%
  \BibitemOpen
  \bibfield  {author} {\bibinfo {author} {\bibfnamefont {A.~M.}\ \bibnamefont
  {Marques}}\ and\ \bibinfo {author} {\bibfnamefont {R.~G.}\ \bibnamefont
  {Dias}},\ }\bibfield  {title} {\bibinfo {title} {Multihole edge states in
  su-schrieffer-heeger chains with interactions},\ }\href
  {https://doi.org/10.1103/PhysRevB.95.115443} {\bibfield  {journal} {\bibinfo
  {journal} {Phys. Rev. B}\ }\textbf {\bibinfo {volume} {95}},\ \bibinfo
  {pages} {115443} (\bibinfo {year} {2017})}\BibitemShut {NoStop}%
\bibitem [{\citenamefont {Nersesyan}(2020)}]{Nersesyan2020}%
  \BibitemOpen
  \bibfield  {author} {\bibinfo {author} {\bibfnamefont {A.~A.}\ \bibnamefont
  {Nersesyan}},\ }\bibfield  {title} {\bibinfo {title} {Phase diagram of an
  interacting staggered su-schrieffer-heeger two-chain ladder close to a
  quantum critical point},\ }\href
  {https://doi.org/10.1103/PhysRevB.102.045108} {\bibfield  {journal} {\bibinfo
   {journal} {Phys. Rev. B}\ }\textbf {\bibinfo {volume} {102}},\ \bibinfo
  {pages} {045108} (\bibinfo {year} {2020})}\BibitemShut {NoStop}%
\bibitem [{\citenamefont {Jin}\ \emph {et~al.}(2023)\citenamefont {Jin},
  \citenamefont {Ruggiero},\ and\ \citenamefont {Giamarchi}}]{Giamarchi2023}%
  \BibitemOpen
  \bibfield  {author} {\bibinfo {author} {\bibfnamefont {T.}~\bibnamefont
  {Jin}}, \bibinfo {author} {\bibfnamefont {P.}~\bibnamefont {Ruggiero}},\ and\
  \bibinfo {author} {\bibfnamefont {T.}~\bibnamefont {Giamarchi}},\ }\bibfield
  {title} {\bibinfo {title} {Bosonization of the interacting
  su-schrieffer-heeger model},\ }\href
  {https://doi.org/10.1103/PhysRevB.107.L201111} {\bibfield  {journal}
  {\bibinfo  {journal} {Phys. Rev. B}\ }\textbf {\bibinfo {volume} {107}},\
  \bibinfo {pages} {L201111} (\bibinfo {year} {2023})}\BibitemShut {NoStop}%
\bibitem [{\citenamefont {Manmana}\ \emph {et~al.}(2012)\citenamefont
  {Manmana}, \citenamefont {Essin}, \citenamefont {Noack},\ and\ \citenamefont
  {Gurarie}}]{Gurarie2012}%
  \BibitemOpen
  \bibfield  {author} {\bibinfo {author} {\bibfnamefont {S.~R.}\ \bibnamefont
  {Manmana}}, \bibinfo {author} {\bibfnamefont {A.~M.}\ \bibnamefont {Essin}},
  \bibinfo {author} {\bibfnamefont {R.~M.}\ \bibnamefont {Noack}},\ and\
  \bibinfo {author} {\bibfnamefont {V.}~\bibnamefont {Gurarie}},\ }\bibfield
  {title} {\bibinfo {title} {Topological invariants and interacting
  one-dimensional fermionic systems},\ }\href
  {https://doi.org/10.1103/PhysRevB.86.205119} {\bibfield  {journal} {\bibinfo
  {journal} {Phys. Rev. B}\ }\textbf {\bibinfo {volume} {86}},\ \bibinfo
  {pages} {205119} (\bibinfo {year} {2012})}\BibitemShut {NoStop}%
\bibitem [{\citenamefont {Tsuchiizu}\ and\ \citenamefont
  {Furusaki}(2004)}]{Tsuchiizu2004}%
  \BibitemOpen
  \bibfield  {author} {\bibinfo {author} {\bibfnamefont {M.}~\bibnamefont
  {Tsuchiizu}}\ and\ \bibinfo {author} {\bibfnamefont {A.}~\bibnamefont
  {Furusaki}},\ }\bibfield  {title} {\bibinfo {title} {Ground-state phase
  diagram of the one-dimensional half-filled extended hubbard model},\ }\href
  {https://doi.org/10.1103/PhysRevB.69.035103} {\bibfield  {journal} {\bibinfo
  {journal} {Phys. Rev. B}\ }\textbf {\bibinfo {volume} {69}},\ \bibinfo
  {pages} {035103} (\bibinfo {year} {2004})}\BibitemShut {NoStop}%
\bibitem [{\citenamefont {Benthien}\ \emph {et~al.}(2006)\citenamefont
  {Benthien}, \citenamefont {Essler},\ and\ \citenamefont
  {Grage}}]{Benthien2006}%
  \BibitemOpen
  \bibfield  {author} {\bibinfo {author} {\bibfnamefont {H.}~\bibnamefont
  {Benthien}}, \bibinfo {author} {\bibfnamefont {F.~H.~L.}\ \bibnamefont
  {Essler}},\ and\ \bibinfo {author} {\bibfnamefont {A.}~\bibnamefont
  {Grage}},\ }\bibfield  {title} {\bibinfo {title} {Quantum phase transition in
  the one-dimensional extended peierls-hubbard model},\ }\href
  {https://doi.org/10.1103/PhysRevB.73.085105} {\bibfield  {journal} {\bibinfo
  {journal} {Phys. Rev. B}\ }\textbf {\bibinfo {volume} {73}},\ \bibinfo
  {pages} {085105} (\bibinfo {year} {2006})}\BibitemShut {NoStop}%
\bibitem [{\citenamefont {Ejima}\ \emph {et~al.}(2016)\citenamefont {Ejima},
  \citenamefont {Essler}, \citenamefont {Lange},\ and\ \citenamefont
  {Fehske}}]{Ejima2016}%
  \BibitemOpen
  \bibfield  {author} {\bibinfo {author} {\bibfnamefont {S.}~\bibnamefont
  {Ejima}}, \bibinfo {author} {\bibfnamefont {F.~H.~L.}\ \bibnamefont
  {Essler}}, \bibinfo {author} {\bibfnamefont {F.}~\bibnamefont {Lange}},\ and\
  \bibinfo {author} {\bibfnamefont {H.}~\bibnamefont {Fehske}},\ }\bibfield
  {title} {\bibinfo {title} {Ising tricriticality in the extended hubbard model
  with bond dimerization},\ }\href {https://doi.org/10.1103/PhysRevB.93.235118}
  {\bibfield  {journal} {\bibinfo  {journal} {Phys. Rev. B}\ }\textbf {\bibinfo
  {volume} {93}},\ \bibinfo {pages} {235118} (\bibinfo {year}
  {2016})}\BibitemShut {NoStop}%
\bibitem [{\citenamefont {Ejima}\ \emph {et~al.}(2018)\citenamefont {Ejima},
  \citenamefont {Lange}, \citenamefont {Essler},\ and\ \citenamefont
  {Fehske}}]{Ejima2018}%
  \BibitemOpen
  \bibfield  {author} {\bibinfo {author} {\bibfnamefont {S.}~\bibnamefont
  {Ejima}}, \bibinfo {author} {\bibfnamefont {F.}~\bibnamefont {Lange}},
  \bibinfo {author} {\bibfnamefont {F.~H.}\ \bibnamefont {Essler}},\ and\
  \bibinfo {author} {\bibfnamefont {H.}~\bibnamefont {Fehske}},\ }\bibfield
  {title} {\bibinfo {title} {Critical behavior of the extended hubbard model
  with bond dimerization},\ }\href
  {https://doi.org/https://doi.org/10.1016/j.physb.2017.09.001} {\bibfield
  {journal} {\bibinfo  {journal} {Physica B: Condensed Matter}\ }\textbf
  {\bibinfo {volume} {536}},\ \bibinfo {pages} {474} (\bibinfo {year}
  {2018})}\BibitemShut {NoStop}%
\bibitem [{\citenamefont {Kainaris}\ and\ \citenamefont
  {Carr}(2015)}]{Kainaris2015}%
  \BibitemOpen
  \bibfield  {author} {\bibinfo {author} {\bibfnamefont {N.}~\bibnamefont
  {Kainaris}}\ and\ \bibinfo {author} {\bibfnamefont {S.~T.}\ \bibnamefont
  {Carr}},\ }\bibfield  {title} {\bibinfo {title} {Emergent topological
  properties in interacting one-dimensional systems with spin-orbit coupling},\
  }\href {https://doi.org/10.1103/PhysRevB.92.035139} {\bibfield  {journal}
  {\bibinfo  {journal} {Phys. Rev. B}\ }\textbf {\bibinfo {volume} {92}},\
  \bibinfo {pages} {035139} (\bibinfo {year} {2015})}\BibitemShut {NoStop}%
\bibitem [{\citenamefont {Kainaris}\ \emph {et~al.}(2017)\citenamefont
  {Kainaris}, \citenamefont {Santos}, \citenamefont {Gutman},\ and\
  \citenamefont {Carr}}]{KainarisCarrGutman2017}%
  \BibitemOpen
  \bibfield  {author} {\bibinfo {author} {\bibfnamefont {N.}~\bibnamefont
  {Kainaris}}, \bibinfo {author} {\bibfnamefont {R.~A.}\ \bibnamefont
  {Santos}}, \bibinfo {author} {\bibfnamefont {D.~B.}\ \bibnamefont {Gutman}},\
  and\ \bibinfo {author} {\bibfnamefont {S.~T.}\ \bibnamefont {Carr}},\
  }\bibfield  {title} {\bibinfo {title} {Interaction induced topological
  protection in one-dimensional conductors},\ }\href
  {https://doi.org/https://doi.org/10.1002/prop.201600054} {\bibfield
  {journal} {\bibinfo  {journal} {Fortschritte der Physik}\ }\textbf {\bibinfo
  {volume} {65}},\ \bibinfo {pages} {1600054} (\bibinfo {year}
  {2017})}\BibitemShut {NoStop}%
\bibitem [{\citenamefont {Santos}\ \emph {et~al.}(2016)\citenamefont {Santos},
  \citenamefont {Gutman},\ and\ \citenamefont {Carr}}]{SantosGutmanCarr2016}%
  \BibitemOpen
  \bibfield  {author} {\bibinfo {author} {\bibfnamefont {R.~A.}\ \bibnamefont
  {Santos}}, \bibinfo {author} {\bibfnamefont {D.~B.}\ \bibnamefont {Gutman}},\
  and\ \bibinfo {author} {\bibfnamefont {S.~T.}\ \bibnamefont {Carr}},\
  }\bibfield  {title} {\bibinfo {title} {Phase diagram of two interacting
  helical states},\ }\href {https://doi.org/10.1103/PhysRevB.93.235436}
  {\bibfield  {journal} {\bibinfo  {journal} {Phys. Rev. B}\ }\textbf {\bibinfo
  {volume} {93}},\ \bibinfo {pages} {235436} (\bibinfo {year}
  {2016})}\BibitemShut {NoStop}%
\bibitem [{\citenamefont {Kainaris}\ \emph {et~al.}(2018)\citenamefont
  {Kainaris}, \citenamefont {Carr},\ and\ \citenamefont {Mirlin}}]{Mirlin2018}%
  \BibitemOpen
  \bibfield  {author} {\bibinfo {author} {\bibfnamefont {N.}~\bibnamefont
  {Kainaris}}, \bibinfo {author} {\bibfnamefont {S.~T.}\ \bibnamefont {Carr}},\
  and\ \bibinfo {author} {\bibfnamefont {A.~D.}\ \bibnamefont {Mirlin}},\
  }\bibfield  {title} {\bibinfo {title} {Transmission through a potential
  barrier in luttinger liquids with a topological spin gap},\ }\href
  {https://doi.org/10.1103/PhysRevB.97.115107} {\bibfield  {journal} {\bibinfo
  {journal} {Phys. Rev. B}\ }\textbf {\bibinfo {volume} {97}},\ \bibinfo
  {pages} {115107} (\bibinfo {year} {2018})}\BibitemShut {NoStop}%
\bibitem [{\citenamefont {Keselman}\ and\ \citenamefont
  {Berg}(2015)}]{Keselman2015}%
  \BibitemOpen
  \bibfield  {author} {\bibinfo {author} {\bibfnamefont {A.}~\bibnamefont
  {Keselman}}\ and\ \bibinfo {author} {\bibfnamefont {E.}~\bibnamefont
  {Berg}},\ }\bibfield  {title} {\bibinfo {title} {Gapless symmetry-protected
  topological phase of fermions in one dimension},\ }\href
  {https://doi.org/10.1103/PhysRevB.91.235309} {\bibfield  {journal} {\bibinfo
  {journal} {Phys. Rev. B}\ }\textbf {\bibinfo {volume} {91}},\ \bibinfo
  {pages} {235309} (\bibinfo {year} {2015})}\BibitemShut {NoStop}%
\bibitem [{Note1()}]{Note1}%
  \BibitemOpen
  \bibinfo {note} {For multiple bands $\phi (k)$ is defined as the complex
  phase $ \det \Delta (k)$.}\BibitemShut {Stop}%
\bibitem [{\citenamefont {Fuchs}\ and\ \citenamefont
  {Pi\'echon}(2021)}]{Fuchs2021}%
  \BibitemOpen
  \bibfield  {author} {\bibinfo {author} {\bibfnamefont {J.-N.}\ \bibnamefont
  {Fuchs}}\ and\ \bibinfo {author} {\bibfnamefont {F.}~\bibnamefont
  {Pi\'echon}},\ }\bibfield  {title} {\bibinfo {title} {Orbital embedding and
  topology of one-dimensional two-band insulators},\ }\href
  {https://doi.org/10.1103/PhysRevB.104.235428} {\bibfield  {journal} {\bibinfo
   {journal} {Phys. Rev. B}\ }\textbf {\bibinfo {volume} {104}},\ \bibinfo
  {pages} {235428} (\bibinfo {year} {2021})}\BibitemShut {NoStop}%
\bibitem [{\citenamefont {Fabrizio}\ and\ \citenamefont
  {Gogolin}(1995)}]{Fabrizio1995}%
  \BibitemOpen
  \bibfield  {author} {\bibinfo {author} {\bibfnamefont {M.}~\bibnamefont
  {Fabrizio}}\ and\ \bibinfo {author} {\bibfnamefont {A.~O.}\ \bibnamefont
  {Gogolin}},\ }\bibfield  {title} {\bibinfo {title} {Interacting
  one-dimensional electron gas with open boundaries},\ }\href
  {https://doi.org/10.1103/PhysRevB.51.17827} {\bibfield  {journal} {\bibinfo
  {journal} {Phys. Rev. B}\ }\textbf {\bibinfo {volume} {51}},\ \bibinfo
  {pages} {17827} (\bibinfo {year} {1995})}\BibitemShut {NoStop}%
\bibitem [{\citenamefont {Mattsson}\ \emph {et~al.}(1997)\citenamefont
  {Mattsson}, \citenamefont {Eggert},\ and\ \citenamefont
  {Johannesson}}]{Mattsson1997}%
  \BibitemOpen
  \bibfield  {author} {\bibinfo {author} {\bibfnamefont {A.~E.}\ \bibnamefont
  {Mattsson}}, \bibinfo {author} {\bibfnamefont {S.}~\bibnamefont {Eggert}},\
  and\ \bibinfo {author} {\bibfnamefont {H.}~\bibnamefont {Johannesson}},\
  }\bibfield  {title} {\bibinfo {title} {Properties of a luttinger liquid with
  boundaries at finite temperature and size},\ }\href
  {https://doi.org/10.1103/PhysRevB.56.15615} {\bibfield  {journal} {\bibinfo
  {journal} {Phys. Rev. B}\ }\textbf {\bibinfo {volume} {56}},\ \bibinfo
  {pages} {15615} (\bibinfo {year} {1997})}\BibitemShut {NoStop}%
\bibitem [{\citenamefont {Cazalilla}(2002)}]{Cazalilla2002}%
  \BibitemOpen
  \bibfield  {author} {\bibinfo {author} {\bibfnamefont {M.~A.}\ \bibnamefont
  {Cazalilla}},\ }\bibfield  {title} {\bibinfo {title} {Low-energy properties
  of a one-dimensional system of interacting bosons with boundaries},\ }\href
  {https://doi.org/10.1209/epl/i2002-00112-5} {\bibfield  {journal} {\bibinfo
  {journal} {Europhysics Letters}\ }\textbf {\bibinfo {volume} {59}},\ \bibinfo
  {pages} {793} (\bibinfo {year} {2002})}\BibitemShut {NoStop}%
\bibitem [{\citenamefont {S{\'e}n{\'e}chal}(2004)}]{Senechal2004}%
  \BibitemOpen
  \bibfield  {author} {\bibinfo {author} {\bibfnamefont {D.}~\bibnamefont
  {S{\'e}n{\'e}chal}},\ }\bibinfo {title} {An introduction to bosonization},\
  in\ \href {https://doi.org/10.1007/0-387-21717-7_4} {\emph {\bibinfo
  {booktitle} {Theoretical Methods for Strongly Correlated Electrons}}},\
  \bibinfo {editor} {edited by\ \bibinfo {editor} {\bibfnamefont
  {D.}~\bibnamefont {S{\'e}n{\'e}chal}}, \bibinfo {editor} {\bibfnamefont
  {A.-M.}\ \bibnamefont {Tremblay}},\ and\ \bibinfo {editor} {\bibfnamefont
  {C.}~\bibnamefont {Bourbonnais}}}\ (\bibinfo  {publisher} {Springer New
  York},\ \bibinfo {address} {New York, NY},\ \bibinfo {year} {2004})\ pp.\
  \bibinfo {pages} {139--186}\BibitemShut {NoStop}%
\bibitem [{\citenamefont {Francesco}\ \emph {et~al.}(2012)\citenamefont
  {Francesco}, \citenamefont {Mathieu},\ and\ \citenamefont
  {S{\'e}n{\'e}chal}}]{francesco2012conformal}%
  \BibitemOpen
  \bibfield  {author} {\bibinfo {author} {\bibfnamefont {P.}~\bibnamefont
  {Francesco}}, \bibinfo {author} {\bibfnamefont {P.}~\bibnamefont {Mathieu}},\
  and\ \bibinfo {author} {\bibfnamefont {D.}~\bibnamefont {S{\'e}n{\'e}chal}},\
  }\href@noop {} {\emph {\bibinfo {title} {Conformal field theory}}}\ (\bibinfo
   {publisher} {Springer Science \& Business Media},\ \bibinfo {year}
  {2012})\BibitemShut {NoStop}%
\bibitem [{\citenamefont {Meidan}\ \emph {et~al.}(2014)\citenamefont {Meidan},
  \citenamefont {Romito},\ and\ \citenamefont {Brouwer}}]{Meidan2014}%
  \BibitemOpen
  \bibfield  {author} {\bibinfo {author} {\bibfnamefont {D.}~\bibnamefont
  {Meidan}}, \bibinfo {author} {\bibfnamefont {A.}~\bibnamefont {Romito}},\
  and\ \bibinfo {author} {\bibfnamefont {P.~W.}\ \bibnamefont {Brouwer}},\
  }\bibfield  {title} {\bibinfo {title} {Scattering matrix formulation of the
  topological index of interacting fermions in one-dimensional
  superconductors},\ }\href {https://doi.org/10.1103/PhysRevLett.113.057003}
  {\bibfield  {journal} {\bibinfo  {journal} {Phys. Rev. Lett.}\ }\textbf
  {\bibinfo {volume} {113}},\ \bibinfo {pages} {057003} (\bibinfo {year}
  {2014})}\BibitemShut {NoStop}%
\bibitem [{\citenamefont {Queiroz}\ \emph {et~al.}(2016)\citenamefont
  {Queiroz}, \citenamefont {Khalaf},\ and\ \citenamefont
  {Stern}}]{Queiroz2016}%
  \BibitemOpen
  \bibfield  {author} {\bibinfo {author} {\bibfnamefont {R.}~\bibnamefont
  {Queiroz}}, \bibinfo {author} {\bibfnamefont {E.}~\bibnamefont {Khalaf}},\
  and\ \bibinfo {author} {\bibfnamefont {A.}~\bibnamefont {Stern}},\ }\bibfield
   {title} {\bibinfo {title} {Dimensional hierarchy of fermionic interacting
  topological phases},\ }\href {https://doi.org/10.1103/PhysRevLett.117.206405}
  {\bibfield  {journal} {\bibinfo  {journal} {Phys. Rev. Lett.}\ }\textbf
  {\bibinfo {volume} {117}},\ \bibinfo {pages} {206405} (\bibinfo {year}
  {2016})}\BibitemShut {NoStop}%
\bibitem [{\citenamefont {Song}\ and\ \citenamefont
  {Schnyder}(2017)}]{Song2017}%
  \BibitemOpen
  \bibfield  {author} {\bibinfo {author} {\bibfnamefont {X.-Y.}\ \bibnamefont
  {Song}}\ and\ \bibinfo {author} {\bibfnamefont {A.~P.}\ \bibnamefont
  {Schnyder}},\ }\bibfield  {title} {\bibinfo {title} {Interaction effects on
  the classification of crystalline topological insulators and
  superconductors},\ }\href {https://doi.org/10.1103/PhysRevB.95.195108}
  {\bibfield  {journal} {\bibinfo  {journal} {Phys. Rev. B}\ }\textbf {\bibinfo
  {volume} {95}},\ \bibinfo {pages} {195108} (\bibinfo {year}
  {2017})}\BibitemShut {NoStop}%
\bibitem [{\citenamefont {Gangadharaiah}\ \emph {et~al.}(2008)\citenamefont
  {Gangadharaiah}, \citenamefont {Sun},\ and\ \citenamefont
  {Starykh}}]{Starykh2008}%
  \BibitemOpen
  \bibfield  {author} {\bibinfo {author} {\bibfnamefont {S.}~\bibnamefont
  {Gangadharaiah}}, \bibinfo {author} {\bibfnamefont {J.}~\bibnamefont {Sun}},\
  and\ \bibinfo {author} {\bibfnamefont {O.~A.}\ \bibnamefont {Starykh}},\
  }\bibfield  {title} {\bibinfo {title} {Spin-orbital effects in magnetized
  quantum wires and spin chains},\ }\href
  {https://doi.org/10.1103/PhysRevB.78.054436} {\bibfield  {journal} {\bibinfo
  {journal} {Phys. Rev. B}\ }\textbf {\bibinfo {volume} {78}},\ \bibinfo
  {pages} {054436} (\bibinfo {year} {2008})}\BibitemShut {NoStop}%
\bibitem [{\citenamefont {Carr}\ \emph {et~al.}(2013)\citenamefont {Carr},
  \citenamefont {Narozhny},\ and\ \citenamefont {Nersesyan}}]{Carr2013}%
  \BibitemOpen
  \bibfield  {author} {\bibinfo {author} {\bibfnamefont {S.~T.}\ \bibnamefont
  {Carr}}, \bibinfo {author} {\bibfnamefont {B.~N.}\ \bibnamefont {Narozhny}},\
  and\ \bibinfo {author} {\bibfnamefont {A.~A.}\ \bibnamefont {Nersesyan}},\
  }\bibfield  {title} {\bibinfo {title} {Spinful fermionic ladders at
  incommensurate filling: Phase diagram, local perturbations, and ionic
  potentials},\ }\href
  {https://doi.org/https://doi.org/10.1016/j.aop.2013.08.007} {\bibfield
  {journal} {\bibinfo  {journal} {Annals of Physics}\ }\textbf {\bibinfo
  {volume} {339}},\ \bibinfo {pages} {22} (\bibinfo {year} {2013})}\BibitemShut
  {NoStop}%
\bibitem [{\citenamefont {Sun}\ \emph {et~al.}(2007)\citenamefont {Sun},
  \citenamefont {Gangadharaiah},\ and\ \citenamefont {Starykh}}]{Starykh2007}%
  \BibitemOpen
  \bibfield  {author} {\bibinfo {author} {\bibfnamefont {J.}~\bibnamefont
  {Sun}}, \bibinfo {author} {\bibfnamefont {S.}~\bibnamefont {Gangadharaiah}},\
  and\ \bibinfo {author} {\bibfnamefont {O.~A.}\ \bibnamefont {Starykh}},\
  }\bibfield  {title} {\bibinfo {title} {Spin-orbit-induced spin-density wave
  in a quantum wire},\ }\href {https://doi.org/10.1103/PhysRevLett.98.126408}
  {\bibfield  {journal} {\bibinfo  {journal} {Phys. Rev. Lett.}\ }\textbf
  {\bibinfo {volume} {98}},\ \bibinfo {pages} {126408} (\bibinfo {year}
  {2007})}\BibitemShut {NoStop}%
\bibitem [{\citenamefont {Moroz}\ and\ \citenamefont
  {Barnes}(1999)}]{Moroz1999}%
  \BibitemOpen
  \bibfield  {author} {\bibinfo {author} {\bibfnamefont {A.~V.}\ \bibnamefont
  {Moroz}}\ and\ \bibinfo {author} {\bibfnamefont {C.~H.~W.}\ \bibnamefont
  {Barnes}},\ }\bibfield  {title} {\bibinfo {title} {Effect of the spin-orbit
  interaction on the band structure and conductance of quasi-one-dimensional
  systems},\ }\href {https://doi.org/10.1103/PhysRevB.60.14272} {\bibfield
  {journal} {\bibinfo  {journal} {Phys. Rev. B}\ }\textbf {\bibinfo {volume}
  {60}},\ \bibinfo {pages} {14272} (\bibinfo {year} {1999})}\BibitemShut
  {NoStop}%
\bibitem [{\citenamefont {Verresen}\ \emph {et~al.}(2018)\citenamefont
  {Verresen}, \citenamefont {Jones},\ and\ \citenamefont
  {Pollmann}}]{Verresen2018}%
  \BibitemOpen
  \bibfield  {author} {\bibinfo {author} {\bibfnamefont {R.}~\bibnamefont
  {Verresen}}, \bibinfo {author} {\bibfnamefont {N.~G.}\ \bibnamefont
  {Jones}},\ and\ \bibinfo {author} {\bibfnamefont {F.}~\bibnamefont
  {Pollmann}},\ }\bibfield  {title} {\bibinfo {title} {Topology and edge modes
  in quantum critical chains},\ }\href
  {https://doi.org/10.1103/PhysRevLett.120.057001} {\bibfield  {journal}
  {\bibinfo  {journal} {Phys. Rev. Lett.}\ }\textbf {\bibinfo {volume} {120}},\
  \bibinfo {pages} {057001} (\bibinfo {year} {2018})}\BibitemShut {NoStop}%
\bibitem [{\citenamefont {Delfino}\ \emph {et~al.}(1996)\citenamefont
  {Delfino}, \citenamefont {Mussardo},\ and\ \citenamefont
  {Simonetti}}]{Delfino1996}%
  \BibitemOpen
  \bibfield  {author} {\bibinfo {author} {\bibfnamefont {G.}~\bibnamefont
  {Delfino}}, \bibinfo {author} {\bibfnamefont {G.}~\bibnamefont {Mussardo}},\
  and\ \bibinfo {author} {\bibfnamefont {P.}~\bibnamefont {Simonetti}},\
  }\bibfield  {title} {\bibinfo {title} {Non-integrable quantum field theories
  as perturbations of certain integrable models},\ }\href
  {https://doi.org/https://doi.org/10.1016/0550-3213(96)00265-9} {\bibfield
  {journal} {\bibinfo  {journal} {Nuclear Physics B}\ }\textbf {\bibinfo
  {volume} {473}},\ \bibinfo {pages} {469} (\bibinfo {year}
  {1996})}\BibitemShut {NoStop}%
\bibitem [{\citenamefont {Ludwig}(2015)}]{Ludwig2016}%
  \BibitemOpen
  \bibfield  {author} {\bibinfo {author} {\bibfnamefont {A.~W.~W.}\
  \bibnamefont {Ludwig}},\ }\bibfield  {title} {\bibinfo {title} {Topological
  phases: classification of topological insulators and superconductors of
  non-interacting fermions, and beyond},\ }\href
  {https://doi.org/10.1088/0031-8949/2015/T168/014001} {\bibfield  {journal}
  {\bibinfo  {journal} {Physica Scripta}\ }\textbf {\bibinfo {volume} {2016}},\
  \bibinfo {pages} {014001} (\bibinfo {year} {2015})}\BibitemShut {NoStop}%
\bibitem [{Note2()}]{Note2}%
  \BibitemOpen
  \bibinfo {note} {Namely, we rotated the band left-moving and right-moving
  fermionic operators (\ref {LR_bosonisation}) such that $\sigma _z \rightarrow
  \sigma _x$, we also shifted the bosonic fields $\protect \sqrt {2\pi }\phi
  _{s,c} \rightarrow \protect \sqrt {2\pi }\phi _{s,c} + \pi /2$. With this
  shift the single-particle terms take the form studied earlier in the Section
  \ref {Sec:uncoupledSSH}}\BibitemShut {NoStop}%
\end{thebibliography}%
\end{document}